\documentclass[a4paper,11pt]{article}

\pdfoutput=1
\usepackage{comment}
\usepackage{fixltx2e}
\usepackage[table,xcdraw]{xcolor}
\usepackage{jcappub} 
\usepackage{float} 
\usepackage{bm}
\usepackage{stmaryrd}
\usepackage[normalem]{ulem}
\usepackage{amsmath}

\def\be{\begin{equation}}
\def\ee{\end{equation}}
\def\bea{\begin{eqnarray}}
\def\eea{\end{eqnarray}}
\newcommand{\vs}{\nonumber\\}
\def\ba#1\ea{\begin{align}#1\end{align}}
\def\bg#1\eg{\begin{gather}#1\end{gather}}

\newcommand{\s}{\sigma}

\newcommand{\refeq}[1]{Eq.~(\ref{eq:#1})}          
\newcommand{\refeqs}[2]{Eqs.~(\ref{eq:#1})--(\ref{eq:#2})}          
          
\newcommand{\reffig}[1]{Fig.~\ref{fig:#1}}          
          
\newcommand{\reftab}[1]{Tab.~\ref{tab:#1}}          
\newcommand{\refsec}[1]{Sec.~\ref{sec:#1}}          
\newcommand{\refapp}[1]{App.~\ref{app:#1}}
          
\newcommand{\shellPT}[1]{{(#1)_{\rm shell}}}

\def\Pshell{P_{\rm shell}}
\def\PshellPsi{P_{\rm shell}^{\Psi}}
\def\Plin{P_{\rm L}}
\def\PlinL{P_{\rm L,\Lambda}}

\def\vphi{\varphi}
\def\vphiG{\varphi_{\rm G}}

\renewcommand{\v}[1]{\bm{#1}}

\renewcommand{\emph}[1]{\textit{#1}}

\newcommand{\vx}{\v{x}}

\newcommand{\vk}{\v{k}}
\newcommand{\vq}{\v{q}}
\newcommand{\vdisp}{\v{s}}
\newcommand{\vp}{\v{p}}

\newcommand{\<}{\langle}
\renewcommand{\>}{\rangle}
\DeclareMathOperator{\Tr}{Tr}

\renewcommand{\d}{\delta}
\newcommand{\D}{\Delta}

\newcommand{\Z}{\mathcal{Z}}
\newcommand{\G}{\mathcal{G}}

\newcommand{\Shell}{\mathcal{S}}

\newcommand{\fnl}{f_{\rm NL}}
\newcommand{\Knl}{K_{\rm NL}}

\def\L{\Lambda}

\def\dirac{\delta_{\rm D}}
\def\diracpi{\hat{\delta}_{\rm D}}

\def\leps{\lambda}

\def\dlin{\delta^{(1)}}
\def\dlinshell{\dlin_{\rm shell}}

\def\L{\Lambda}

\def\P{\mathcal{P}}

\def\O{\mathcal{O}}
\def\Del{\mathcal{D}}

\newcommand{\perm}[1]{ \expandafter\ifstrempty\expandafter{#1} {\mbox{perm.}} {\mbox{$#1$ perm.}} }

\makeatletter
\newlength{\apb@width}
\newcommand{\autoparbox}[2][c]{\settowidth{\apb@width}{#2}\parbox[#1]{\apb@width}{#2}}
\newcommand{\includegraphicsbox}[2][]{\autoparbox{\includegraphics[#1]{#2}}}
\makeatother

\def\dirac{\delta_{\rm D}}
\def\diracpi{\hat{\delta}_{\rm D}}

\renewcommand{\comment}[1]{}

\usepackage[T1]{fontenc}
\usepackage{graphbox}
\usepackage{subfig}
\usepackage{booktabs}
\subheader{TUM-HEP- 1510/24}
\title{The Renormalization Group for Large-Scale Structure: Primordial non-Gaussianities}
\usepackage{amsmath}
\author[a, b]{Charalampos Nikolis,}
\author[c]{Henrique Rubira,}
\author[b]{Fabian Schmidt}
\affiliation[a]{Fakultät f\"{u}r Physik, Ludwig-Maximilians-Universit\"at M\"unchen,\\ Geschwister-Scholl-Platz 1, D-80539 M\"unchen, Germany}
\affiliation[b]{Max-Planck-Institut f\"{u}r Astrophysik,\\ 
Karl-Schwarzschild-Str. 1, 85748 Garching, Germany}
\affiliation[c]{Physik Department T31, Technische Universit\"at M\"unchen,\\
James-Franck-Stra{\ss}e 1, D-85748 Garching, Germany}

\emailAdd{cnikolis@mpa-garching.mpg.de}
\emailAdd{henrique.rubira@tum.de}
\emailAdd{fabians@mpa-garching.mpg.de}

\abstract{The renormalization group for large-scale structure (RG-LSS) describes the evolution of galaxy bias and stochastic parameters as a function of the cutoff $\Lambda$.  In this work, we introduce interaction vertices that describe primordial non-Gaussianity into the Wilson-Polchinski framework, thereby extending the free theory to the interacting case.
The presence of these interactions forces us to include new operators and bias coefficients to the bias expansion to ensure closure under renormalization.
We recover the previously-derived ``scale-dependent bias'' contributions, as well as a new (subdominant) stochastic contribution.
We derive the renormalization group equations governing the RG-LSS for a large class of interactions which account for vertices at linear order in $f_{\rm NL}$ that parametrize interacting scalar and massive spinning fields during inflation.
Solving the RG equations, we show the evolution of the non-Gaussian contributions to galaxy clustering as a function of scale.
}


\begin{document}

\maketitle
\flushbottom

%%%%%%%%%%%%%%%%%%%%%%%%%%%%%%%%%%%%%%%%%%%%%%%%%%%%%%%%%%%%%%%%%%%%%%%%%%%
%%%%%%%%%%%%%%%%%%%%%%%%%%%%%%%%%%%%%%%%%%%%%%%%%%%%%%%%%%%%%%%%%%%%%%%%%%%
\section{Introduction}\label{sec:intro}

Primordial non-Gaussianities (PNG) in the initial conditions of the Universe are a direct probe of distinct inflationary scenarios \cite{Guth:1980zm,Linde:1981mu,Guth:1982ec,PhysRevD.42.3936,maldacena:2003,Cheung:2007st,Sohn:2024xzd}. 
Observations from the cosmic microwave background (CMB) have put strong constraints on the magnitude of the non-Gaussianities \cite{Planck:2018jri}.
Interest in PNG has been renewed due to recent progress in calculating cosmological correlators via the ``cosmological bootstrap'' program \cite{arkanihamed2019cosmological} (see \cite{baumann2022snowmass} and references therein).

The large-scale structure (LSS) of the Universe provides an important probe to search for PNG on complementary scales to those probed by the CMB. 
Based on the framework of the effective field theory of LSS (EFTofLSS) \cite{Baumann:2010tm,carrasco/etal:2012,Carroll:2013oxa,abolhasani/etal,Konstandin:2019bay,DAmico:2022ukl} (for applications of PNG in the context of EFTofLSS see \cite{baldauf2016lss,Welling_2016,Sprenger_2019,Karagiannis_2018,deputter2018primordial, Assassi:2015EFT,Vasudevan_2019,DAmico:2022gki,Cabass:2024wob,Cabass:2022wjy,Cabass:2022ymb}), we can expand the galaxy density as a sum of all operators $O$ allowed by the symmetries of non-relativistic General Relativity, and assign to each of them a bias (or Wilson) coefficient $b_O$ \cite{Desjacques:2016bnm}
\begin{equation}
\label{eq:bias-expansion}
    \d_g(\vk,\tau)=\sum_O b_O^\L(\tau) O[\dlin_\L](\vk,\tau)\,.
\end{equation}
This relation implies that physical processes inherently non-local in time such as galaxy formation \cite{MSZ,senatore:2014} can be factorized in the time-dependent bias coefficients. This is made possible by properties of the perturbation theory for matter and the gravitational potential at each order \cite{matsubara:2015,Schmidt:2020ovm}. 

A crucial point of the bias expansion \refeq{bias-expansion} is that it contains {\it composite} operators, meaning that it suffers from small-scale Fourier modes back-reacting onto large scales  \cite{McDonald:2006mx,Assassi2014,Assassi:2015fma}. Therefore, one has to smooth it with a cutoff $\L$ that regularizes the theory.
A central piece of the EFTofLSS and the bias expansion above is that it allows us to treat the dark matter and galaxy density perturbatively, while allowing for a consistent way to parametrize non-linear non-perturbative modes in the theory by imposing this cutoff $\L$ that smooths out small-scale modes. %The physical information from small-scale modes is kept via additional counter-terms.

In the presence of PNG, the bias expansion needs a non-trivial generalisation \cite{2008PhRvD..77l3514D,2008ApJ...677L..77M,2008JCAP...08..031S,Schmidt:2010gw}. The reason is that the horizon-scale dynamics during inflation that generated PNG correspond to large-scale correlations in the late universe that cannot be generated by subhorizon gravitational or non-gravitational forces.
  Since the statistics of galaxies which we want to describe via the EFT involve much larger scales than the scales that govern local galaxy formation, the additional galaxy bias terms depend on specific kinematic regimes of primordial interactions, in particular those where one of the wavelengths is soft compared to the others, commonly called the {\it squeezed limit}. The unique new bias operators
  offer the opportunity to robustly detect signatures from different models of primordial interactions. However, these signatures are multiplied by additional Wilson coefficients $b_{O_{\rm NG}}$.
  Therefore, there is strong motivation to put independent constraints on the values of these coefficients, as they are largely degenerate with the amplitude of PNG \cite{2008JCAP...08..031S,2022JCAP...11..013B}.
  Moreover, the values of these bias coefficients are found to depend sensitively on the tracer population considerered \cite{2010JCAP...07..013R,2020JCAP...12..013B,2023JCAP...01..023L}, at least for the leading bias parameter introduced by local-type PNG. 
  Our goal here in particular is to derive how the $b_{O_{\rm NG}}$ run from one scale $\Lambda$ to another. While this does not tell us about the fully non-perturbative physics setting the absolute scale of the bias coefficients, it can be compared with measurements of the bias coefficients as a function of scale (for example in simulations), and thus provides another validation test of the theory.

Different renormalization schemes have been proposed for the bias expansion. The most used option so far is to renormalize the bias operators at the $n$-point function level \cite{Assassi2014,Assassi:2015fma}, defining the (renormalized) bias operators deep in the IR via the corresponding $n$-point function for which they appear at tree level. 
In this work, we follow the path-integral approach of \cite{Carroll:2013oxa,Rubira:2023vzw,Rubira:2024tea}, keeping the cutoff $\L$ explicitly. We generalize the renormalization of bias parameters of \cite{Rubira:2023vzw,Rubira:2024tea} to non-Gaussian initial conditions. We use the Wilson-Polchinski approach to derive the renormalization group (RG) equations for the bias parameters including PNG. The RG flow approach in principle allows for the resummation of Feynman diagrams, and therefore there are some grounds to expect that additional information can be extracted mainly when adding PNG interactions.

Our Renormalization Group Equations (RGE)  describe a large class of cubic interactions during inflation. The equations are general enough to depend {\it only} on the characteristic squeezed limit of the non-Gaussian interaction $\sim k^\D$. This already enables us to calculate the RG flow for different known isotropic (spin zero) and non-isotropic (spin 2) shapes of PNG. More importantly,  novel calculations for bispectra (e.g from the bootstrap programme) could lead to shapes of non-Gaussianity with different characteristic soft scalings than previously known. These shapes can directly be implemented in the RG-flow we present in this work. Thus, the effect of new PNG shapes in the galaxy statistics can easily be retrieved from our general RG equations.\footnote{For other work on the different regularization schemes for PNG see \cite{Patrone:2023cqe}.}

The outline of the paper is the following. In the rest of this section we set the notation. In \refsec{nongauss_theory} we introduce PNG, generalize the operator basis to include them and derive the non-Gaussian partition function. In \refsec{running} we derive the running of the bias parameters via the Wilson-Polchinski framework. We present the results in \refsec{results} and conclude in \refsec{disc}. We have four appendices: \refapp{couplingtoJ} is dedicated to an alternative formulation of PNG in the partition function, coupling them directly to the current $J$; \refapp{pkbk} for examples on how to derive $n$-point functions from the partition function; \refapp{cubic} discussing the interacting PNG vertex and \refapp{SOeval} for explicit evaluations of the bias renormalization.

For numerical results we use the transfer function as well as the linear matter power spectrum computed by \texttt{CLASS} \cite{Blas:2011rf} with Planck 2018 Euclidian $\L$CDM cosmology \cite{Planck:2018vyg}.

%%%%%%%%%%%%%%%%%%%%%%%%%%%%%%%%%%%%%%%%%%%%%%%%%%%%%%%%%%%%%%%%%%%%%%%%%%%
%%%%%%%%%%%%%%%%%%%%%%%%%%%%%%%%%%%%%%%%%%%%%%%%%%%%%%%%%%%%%%%%%%%%%%%%%%%
\subsection*{Notation and conventions}\label{sec:notation}
We review now the notation adopted in \cite{Rubira:2023vzw}.

\paragraph{Momentum conventions.} We use $\vk,\vp$ for momenta variables where with bold letters we denote Euclidean vectors. We adopt a shorthand notation $\vp_{1 \dots n} = \vp_{1} + \dots + \vp_{n}$ for the sum of vectors.
The momentum-space integrals are written as
\be
\int_{\vp_1,\dots,\vp_n} = \int \frac{d^3p_1}{(2\pi)^3} \dots \int \frac{d^3p_n}{(2\pi)^3} \,.
\ee
We keep the variable $\vq$ for the Lagrangian position.

\paragraph{Cutoff filters.}
We can coarse-grain a field $f$ by applying a sharp filter $W$ on a length scale $1/\L$. In Fourier space this is simply a product
    \be
        f_{\L}(\vk) = \,W_\L(\vk) f(\vk)\,.
    \ee

\paragraph{The kernels.}
We consider three types of kernels and vertex diagrams (beyond the non-Gaussian kernel presented below). 
First, we write the operators $O$ of the bias expansion \refeq{bias-expansion} in terms of the dark matter overdensity field $\d(\vp_i)$ as
\be
\label{eq:oper-convolution}
O[\dlin_\L](\vk)=\int_{\vp_1,...,\vp_n} \diracpi(\vk-\vp_{12\dots n})S_{O}(\vp_1,...,\vp_n)\d(\vp_1)...\d(\vp_n) \,,
\ee
with $\diracpi=(2\pi)^3\dirac$ for the Dirac delta function. 
These operators can be classified according to their perturbative order, so in this case, we have $O=O^{[n]}$. 
Second, we expand the matter overdensity with the usual Eulerian kernels $F_{n}$ \cite{lssreview}
\ba
\label{eq:matter-expansion}
&\d[\dlin_\L](\vk)=\sum_{n=1}^{\infty} \int_{\vp_1,...,\vp_n} \diracpi(\vk-\vp_{12..n})F_n(\vp_1,...,\vp_n)\dlin_\L(\vp_1)...\dlin_\L(\vp_n)\;,
\ea
where $\dlin_\L$ is the smoothed linear density field. 
The third type of vertex $K$ represents a union of the former two
    \be
    O(\vk) = \sum_{\ell={\rm order}(O)}^\infty\int_{\vp_1,\ldots,\vp_\ell} \diracpi(\vk-\vp_{1\ldots \ell}) K_O^{(\ell)}(\vp_1,\ldots \vp_\ell) \dlin(\vp_1) \cdots \dlin(\vp_\ell) \,,
    \label{eq:Ok}
    \ee
in which we expanded the operators in terms of linear legs. Notice it is possible to derive the $K^\ell_O$ by inserting \refeq{matter-expansion} into \refeq{oper-convolution}. Moreover, we define $K_O = \sum_{\ell={\rm order}(O)}^\infty K^{(\ell)}_O$ and we assume $K^{(\ell)}_O$ to be symmetrized in its arguments. 
The three types of vertices are represented by 
    \be
    \raisebox{-0.0cm}{\includegraphicsbox[scale=1]{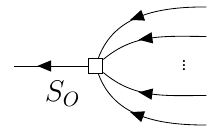}}
    \raisebox{-0.0cm}{\includegraphicsbox[scale=1]{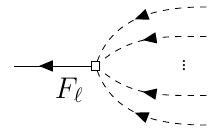}}
    \raisebox{-0.0cm}{\includegraphicsbox[scale=1]{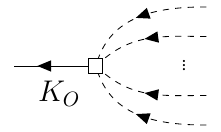}} \,,
    \quad
    \ee
with the differentiation between large boxes and the small boxes, and also the dashed lines representing linear legs.

\paragraph{Power spectra and field variance.}
From the linear matter overdensity, we can define its power spectrum (i.e., the linear propagator)
\be  \label{eq:Plindef}
P_{\rm L}^\L (k) =\left<\dlin_\L(\vk)\dlin_\L(\vk')\right>' = \raisebox{-0.0cm}{\includegraphicsbox[scale=1.0]{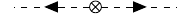}} \,,
\ee
where the prime notation in $\langle \dots \rangle'$ indicates that the  momentum-conservation factor $\diracpi$ has been dropped. Notice that we imposed momentum restrictions on the propagators, cutting them off at $\L$. If we shift the cutoff to $\Lambda'=\Lambda+\lambda$, we can separate the contributions into one with support up to wavenumber $\L$ and one with support in the {\it momentum shell} $k\in [\L,\L+\lambda]$,
\ba \label{eq:Pshelldef}
\Pshell (k) &=\left<\dlinshell(\vk)\dlinshell(\vk')\right>' = \raisebox{-0.0cm}{\includegraphicsbox[scale=1.0]{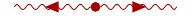}} \,.
\ea
Furthermore, from the matter linear propagator, we can calculate the matter overdensity variance 
\ba \label{eq:var}
\sigma^2_{\L}=\int_{\vp}^\L \left<\dlin_\L(\vp)\dlin_\L(\vp')\right>' =  \int_{\vp}^\L \Plin^\L(p)\,.
\ea

\paragraph{Gaussian partition function.}
For the case of Gaussian initial conditions we can generate all $n$-point functions of matter ($m$) and galaxies ($g$) from a partition function
\ba
\Z_{\rm G}[J_{ m,\L},J_{ g,\L}]=\int\Del\dlin_\L \P[\dlin_\L]\exp\left\{{\int_{\vk}J_{ m,\L}(\vk)\delta[\dlin_\L](-\vk)+\int_{\vk}J_{g,\L}(\vk)\delta_{g}[\dlin_\L](-\vk)}\right\}\;, \label{eq:gausspartition_withmatter}
\ea
with the full nonlinear solutions for the dark matter density field and the galaxy density fields coupled to their respective currents $J_{m,\L},\;J_{g,\L}$.
This partition function generates the matter, galaxy and cross $n$-point correlations (see discussion in \refapp{pkbk}). 
Here we have a path-integral over all initial configurations of the linear overdensity $\dlin_\L$ with support over wavenumbers up to $\L$. $\P[\dlin_\L]$ refers to the Gaussian measure:
\be
\P[\dlin_\L]=\left(\prod_{\vk}^\L 2\pi \PlinL(k)\right)^{-1/2}\exp{\left[-\frac{1}{2}\int_{\vk}^\L \frac{|\dlin_\L|^2}{\PlinL(k)}\right]}\;, \label{eq:Gaussmeasure}
\ee
and $\d[\dlin_\L]$ is the perturbative expansion of the dark-matter \refeq{matter-expansion}. For other constructions of the large-scale structure partition function, see \cite{matarrese/pietroni,floerchinger/etal,Blas:2015qsi,Blas:2016sfa,Cabass:2019lqx}.

This enables us to reproduce all the $n$-point statistics for the matter and the galaxies by taking derivatives with respect to the matter or galaxy currents $J_{m}\;,J_{g}$ (see e.g. \cite{Cabass:2019lqx}).
For example, the Fourier-space $m$-point correlation function of a given biased tracer is given by 
\ba
\langle \d_g(\vk_1) \dots \d_g(\vk_m) \rangle &= \int \Del\dlin_\L \P[\dlin_\L] \, \d_g(\vk_1) \dots \d_g(\vk_m)\,,
\ea
which can be obtained by taking derivatives of the partition function w.r.t.~$J_{g,\L}$ 
\be
\langle \d_g(\vk_1) \dots \d_g(\vk_m) \rangle =  \frac{1}{\Z[0]} \frac{\delta^m \Z }{\delta J_{g,\L}(\vk_1) \dots \delta J_{g,\L}(\vk_m)}\Big|_{J_{m,\L}=0,J_{g,\L}=0}\;,
\ee
and we can also have cross-correlations, like the matter-galaxy cross power spectrum
\be
\langle \d(\vk_1)\d_g(\vk_2) \rangle =   \frac{1}{\Z[0]}\frac{\delta^2\Z }{\delta J_{m,\L}(\vk_1)\delta J_{g,\L}(\vk_2)}\Big|_{J_{m,\L}=0,J_{g,\L}=0} \,.
\ee
We discuss some basic examples in \refapp{pkbk}.

\paragraph{List of Gaussian operators.}
The basis of of Gaussian operators is given by
 \ba
 \label{eq:Gauss-ops}
    \mbox{First order:}&\quad \d;\vs
    \mbox{Second order:}&\quad \d^2,\, \G_2;\; \vs
    \mbox{Third order:}&\quad \d^3,\;  \d\G_2,\; \Gamma_3,\;  \G_3 \,,
    \ea
with the Galileon operators defined as
    \bea
    \G_2( \Phi_g) &\equiv& (\nabla_i\nabla_j\Phi_g)^2 - (\nabla^2  \Phi_g)^2 \label{eq:galileon2}\,, \\
    \G_3( \Phi_g) &\equiv& -\frac{1}{2}\left[2\nabla_i\nabla_j  \Phi_g \nabla^j\nabla_k  \Phi_g\nabla^k\nabla^i  \Phi_g + (\nabla^2  \Phi_g)^3-3(\nabla_{i}\nabla_{j} \Phi_g)^2\nabla^2  \Phi_g\right]\,, \label{eq:galileon3}
    \eea
    with $\Phi_g \equiv \nabla^{-2}\d$ being the scaled gravitational potential. We also have
    \bea
    \Gamma_3( \Phi_g,\Phi_v) \equiv \G_2(\Phi_g) - \G_2(\Phi_v) \,,
    \eea
    with $\Phi_v \equiv \nabla^{-2} \bm{\nabla}\cdot\v{v}$, the  velocity potential. 
Moreover, we define as 
\be
\s^2_{\vk_1,\vk_2}  = \left( \vk_1 \cdot \vk_2/k_1k_2 \right)^2 - 1 \,,
\ee 
the Fourier-transformed coefficient of $\G_2$, such that in the notation of \refeq{oper-convolution}, we can clearly see that $\s^2_{\vk_1,\vk_2}\equiv S_{\G_2}$.
Finally, we normalize the operators as
    \be \label{eq:Onorm}
    O(\vk) \to O(\vk) - \langle O(\vk=0)\rangle \,,
    \ee
to avoid tadpole contributions.

%%%%%%%%%%%%%%%%%%%%%%%%%%%%%%%%%%%%%%%%%%%%%%%%%%%%%%%%%%%%%%%%%%%%%%%%%%%

%%%%%%%%%%%%%%%%%%%%%%%%%%%%%%%%%%%%%%%%%%%%%%%%%%%%%%%%%%%%%%%%%%%%%%%%%%%
\section{The generalized galaxy-bias partition function with non-Gaussianity} \label{sec:nongauss_theory}
%%%%%%%%%%%%%%%%%%%%%%%%%%%%%%%%%%%%%%%%%%%%%%%%%%%%%%%%%%%%%%%%%%%%%%%%%%%

In this section we start by discussing the connection of the partition function formalism for the bias expansion to the usual textbook interactions in quantum field theory (QFT). Next, we present how to introduce non-Gaussianities in cosmology and also fix the notation for the new interacting vertex. Later we describe how to modify the galaxy bias expansion to include PNG. Finally, we derive the partition function for the galaxy bias including non-Gaussianities. 

\subsection{Parallelism with QFT}

We begin with a QFT example, that will serve as basis for our later analysis. Consider a real scalar field with a cubic interaction of the type
\ba
\mathcal{L} = -\frac12 \phi (\square + m^2)\phi - \frac{g}{3!}\phi^3  = \mathcal{L}_0 + \mathcal{L}_{\rm int}\,.
\ea
We define the \emph{free} partition function as
\ba
\Z_0[J] &= \int \Del \phi \, e^{iS_0[\phi] + i\int d^4x J\phi } \equiv \int \Del \phi \P[\phi]\, e^{i\int d^4x J\phi } \,,
\ea
where $\P[\phi]$ denotes the free (Gaussian) measure of the path-integral.
The textbook approach of handling QFT interactions is to set up a perturbative expansion of the interaction vertices, and find quantum corrections to the free theory, that will induce deviations from the free-Gaussian case
\ba \label{eq:ZQFT}
\Z[J] &= \int \Del \phi \, e^{iS_0[\phi] + iS_{\rm int}[\phi] + i\int d^4x J\phi} = \int \Del \phi \P[\phi] \, e^{+ i\int d^4x J\phi}
\vs
&\quad \times \left( 1 - \frac{ig}{3!}\int_{\vx_1}\phi^3(\vx_1) +\frac{1}{2} \left(\frac{-ig}{3!}\right)^2 \int_{\vx_1, \vx_2}\phi^3(\vx_1)\phi^3(\vx_2) +  \dots	\right) \,.
\ea
An important difference between the (usual) QFT case and the EFTofLSS, to be explored below, is the canonical linear coupling of the current to the field. 

The $n$-point function for the interacting theory are calculated as the perturbative expansion of the interaction exponential
\ba
&\langle \phi(\vk_1) \dots \phi(\vk_n) \rangle = \frac{1}{\Z[0]} \frac{\delta^n \Z }{\delta J_i \dots \delta J_n}\Big|_{J=0} = \int \Del \phi \P[\phi] \, \phi(\vk_1) \dots \phi(\vk_n) \, e^{iS_{\rm int}[\phi]}\vs &=\int \Del \phi \P[\phi] \, \phi(\vk_1) \dots \phi(\vk_n)\left(1+iS_{\rm int}[\phi]+\frac{i^2}{2}S_{\rm int}[\phi]^2 \dots \right) \,.
\ea
Corrections from the Gaussian free field are sourced by the interactions in the Lagrangian, or in other words by the $J^0$ term in the partition function. Moreover, a linear coupling of the current with the field is added as a tool to calculate $n$-point interactions.

An alternative way of formulating interactions is to couple the current to nonlinear operators of the (free or interacting) field. These are termed \textit{composite operators}. As we show now, in general one can switch between the two formulations via a nonlinear field redefinition, where we restrict ourselves to linear order in $g$. 
The classical equation of motion
\ba
\left(\square+m^2\right)\phi=-\frac{g}{2}\phi^2 \,,
\ea
admits a formal solution of the form
\ba
\phi_{\rm cl}=\phi-i\frac{g}{2}\int d^4x' G\left(x,x'\right)\phi_{\rm cl}^2(x')=\phi-i\frac{g}{2}\int d^4x' G\left(x,x'\right)\phi^2(x')+\O(g^2) \,,
\ea
with $G(x,x')$ the Green's function of the free equation of motion,
\ba
\left(\square+m^2\right)G(x,x')=-i\dirac(x-x')\,,
\ea
and $\phi$ the free-field solution. This motivates us to do the following field redefinition 
\ba \label{eq:fieldredef}
\Tilde{\phi}=\phi+i\frac{g}{3!}\int d^4x' G\left(x,x'\right)\phi^2(x') \,,
\ea
then
\ba
\phi=\Tilde{\phi}-i\frac{g}{3!}\int d^4x' G\left(x,x'\right)\Tilde{\phi}^2(x')+\O(g^2) \,.
\ea
Plugging it back to the Lagrangian we get
\ba
\mathcal{L}=-\frac{1}{2}\Tilde{\phi}\left(\square+m^2\right)\Tilde{\phi}+J\Tilde{\phi}\,,
\ea
where the interaction term has been cancelled and the current is coupled to the full solution of the field. The composite-operator formulation is useful when no fundamental interaction Lagrangian is known, but one can still write down the equations of motion.

For the case of LSS, the tracer field $\d_g[\dlin]$ in \refeq{bias-expansion} is the main object from which we want to calculate correlators and therefore have a partition function for. We differentiate between two types of interactions that appear for matter tracers when constructing the partition function: the interactions present in the initial conditions due to inflation [imprinted in $\dlin$ as e.g. \refeq{nonGoverdensity} below], and
the non-linearities introduced by the gravitational evolution [both in the operator definition \refeq{oper-convolution} and in the $F$ kernels of \refeq{matter-expansion}].

For the first type of interactions, we can directly formulate the inflationary initial conditions as nonlocal interactions of $\dlin$
[see e.g.~\refeq{nonGoverdensity} below]. 
Therefore, it is convenient to include those interactions as new vertices in the $J$-independent (interaction) term, following the canonical steps outlined above leading to \refeq{ZQFT}.

For the interactions from nonlinear evolution, on the other hand,
the composite-operator formulation is useful since the forward evolution of the density field can be described at the level of equations of motion via the EFTofLSS framework (whereas no fundamental Lagrangian is known for this problem). We then couple the current to the (classical) solution of the equations of motion.\footnote{In general, the renormalization of composite operators requires contributions to the effective action that are higher-order in the current \cite{SHORE199185,Polonyi_2001}; for a general treatment of composite operators and renormalization methods see \cite{Zinn-Justin:1989rgp}, and for the application to LSS see \cite{Rubira:2024tea}. 
  If these have a local form, such as $\int_{\vx} J^2(\vx) \cdots$, they are known
  as \emph{contact terms}. 
  In the case of the partition function for LSS, terms coupled to higher powers of the current are in one-to-one correspondence with stochastic (noise) contributions to tracer $n$-point functions \cite{Cabass:2019lqx,Cabass:2020nwf,Rubira:2024tea}.}

In summary, for constructing the partition function we {\it add the non-linearities from non-Gaussian initial condition in the interacting $J$-independent term, while the non-linearities from the later evolution couple to interacting $J$-dependent terms}.

As a final note, one can also consider a formulation where both types of interactions are represented in terms of composite operators; that is, moving the non-Gaussian initial conditions from the Lagrangian to the current coupling via a field redefinition such as \refeq{fieldredef}. We provide one example for this in \refapp{couplingtoJ} and \refapp{Sint2shell}. In fact, such reformulation in general offers independent physical insights.

\subsection{Primordial non-Gaussianities}

We now introduce non-Gaussianities in the context of cosmology. In the usual approach \cite{Schmidt:2010gw,Assassi:2015fma}, weakly non-Gaussian initial conditions are defined by expanding around a Gaussian field $\varphi_G$,
\ba  \label{eq:PNGforphi}
	\vphi (\vk) &= {\vphiG} (\vk) + \fnl \int_{\vp_1, \vp_2} \diracpi(\vk-\vp_{12}) \Knl(\vp_1, \vp_2){\vphiG} (\vp_1){\vphiG} (\vp_2)+\O[f_{\rm NL}^2,g_{\rm NL}]\,,
\ea
with $\Knl$ the PNG kernel that depends on the structure of interactions during inflation.
Here, $\vphi$ is the superhorizon gravitational potential during matter domination, which is directly related to the curvature perturbation in comoving gauge $\mathcal{R}$ by $\vphi = (3/5) \mathcal{R}$.
This leads to a non-vanishing bispectrum for the primordial fluctuations,
\bea
\label{eq:non-vanishingB}
\left<\vphi(\vk_1)\vphi(\vk_2)\vphi(\vk_3)\right>' \equiv B_\vphi\left(k_1,k_2,k_3\right)= 2\fnl\Knl(\vk_1,\vk_2)P_{\vphi}(k_1)P_{\vphi}(k_2)+2\,\text{perms}\,,
\eea
where two cyclic permutations of the three momenta $k_1,k_2,k_3$ shall be considered.
An important limit for galaxy clustering is the {\it squeezed limit} defined as
\ba
&B_\vphi\left(k_\ell,|\vk_s-\frac12\vk_\ell|,|\vk_s+\frac12\vk_\ell|\right)\xrightarrow{k_\ell\ll k_s}
\vs
&\hspace{2cm}2\fnl\Big[\Knl(\vk_s,\vk_\ell)P_\vphi(k_s)P_\vphi(k_\ell)+\Knl(-\vk_s,\vk_\ell)P_\vphi(k_s)P_\vphi(k_\ell)\Big]
\vs
&\hspace{2cm}+2\fnl\Knl(-\vk_s,\vk_s)\left[P_\vphi(k_s)\right]^2+\O\left(\frac{k^2_\ell}{k^2_s}\right) \,.
\ea
The first two terms in the RHS are what we usually refer to as the squeezed limit of the kernel, which can be uniquely determined by the bispectrum. We refer to those terms as the {\it scattering interaction}, for reasons that will become clear in \refsec{interaction}. The last term is what we refer to as the {\it annihilation interaction} which will contribute to the stochastic bias operators.

We can immediately translate the relation \refeq{PNGforphi} to one involving the linearly evolved matter overdensity
\begin{equation}\label{eq:nonGoverdensity}
\boxed{
\begin{aligned}
\dlin(\vk)&=\dlin_{\rm G}(\vk)+\fnl\int_{\vp_1,\vp_2}\diracpi(\vk-\vp_{12}) \Knl(\vp_1,\vp_2)\frac{M(|\vp_1+\vp_2|)}{M(p_1)M(p_2)}\dlin_{\rm G}(\vp_1)\dlin_{\rm G}(\vp_2)
\\
&\hspace{5cm}+\O[f_{\rm NL}^2,g_{\rm NL}],
\end{aligned}
}
\end{equation}
with $g_{\rm NL}$ the quartic interaction and
\be 
\dlin(\vk, \tau)=M(k,\tau)\varphi(\vk)\,,
\ee
and
\be
M(k,\tau)=\frac{2}{3}\frac{k^2T(k)D(\tau)}{H_0^2 \Omega_{m,0}}\,,
\label{eq:Mdef}
\ee
where $T(k)$ is the {\it transfer function} describing linear evolution through horizon crossing, normalized such that $T(k)\to 1$ for $k\ll k_{\rm eq}$, $H_0$ is the value of the Hubble constant today, $\Omega_{m,0}$ the fractional matter density and $D$ the growth factor. 
For simplicity, we have dropped in \refeq{nonGoverdensity} the time argument $\tau$. For the rest of the text it should be kept in mind that the transfer function should be evaluated at the time of observation $\tau$.
Finally, $\dlin_{\rm G}$ is the Gaussian overdensity acquired from acting with $M(k,\tau)$ on the Gaussian counterpart of the primordial fluctuation $\varphi_{\rm G}$.

Assuming that the initial conditions are homogeneous and isotropic, we can deduce that $\Knl(\vk_1,\vk_2)$ depends only on $k_1 = |\vk_1|$, $k_2 = |\vk_2|$ and the angle between the two vectors $\hat{\vk}_1\cdot\hat{\vk}_2$ \cite{Schmidt:2010gw}. This allows us to express the squeezed limit\footnote{A point worth mentioning is that the scaling of the kernel $\Knl$ in a regime different than the squeezed limit (such as for the {\it annihilation interaction}) is ambiguous and depends on the complete form of the kernel. Since we cannot completely define this form from the bispectrum alone, studies investigating the kernel have reversed-engineered an expression \cite{Schmidt:2010gw,Scoccimarro_2012} and find that this term contributes with a non trivial scaling for momentum regions different than the squeezed limit even for the equilateral case, while preserving the squeezed limit behaviour as expected. } ($k_l\ll k_s$) of the kernel with Legendre polynomials of even order $\P_L$. If we make the further assumption of a scale invariant bispectrum, we expect only dimensionless quantities in the expansion, and we have \cite{Assassi:2015fma}
\begin{equation}
\label{eq:squeezed_limit}
\Knl(\vk_s,\vk_l)\xrightarrow{k_l\ll k_s}\sum_{L,i} a_{L,i}\left(\frac{k_l}{k_s}\right)^{\Delta_i} \mathcal{P}_L\left(\hat{\vk}_{s}\cdot\hat{\vk}_l\right) \,,
\end{equation}
where $i$ refers to the different non-Gaussianities, $L$ to the rank (spin) of the multipole expansion, and $a_{L,i}$ are dimensionless constants. In the following, we always focus on one fixed $\Delta_i$ and hence keep a single coefficient $a_{L}$ for the amplitude, which moreover we fix to $a_L = 1$ in our numerical results.
This is a way to express a class of different micro-physical models\footnote{Even though it cannot account for recent non-perturbative models of non-Gaussianity such as \cite{creminelli2024nonperturbative}.} such as:
\begin{itemize}
    \item Spin-0 non-Gaussianity ($L=0$): In that case, we have $\P_0=1$ and we can recover the well known {\it local} $\D=0$ \cite{Komatsu:2001rj}, and the {\it equilateral} ($\D=2$) cases \cite{Alishahiha:2004eh,Cheung:2007st}. In different inflationary scenarios we can have intermediate values of $\D$ \cite{Chen_2010,Green_2013}.
    \item Higher-spin non-Gaussianity ($L=2,4\dots$): This type of non-Gaussianity can be a smoking gun of massive spinning fields during inflation (see e.g \cite{arkanihamed2015cosmological,Lee_2016,Flauger_2017}).
\end{itemize}

%%%%%%%%%
\subsection{The non-Gaussian bias basis} \label{sec:non-G-basis}

The non-Gaussianity in the initial conditions will lead to new terms appearing in the bias expansion, as discussed in Sec.~\ref{sec:running} and was originally pointed out for local non-Gaussianity in \cite{2008PhRvD..77l3514D,McDonald_2008}, and for non-local shapes in \cite{Schmidt:2010gw}. 
In this part we discuss modification in the basis of the operators to account for those new terms. While we appear to introduce these terms {\it ad-hoc} here, we will later show that it is precisely these terms that are generated by the RG flow in the presence of PNG.

\paragraph{Spin-0.} Assuming we have only one source of non-Gaussianity ($i=1$), for the spin-0 ($L=0$) case we add to the bias expansion a linear operator $\Psi$ with the following form
\begin{equation}
\label{eq:Psi}
a_0\fnl\Psi_\L(\vk)=a_0\fnl\left(\frac{k}{k_{\rm NG}}\right)^\Delta \frac{\dlin_\L(\vk)}{M(k)}\equiv a_0\fnl\left(\frac{k}{k_{\rm NG}}\right)^\Delta \varphi_\L(\vk)\;,
\end{equation}
with $k_{\rm NG}$ an arbitrary momentum scale to make the operator dimensionless, so we can keep the bias parameters as dimensionless constants.
If we choose $k_{\rm NG} \sim k_* = R_*^{-1}$ to be of the order of the characteristic spatial length scale governing the formation of the tracer considered, then we expect the associated bias coefficients to be of order 1 \cite{Desjacques:2016bnm} at the scale $\L \to 0$. Different choices of $k_{\rm NG}$ can always be absorbed in a rescaling of the bias coefficients.
For the local ($\Delta = 0$) spin-zero type of non-Gaussianities, we then find 
\begin{equation}
\Psi_\L(\vk)= \varphi_\L(\vk)\;,
\end{equation}
which is usually referred in the literature with the notation $b_\Psi \to b_\varphi$ (or $b_\phi$).
For any $\Delta$, the new basis will consist of operators of the form $\left\{O_{\rm G},O_{\rm G}\Psi\right\}$, where $O_{\rm G}$ refers to the Gaussian basis \refeq{Gauss-ops} and $O_{\rm NG}=O_{\rm G}\Psi$ are the new non-Gaussian operators added, with respective bias coefficients $b_{O_{\rm NG}}$ (recall that we are working to linear order in $\fnl$).
We have now an expanded list of operators due to non-Gaussianity in the bias expansion:

 \ba
 \label{eq:Generalised-Bias-spin-zero}
    \mbox{First order:}&\quad \d, \, \Psi;\vs
    \mbox{Second order:}&\quad \d^2,\, \G_2,\, \d\Psi;\; \vs
    \mbox{Third order:}&\quad \d^3,\, \d^2\Psi,\;  \d\G_2,\; \Psi\G_2,\; \Gamma_3,\;  \G_3 \,.
\ea
For example, the first few terms can be written as \cite{Desjacques:2016bnm} 
\ba
    \d_g[\vk]&=b_\d^\L\d[\dlin_\L](\vk)+ a_0\fnl b_\Psi^\L \Psi_\L [\dlin_\L](\vk)
    \vs
    &\quad +b_{\d^2}^\L\d^2[\dlin_\L](\vk)+b_{\G_2}^\L\G_2[\dlin_\L](\vk)+  a_0\fnl b_{\Psi\d}^\L \Psi_\L[\dlin_\L](\vk)\d[\dlin_\L](\vk)+\dots\;.
    \label{eq:dgPNG}
\ea
Notice that throughout we have assumed that the initial conditions  can be written in terms of a single degree of freedom, highlighted by explicitly writing $\Psi_\L[\dlin_\L]$. 
In multifield inflationary scenarios, the potential perturbations $\vphi$ are in general given by a sum over several independent degrees of freedom. To capture this, our formalism can be straightforwardly generalized to integrate out these fields as well, using additional propagators and interaction terms appropriate to these fields. In that case, there would be additional contributions to \refeq{dgPNG} that are independent of $\dlin_\L$ \cite{2010PhRvD..82d3531T,2013JCAP...05..001B}.

Beyond the linear matter propagators defined in \refeq{Plindef} and \refeq{Pshelldef}, we now also have propagators defined for the new non-Gaussian (linear) operator $\Psi$
\ba 
P^\L_{\rm L,\Psi}(k)&=\left<\dlin_\L(\vk)\Psi_\L(\vk')\right>'=\left(\frac{k}{k_{\rm NG}}\right)^\Delta\frac{1}{M(k)}P^\L_{\rm L}(k) \,,
\vs
\PshellPsi (k)&=\left<\dlinshell(\vk)\Psi_{\rm shell}(\vk')\right>' =\left(\frac{k}{k_{\rm NG}}\right)^\Delta\frac{1}{M(k)}\Pshell(k) \,.
\ea
The propagators including two instances of $\Psi$, such as $\langle\Psi_\L(\vk)\Psi_\L(\vk')\rangle'$, contribute at $\O(\fnl^2)$ and are not included in this work. Including those terms would also force us to include new terms in the interaction of order $g_{\rm NL}$ (quartic interactions). We expect both contributions to be strongly suppressed by the fact that $\varphi\sim 10^{-5}$ \cite{Assassi:2015fma}, although this of course depends on the value of the coefficients $\fnl,g_{\rm NL}$. However, even though terms proportional to $(\fnl)^n$ are ignored in the evaluation of the RG flow, they are generated at the $n$-point function level even from a linear in $\fnl$ partition function and can be significant at sufficiently large scales if one of the modes is of order $aH$ \cite{Desjacques:2016bnm}. 
Similar to \refeq{var}, we also have the variance due to the mixed propagator
\ba
\sigma^2_{\Psi,\L}=\int_{\vp}^\L \left<\dlin_\L(\vk)\Psi_\L(\vk')\right>' =  \int_{\vp}^\L P_{{\rm L},\Psi}^\L\,.
\ea    

\paragraph{Spin-2.} For spin-2 ($L=2$) non-Gaussianity, the squeezed limit of the kernel motivates the introduction of a trace-free two-tensor
\be
\Psi^\L_{ij}(\vk)=\frac{3}{2}\left(\frac{k_ik_j}{k^2}-\frac{\d_{ij}}{3}\right)\left(\frac{k}{k_{\rm NG}}\right)^\D\vphi_\L(\vk) \,.
\ee
The non-Gaussian bias operator at leading order will be $\Psi^{ij}\Pi_{ij}^{[1]}=\Tr\left[\Psi\Pi^{[1]}\right]$, with \cite{Desjacques:2016bnm}
\be
\Pi_{ij}^{[1]}(\vk)= \frac{k_ik_j}{k^2} \d \,.
\ee
For spin-2 non-Gaussianity, the operator basis becomes:
\ba
 \label{eq:Generalised-Bias-spin-two}
    \mbox{First order:}&\quad \d;\vs
    \mbox{Second order:}&\quad \d^2,\; \G_2,\; \Tr\left[\Psi\Pi^{[1]}\right];\; \vs
    \mbox{Third order:}&\quad \d^3,\; \d\G_2,\; \d\;\Tr\left[\Psi\Pi^{[1]}\right],\; \Gamma_3,\;  \G_3, \; \Tr\left[\Psi\Pi^{[2]}\right],
\ea
with 
    \be
   \Pi^{ij}_{[2]}=\partial^i\partial_k\Phi_g\partial^k\partial^j\Phi_g-\frac{5}{7}\frac{\partial^i\partial^j}{\nabla^2}\G_2 \,.
    \ee  
Notice that spin-2 PNG contributes to the galaxy bias expansion only at second order in perturbation theory. For non-scalar tracers such as galaxy shapes, this changes however: in that case there is a leading-order contribution $\propto \Psi_{ij}$ \cite{2015JCAP...10..032S,2021JCAP...03..060K}.
Scalar contractions of order $(\Psi_{ij})^n$ with $n>1$
should in principle be included, but those terms are $\O(f_{\rm NL}^n)$ which are beyond the scope of this work.

One final important point to note is that the fields $\Psi, \Psi_{ij}$ are to be evaluated at the Lagrangian position $\v{q}$ corresponding to the Eulerian position of observation $\vx$  (see \refapp{SOeval} for a more detailed justification).

\subsection{The non-Gaussian partition function for the galaxy bias}

We now generalize the Gaussian partition function to include PNG as interactions which will then be translated into non-Gaussian contributions to $n$-point functions. 
The guiding point for this is to construct a partition function that reproduces the primordial curvature bispectrum at first order of the nonlinear kernels. We of course want to keep the integrals to be over the Gaussian initial conditions after expanding the interactions, such as \refeq{ZQFT}, such that we can still use QFT tools. 
We can then modify the Gaussian measure to account for slightly non-Gaussian fields. Inserting \refeq{nonGoverdensity} into the kinetic term of the action, we get
\bea
&&-\frac{1}{2}\int_{\vk}^\L \frac{|\dlin_\L|^2}{\PlinL(k)}\rightarrow-\frac{1}{2}\int_{\vk}^\L \frac{|\dlin_{\L}|^2}{\PlinL(k)}-\fnl \int_{\vp_1,\vp_2,\vp_3}\left[ \diracpi(\vp_{123})  \right. \\
&&\hspace{5cm}\times \left.\Knl(\vp_2,\vp_3)\frac{M(p_1)}{M(p_2)M(p_3)}\frac{\dlin_{\L}(\vp_1)}{\Plin^\L(p_1)}\dlin_{\L}(\vp_2)\dlin_{\L}(\vp_3) \right]\;. \nonumber
\eea
Notice that the new cubic kernel includes a $(\Plin^\L)^{-1}$ factor, which has an important role when reproducing $n$-point statistics.
We hereafter drop the subscript ``G'' in $\dlin_{\rm \L,G}$. The equation above implies that we can add an interaction term to the Lagrangian which is equivalent (at linear order in $\fnl$) to modifying the Gaussian measure in the partition function, therefore it will induce non-Gaussian statistics (i.e non-vanishing bispectrum).

Our modified partition function is therefore given by
\be
\label{eq:non-GaussZ}
\Z[J_{m,\L},J_{g,\L}]=\int\Del\dlin_\L \P[\dlin_\L]\exp\left\{{S_{\rm int}+\int_{\vk}J_{m,\L}(\vk)\delta[\dlin_\L](-\vk)+\int_{\vk}J_{g,\L}(\vk)\delta_{g}[\dlin_\L]}(-\vk)\right\}\,,
\ee
with $S_{\rm int}$  a cubic interaction of the form
\begin{equation}
\boxed{
\begin{aligned}
S_{\rm int}[\dlin_\L] &=\fnl\int_{\vp_1,\vp_2,\vp_3}\diracpi(\vp_{123})\Knl(\vp_2,\vp_3)\frac{M(p_1)}{M(p_2)M(p_3)}\frac{\dlin_\L(\vp_1)}{\PlinL(p_1)} \dlin_\L(\vp_2)\dlin_\L(\vp_3) \label{eq:Sintdef}
\\
  &\hspace{4cm}= \raisebox{-0.0cm}{\includegraphicsbox[scale=.7]{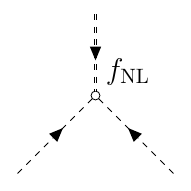}} \,,
\end{aligned}
}
\end{equation}
where in the diagrammatic representation of the $\fnl$ vertex the dashed lines represent the $\dlin_\L$ legs and the doubled-dashed line corresponds to the leg that is divided by a propagator, $\dlin_\L(\vp)/\PlinL(p)$, while the momenta are all incoming due to the Dirac-delta function $\diracpi(\vp_{123})$. We stress again that the terms coupled to the current obey the Gaussian equations of motion, since all non-Gaussianites are induced from the interaction term.
From this we can reproduce all the non-Gaussian matter and bias $n$-point statistics, by taking derivatives with respect to the different currents of the partition function. For example, for the matter bispectrum
\ba
\langle \delta(\vk_1)\delta(\vk_2)\delta(\vk_3)\rangle 
&= \frac{1}{\Z_\L[0]}\frac{\d^3\Z}{\d J_{m, \L}\d J_{m, \L}\d J_{m,\L}}\Big{|}_{J_{m,\L}=0,J_{g,\L}=0}
\vs
&=\int \Del \dlin_\L \P[\dlin_\L ] e^{S_{\rm int}}\; \delta[\dlin_\L](\vk_1)\delta[\dlin_\L](\vk_2)\delta[\dlin_\L](\vk_3)\;.
\ea
We see that by taking three derivatives, we attach a propagator to each of the three incoming lines. The double dashed line, which represents the $\dlin_\L(\vp)/\PlinL(p)$, signals that this line will be cut, resulting in the usual form of the bispectrum with the two propagators (power spectra) attached to it [see also \refeq{non-vanishingB}]
\be
\raisebox{-0.0cm}{\includegraphicsbox[scale=.7]{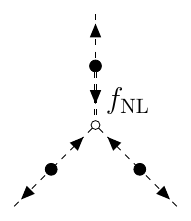}}\,. 
\ee
We discuss some specific examples in \refapp{pkbk}, showing that this partition function reproduces the known results for the matter and galaxy bispectra.

Finally, we can see that when calculating the expectation value of an operator
\ba
\label{eq:oper-expvalues-non-gauss}
\langle O \rangle \equiv \int \Del \dlin \P[\dlin ]\; O\;e^{S_{\rm int}} &=\int \Del \dlin \P[\dlin ]\;\left[ O + O\times S_{\rm int}+ \O\left(S_{\rm int}^2\right)\right]
\vs
&\equiv \langle O \rangle_{\rm free}+\langle O \rangle_{\rm int}\;,
\ea
where we have split the terms into a free and an interacting part, with
\ba
\langle O \rangle_{\rm free} &\equiv \int \Del \dlin \P[\dlin ]\;O = \langle O \rangle_{\rm G}  \,, 
\vs
\langle O \rangle_{\rm int} &\equiv \langle O \times S_{\rm int}\rangle_{\rm G}+\O\left(S_{\rm int}^2\right) \,,
\ea
where we use the notation $\langle \dots \rangle_{\rm G}$ to represent expectation values w.r.t the Gaussian measure.
Hereafter, to not clutter the text, we drop the contributions quadratic in $\fnl$ from $\O(S_{\rm int}^2)$.
In summary, we can separate expectation values into the usual expectation value for Gaussian initial conditions and a new term induced purely by the non-Gaussian interaction $S_{\rm int}$.

%%%%%%%%%%%%%%%%%%%%%%%%%%%%%%%%%%%%%%%%%%%%%%%%%%%%%%%%%%%%%%%%%%%%%%%%%%%
%%%%%%%%%%%%%%%%%%%%%%%%%%%%%%%%%%%%%%%%%%%%%%%%%%%%%%%%%%%%%%%%%%%%%%%%%%%
\section{RG flow via the partition function} \label{sec:running}
%%%%%%%%%%%%%%%%%%%%%%%%%%%%%%%%%%%%%%%%%%%%%%%%%%%%%%%%%%%%%%%%%%%%%%%%%%%
%%%%%%%%%%%%%%%%%%%%%%%%%%%%%%%%%%%%%%%%%%%%%%%%%%%%%%%%%%%%%%%%%%%%%%%%%%%

Now that we have the non-Gaussian partition function \refeq{non-GaussZ} for the galaxy bias, we proceed to integrate out the shell modes similarly to \cite{Rubira:2023vzw}. 
We start with the Wilson-Polchinski approach for the shell integrals in the presence of PNG. Later we describe the contribution of the new interaction vertex to the running of bias terms and then we derive the RGE for the bias parameters.

%%%%%%%%%%%%%%%%%%%%%%%%%%%%%%%%%%%%%%%%%%%%%%%%%%%%%%%%%%%%%%%%%%%%%%%%%%%
\subsection{The Wilson-Polchinski shell integrals}
%%%%%%%%%%%%%%%%%%%%%%%%%%%%%%%%%%%%%%%%%%%%%%%%%%%%%%%%%%%%%%%%%%%%%%%%%%%
We want to consider the running over an infinitesimal momentum difference $\L=\L'-\lambda$ integrating out field modes that have support in the shell with infinitesimal width $\lambda$.
This means that the modes can be split in
\be 
\label{eq:shelldef}
\dlin_{\L'}(\vk) = \dlin_\L(\vk) + \dlinshell(\vk)\,.
\ee
We then start from the partition function for the bias expansion making explicit the split between shell and $\L$ modes 
\begin{align}
&\Z[J_{\L}] =\int \Del\dlin_\L \P[\dlin_\L]\int\Del\dlinshell \P[\dlinshell]\exp\left(S_{\rm int} [\dlin_\L+\dlinshell]]\right)\\
&\hspace{4cm} \times \exp\left( \int_{\vk}J_\L(\vk)[\sum_Ob_O^{\L'}O[\dlin_\L+\dlinshell](-\vk)]+\O\left[J^2\right] \right) \,, \nonumber
\end{align}
where the current $J$ has support only up to $\L$, in order to guarantee the orthogonality condition $\int_{\vp} J_\L(\vp)\dlinshell(\vp) = 0$.
We can then expand the operators evaluated at $n^{\rm th}$-order $O^{(n)}[\dlin_\L+\dlinshell]$ in powers of $\dlinshell$ (see Appendix~A.1 of \cite{Rubira:2023vzw} for examples)
\ba \label{eq:Oshell}
&&O^{(n)}[\dlin_\L+\dlinshell] =O^{(n)}[\dlin_\L] + O^{(n),\shellPT{1}}[\dlin_\L, \dlinshell] + O^{(n),\shellPT{2}}[\dlin_\L, \dlinshell] \\
&& \hspace{5cm}+ \ldots + O^{(n),\shellPT{n-1}}[\dlin_\L, \dlinshell] + O^{(n)}[\dlinshell]\,. \nonumber
\ea
In addition, the partition function with PNG includes the interaction term \refeq{Sintdef} 
\ba
& S_{\rm int} [\dlin_\L+\dlinshell]= \label{eq:intLprime}\\
& \hspace{1cm}  \fnl\int_{\vp_1,\vp_2, \vp_3}\diracpi(\vp_{123}) \Knl\left(\vp_2,\vp_3\right)\frac{M(\vp_1)}{M\left(\vp_3\right)M(\vp_2)}\frac{\dlin_{\L'}\left(\vp_1\right)}{P_{{\rm L},\L'}\left(\vp_1\right)} \dlin_{\L'}\left(\vp_2\right)\dlin_{\L'}\left(\vp_3\right)\,. \nonumber
\ea
Hence, on top of the expansion \refeq{Oshell} for the bias operators, this interaction term also has to be expanded in powers of $\dlinshell$, as we discuss below in \refsec{interaction} and with more details in \refapp{cubic}. 

After integrating out the shell modes, we find the effective partition function
\ba
\Z[J_{\L}] &= \int \Del\dlin_\L \P[\dlin_\L]   \exp\left(S_{\rm int} [\dlin_\L]\right)
\vs
&\times \exp \left(\int_{\vk}  J_{\L}(\vk) \sum_O b_O^{\L'} \left[ O[\dlin_\L](-\vk) + \Shell_O[\dlin_\L](-\vk) \right] +\O(J_\L^2) \right) \;,  \label{eq:Z_with_S}
\ea
where $\Shell_O[\dlin_\L]$ is the shell correction, which we can write as
\ba
\mathcal{S}_O(\vk) &\equiv  \left[ \mathcal{S}_O^{2}  \right]_{\rm free}(\vk) +  \left[ \mathcal{S}_O \right]_{\rm int}(\vk) + \O(\lambda^2) \,,
\ea
where we have dropped higher-loops contributions proportional to $\lambda^2$;
as argued in \cite{Rubira:2023vzw}, we can always choose $\lambda$ sufficiently small so that $\O(\lambda^2)$ corrections can be neglected.
We have split into a free and an interacting part similar to \refeq{oper-expvalues-non-gauss}, with the free part being defined as in \cite{Rubira:2023vzw}
\ba \label{eq:freeterm}
\left[ \mathcal{S}_O^{2}  \right]_{\rm free}(\vk) &\equiv \int \Del\dlinshell \P[\dlinshell]\sum_{n\geq2}O^{(n),(2)_{\rm shell}}[\dlin_{\L},\dlinshell](\vk) 
= \sum_{\ell}\raisebox{-0.0cm}{\includegraphicsbox[scale=.9]{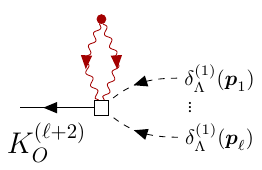}} \,,
\ea
where we see that the contribution from two shells is still the relevant term for the free component, since $\dlinshell$ is still a Gaussian field.
The new interaction arising due to the inclusion of PNG can be written as 
\ba
\left[ \mathcal{S}_O  \right]_{\rm int}(\vk) &\equiv \int \Del\dlinshell \P[\dlinshell] O[\dlin_{\L},\dlinshell](\vk) \left( S_{\rm int} [\dlin_\L+\dlinshell]\right) \,.
\ea
The structure of this interaction term is the topic of next section.

\subsection{The cubic interaction} \label{sec:interaction}

We now discuss the shell expectation values in the presence of cubic interaction PNG, focusing only on the central points. The main conclusion is summarized in \refeq{shell-exp} and we leave a detailed discussion for \refapp{cubic}. 

We start by splitting the interaction term evaluated for $\dlin_{\L'} = \dlin_\L+\dlinshell$ into four components
\ba 
&S_{\rm int}[\dlin_\L+\dlinshell]= S_{\rm int}[\dlin_\L]+S_{\rm int}^{(1)_{\rm shell}}[\dlin_\L,\dlinshell] \label{eq:Sintsplit}
\\
& \hspace{4cm}+S_{\rm int}^{(2)_{\rm shell}}[\dlin_\L,\dlinshell]+S_{\rm int}^{(3)_{\rm shell}}[\dlinshell]\,, \nonumber
\ea
where $S_{\rm int}^{(i)_{\rm shell}}$ refers to the number of (upstairs) legs in \refeq{intLprime} that have support in the shell.\footnote{Notice however that the net shell counting can change by a factor -2 depending whether $\dlin/P_{{\rm L},\L'}$ has support on the shell or not, due the $1/\Pshell$ factor. Moreover, the support of the terms up and downstairs of $\dlin_{\L'}\left(\vp\right)/P_{{\rm L},\L'}\left(\vp\right)$ are always related since   
\be
\frac{\dlin_{\L'}\left(\vp\right)}{P_{{\rm L},\L'}\left(\vp\right)} = \frac{\dlin_{\L}\left(\vp\right)}{P_{{\rm L},\L}\left(\vp \right)} + \frac{\dlinshell\left(\vp \right)}{\Pshell\left(\vp \right)} \,.
\ee
} We find for each one of those terms:
\begin{itemize}
    \item $S_{\rm int}[\dlin_\L]$ captures the interactions of modes that are not integrated out, and therefore does not act as source for the RGE.
    
    \item The contribution with one shell is given by
    \ba
    S_{\rm int}^{(1)_{\rm shell}}[\dlin_\L,\dlinshell] &=  \fnl\int_{\vp_1,\vp_2, \vp_3}\diracpi(\vp_{123}) \Knl\left(\vp_2,\vp_3\right)\frac{M(p_1)}{M(p_2)M(p_3)} \label{eq:ints1_main}
    \\
    &\hspace{-1cm}\times \left[ \frac{\dlinshell(\vp_1)}{P_{\rm shell}(p_1)}\dlin_\L\left(\vp_2\right)\dlin_\L\left(\vp_3\right) + 2\,\frac{\dlin_\L\left(\vp_1\right)}{\PlinL\left(p_1\right)}\dlinshell\left(\vp_2\right)\dlin_\L\left(\vp_3\right) \right]   \nonumber
    \vs
    &= \raisebox{-0.0cm}{\includegraphicsbox[scale=.7]{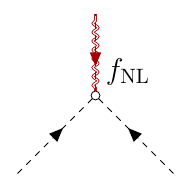}} + \raisebox{-0.0cm}{\includegraphicsbox[scale=.7]{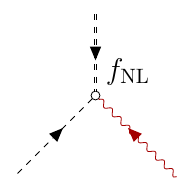}} \,, \nonumber
    \ea
    in which, again, the double lines represent the $\dlinshell(\vp_1)/P_{\rm shell}(p_1)$ (in the left diagram) or $\dlin_\L\left(\vp_1\right)/\PlinL\left(\vp_1\right)$ legs (in the right diagram). The red wiggly lines represent shell modes, whereas dashed lines represent lines with support in $\L$.
    Diagrams of this type are not allowed by the momentum constrain $\diracpi(\vp_{123})$ when taking two of the legs with momentum much smaller than the third one that is on the shell.
    
    \item The contributions with two shell modes demand more attention. They are given by
    \ba
    S_{\rm int}^{(2)_{\rm shell}}[\dlin_\L,\dlinshell] &=  \fnl\int_{\vp_1,\vp_2, \vp_3}\diracpi(\vp_{123}) \Knl\left(\vp_2,\vp_3\right)\frac{M(p_1)}{M(p_2)M(p_3)} 
    \vs
    &\hspace{-1cm}\times \left[2\, \frac{\dlinshell(\vp_1)}{P_{\rm shell}(p_1)}\dlinshell\left(\vp_2\right)\dlin_\L\left(\vp_3\right) + \frac{\dlin_\L\left(\vp_1\right)}{\PlinL\left(p_1\right)}\dlinshell\left(\vp_2\right)\dlinshell\left(\vp_3\right) \right]   
    \vs
    &\equiv  \left[S_{\rm int}^{(3),(2)_{\rm shell}}\right]_{\rm scatter} + \left[S_{\rm int}^{(3),(2)_{\rm shell}}\right]_{\rm annihilation} \label{eq:ints2main}
    \\
    &= \raisebox{-0.0cm}{\includegraphicsbox[scale=.7]{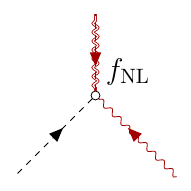}} + \raisebox{-0.0cm}{\includegraphicsbox[scale=.7]{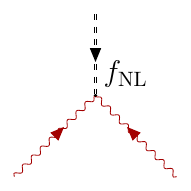}} \,, \nonumber
    \ea
    where we have split into a {\it scattering} part, in which a shell UV mode interacts with a $\L$ IR mode leading to another shell mode, and an {\it annihilation} part, with two shell UV modes interacting to generate a large-scale mode below $\L$. We show (up to our knowledge for the first time in the literature) in \refapp{Sint2shell} that the annihilation modes lead to a stochastic contribution that starts to contribute at $\O(J^2)$ . We leave a full treatment of stochastic contributions in the presence of PNG for a future work (for Gaussian initial conditions, see \cite{Rubira:2024tea}).
    After contracting the scattering interaction with two other shell modes we find
    (see also Appendix~A.2 of \cite{Assassi:2015fma})
    \begin{equation}
    \boxed{
    \begin{aligned} \label{eq:interactingmain}
    \left[ \mathcal{S}_O  \right]_{\rm int}(\vk) &=\int \Del\dlinshell \P[\dlinshell] \sum_{n\geq2}O^{(n),(2)_{\rm shell}}(\vk) \left[S_{\rm int}^{(2)_{\rm shell}}\right]_{\rm scatter}  
    \\
    &= \sum_{n\geq2} 4 \int_{\vp_1,\dots,\vp_n} \diracpi(\vk-\vp_{1\dots n})K_O^{(n)}(\vp_1, \vp_2, \dots, \vp_n) \dlin_{\L}\left(\vp_{1}\right) \dots \dlin_{\L}\left(\vp_{n-2}\right)
    \\
    & \hspace{2cm}\Knl(\vp_{n-1}+\vp_{n},-\vp_n)\frac{\dlin_{\L}\left(\vp_{n-1}+\vp_{n}\right)}{M\left(|\vp_{n-1}+\vp_n|\right)} \Pshell(p_n)
    \\
    &\hspace{3cm}= \sum_{\ell} \raisebox{-0.0cm}{\includegraphicsbox[scale=.7]{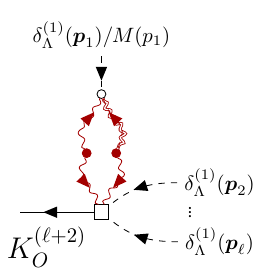}} \,.
    \end{aligned}
    }
    \end{equation}
    We notice two things about the interaction term. First, that it contains an extra external leg compared to the free term \refeq{freeterm}. Second, since in \refeq{interactingmain} $\vp_n$ is constrained to be in the shell and $\vp_{n-1}+\vp_{n}$ is constrained to have support in $\L$, we can use the squeezed limit of the kernel $\Knl(\vp_{n-1}+\vp_{n},-\vp_n)$, \refeq{squeezed_limit}.
    
    \item The contribution with three shell modes
    \ba
    S_{\rm int}^{(3)_{\rm shell}}[\dlinshell] &=  \fnl\int_{\vp_1,\vp_2, \vp_3}\diracpi(\vp_{123}) \Knl\left(\vp_2,\vp_3\right)\frac{M(p_1)}{M(p_2)M(p_3)} \label{eq:ints3_main}
    \\
    &\hspace{4cm} \times  \frac{\dlinshell(\vp_1)}{P_{\rm shell}(p_1)}\dlinshell\left(\vp_2\right)\dlinshell\left(\vp_3\right)   \,. \nonumber
    \vs
     &= \raisebox{-0.0cm}{\includegraphicsbox[scale=.7]{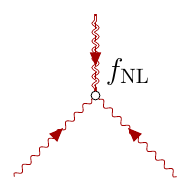}}  \,, \nonumber
    \ea
    only leads to higher-loop or tadpole contributions.
\end{itemize}

In summary, we find that the shell corrections when including both free and interacting terms are represented by two diagrams 
\begin{equation}
\boxed{
\begin{aligned}
\mathcal{S}_O(\vk) &\equiv \left[ \mathcal{S}_O^{2}  \right]_{\rm free}(\vk) +  \left[ \mathcal{S}_O \right]_{\rm int}(\vk) \label{eq:shell-exp} 
\\
&= \sum_{\ell} \raisebox{-0.0cm}{\includegraphicsbox[scale=.9]{figs/diags/diag_S2_ell_legs.pdf}} + \sum_{\ell}\raisebox{-0.0cm}{\includegraphicsbox[scale=.9]{figs/diags/NG_inter.pdf}}\;. 
\end{aligned}
}
\end{equation}
Via the free part, Gaussian operators source other Gaussian operators (as calculated in \cite{Rubira:2023vzw}). Non-Gaussian operators in the new basis \refeq{Generalised-Bias-spin-zero} or \refeq{Generalised-Bias-spin-two} can source either Gaussian and non-Gaussian terms, as we calculate below. 
The new interacting part, originating from the cubic PNG vertex, are only due to the scattering term in \refeq{ints2main}. The interaction also allows for the sourcing of non-Gaussian operators by Gaussian terms, while the contributions from non-Gaussian to Gaussian terms that $S_{\rm int}$ sources are suppressed by $\fnl^2$.

%%%%%%%%%%%%%%%%%%%%%%%%%%%%%%%%%%%%%%%%%%%%%%%
\subsection{On the \texorpdfstring{$\Shell_{O'}$}{SO} corrections to \texorpdfstring{$O$}{O} and the RG equations}
\label{sec:Corrections}

We can now derive the RG equations describing the evolution of the bias parameters following the steps of \cite{Rubira:2023vzw, Rubira:2024tea}.
We compare the terms linear in the current $J_\L$ in \refeq{Z_with_S} finding 
\bea
\sum_Ob_O^\L O[\dlin_\L](\vk) =\sum_Ob_O^{\L'}\left( O[\dlin_\L] (\vk)+\mathcal{S}_O [\dlin_\L ] (\vk)\right) \,.
\eea
For the Gaussian initial condition case in \cite{Rubira:2023vzw}, we could use that the shell correlators $\Shell$ can be written as an operator expansion with coefficients $s_{O'}^O$ 
\be
\Shell_{O'}[\dlin_\L] = \Shell_{O'}^2[\dlin_\L] = \leps  \frac{d\s^2_\L}{d\L}\Big|_{\L} \sum_{O} s_{O'}^O O[\dlin_\L],
\label{eq:SO2exp}
\ee
where we pulled out the linear dependence on the shell width $\lambda$ through 
\be
\int_{\vp} P_{\rm shell}(p) = \int_{\L}^{\L+\leps} \frac{p^2 dp}{2 \pi^2 }\Plin(p) = \frac{d \s^2_\L}{d \L}\Big|_{\L}\leps + \O(\leps^2)\,.
\label{eq:sigmashell}
\ee
See the first block of \reftab{cCoeff} for a summary of the $s_{O'}^O$  coefficients for the Gaussian case.

When including non-Gaussian interactions, on top of the shell integral above, we will also find terms proportional to
\ba
\int_{\vp}\PshellPsi(p)=\int_{\L}^{\L+\lambda}\frac{dp\;p^2}{2\pi^2}P_{L,\Psi}(p)=\frac{d\sigma^2_\Psi}{d\L}\lambda+\O(\lambda^2)\;.
\ea
We can relate this mixed variance to the variance $\sigma_\L^2$ by writing
\ba
\int_{\L}^{\L+\lambda}\frac{dp\;p^2}{2\pi^2}P^{\Psi}_{\rm shell}(p) &\simeq \left(\frac{\Lambda}{k_{\rm NG}}\right)^\D\left(\frac{H_0}{\L}\right)^2\frac{3\;\Omega_m}{2\;T(\L)}\frac{d\sigma_\L^2}{d\L}\Big|_{\L}\lambda\;, \label{eq:Ppsishell}
\ea
using \refeq{Mdef}. From the squeezed limit of the non-Gaussian kernel, we also encounter terms of the form
\ba
\int_{\L}^{\L+\lambda}\frac{dp\;p^2}{2\pi^2}\left(\frac{k_{\rm NG}}{p}\right)^\D P_{\rm shell}(p) &\simeq\left(\frac{k_{\rm NG}}{\Lambda}\right)^\D\frac{d\sigma_\L^2}{d\L}\Big|_{\L}\lambda\;. \label{eq:Pshell2}
\ea
\refeq{Ppsishell} appears when considering contributions from non-Gaussian to Gaussian operators, whereas \refeq{Pshell2} appears when considering the opposite case.
Notice that $(H_0/\Lambda)^2 \ll 1$ acts as a suppression factor. Also since $(\L/k_{\rm NG}) \lesssim 1$ in order to keep the perturbative expansion under control, and $\Delta \geq 0$, we find that this factor acts as a suppression in \refeq{Ppsishell} but as an enhancing factor in  \refeq{Pshell2}. 

The appearence of new types of shell integrals from PNG forces us to slightly modify \refeq{SO2exp} to accommodate these new $\L$-dependent factors via functions $f_{O'}^O(\L)$, writing
\be
\Shell_{O'}[\dlin_\L] = \leps  \frac{d\s^2_\L}{d\L}\Big|_{\L} \sum_{O} s_{O'}^O f_{O'}^O(\L) O[\dlin_\L]\,.
\label{eq:SO2exp_NG}
\ee
Therefore, we can always parametrize the running using $d\sigma_\L^2/d\L$: despite having more than one propagator, they relate to each other in the shell region. 
We then find the ODE describing the running of the bias parameters
\be
\boxed{
\frac{d}{d\L} b_O(\L) = - \frac{d\s^2_\L}{d\L} \sum_{O'} s_{O'}^O \, f_{O'}^O(\L) b_{O'}(\L)\, , \label{eq:S2forallorder}
}
\ee
similar to \cite{Rubira:2023vzw} but now with the extra factor $f_{O'}^O$ that satisfies (at linear order in $\fnl$)
\ba
f_{O'}^O = 1\,, \quad &\textrm{if $O$ and $O'$ are both Gaussian,} 
\vs
f_{O'}^O = a_L \fnl\,, \quad &\textrm{if $O$ and $O'$ are both non-Gaussian,} 
\vs
f_{O'}^O(\L) = a_L \fnl\left(\frac{\Lambda}{k_{\rm NG}}\right)^\D\left(\frac{H_0}{\L}\right)^2 \frac{3\;\Omega_m}{2\;T(\L)}\,, \quad &\textrm{if $O$ is Gaussian and $O'$ non-Gaussian,} 
\vs
f_{O'}^O(\L) = a_L \fnl \left(\frac{k_{\rm NG}}{\Lambda}\right)^\D \,, \quad &\textrm{if $O$ is non-Gaussian and $O'$ Gaussian.}
\label{eq:fOO}
\ea
\refeq{S2forallorder} fully describes the structure of the bias RGE with any second-order PNG at $\O(\fnl)$, generalizing the previously derived Gaussian case which corresponds to the first line of \refeq{fOO}.  In the following, we explicitly derive the RGE for two types of PNG, the spin-0 and spin-2 cases. We focus in this work on corrections to first and second-order operators by operators that are up to third-order.\footnote{For spins higher than two, the procedure to derive the RGE is similar, by adding the appropriate operators to the respective bias expansion. The operators associated to $L>2$ are order higher than 2 and therefore are not considered in this work. Odd ranked spins (e.g., $L=1$) are zero at leading order due to the invariance of the squeezed bispectrum in $\vk_s\rightarrow -\vk_s$  as mentioned in \cite{Assassi:2015fma} (for more details see Appendix~A of \cite{Assassi:2015EFT} and \cite{Shiraishi_2013,Lewis_2011}).}

\paragraph{Spin-0 non-Gaussianity.}

We start by calculating the shell correlators (see \refapp{spin0eval} for a more complete explanation) and deriving the RGE for the spin-0 case. 
The assumption that $k \ll p_{\rm shell}$ allow us to use the squeezed limit \refeq{squeezed_limit} of the $\Knl$ kernels, writing them in terms of indices $\Delta$ and projecting onto the Legendre polynomial of order zero $\mathcal{P}_0  = 1$
\begin{equation}
\label{eq:squeezed_limit0}
\Knl(\vp_{\rm shell},\vk)\xrightarrow{k\ll p_{\rm shell}} a_{0}\left(\frac{k}{p_{\rm shell}}\right)^{\Delta}\,.
\end{equation}
Corrections to this approximation lead to higher-derivative non-Gaussian operators such as $\nabla^2\Psi$, which we will neglect here; see \cite{2013PhRvD..87l3518S,2018JCAP...09..003C} for a discussion and attempts to resum these beyond-squeezed-limit corrections.

For the Gaussian operators, those that we already include in the bias expansion without having non-Gaussianities, we find that their new contributions due to the interaction vertex are
\ba
\left[ \mathcal{S}_{\d^2} \right]_{\rm int} (\vk) &= \left(4\fnl a_0 \Psi(\vk)+\frac{272}{21}\fnl a_0[\Psi\d](\vk)\right)  \int_{\vp}\left(\frac{k_{\rm NG}}{p}\right)^\D P_{\rm shell}(p)+\text{h.d.}\;, \label{Eq:Sd2}
\\
\left[ \mathcal{S}_{\d^3} \right]_{\rm int} (\vk) &=12\fnl a_0[\Psi\d](\vk) \int_{\vp} \left(\frac{k_{\rm NG}}{p}\right)^\Delta P_{\rm shell}(p)+\text{h.d.}\;,
\\
\left[ \mathcal{S}_{\G_2\d} \right]_{\rm int} (\vk) &=-\frac{8}{3}\fnl a_0\left[\Psi\d\right](\vk)\int_{\vp} \left(\frac{k_{\rm NG}}{p}\right)^\D P_{\rm shell}(p)+\text{h.d.}\;,
\ea
in which we omitted higher-derivative (h.d.) terms. 
We can notice a few things. 
First, non-Gaussian operators are sourced by Gaussian operators via the interaction and one therefore has to include operators with $\Psi$ in the basis \refeq{Generalised-Bias-spin-zero} in order to be closed under normalization, as already pointed out by \cite{Assassi:2015fma}. 
Second, as expected, at linear-order in $\fnl$ {\it only non-Gaussian operators} are sourced by Gaussian operators via the interaction term.
Third, also expected, the amplitude of the RG running is proportional to  $\fnl a_0$.
Finally, we notice the factor $\left(k_{\rm NG}/p\right)^\D$ in the integral, which is due to the squeezed limit \refeq{squeezed_limit0} and will lead to developments we comment on below.

For the new non-Gaussian operators, those are already linear order in $\fnl$. Therefore, their (linear-in-$\fnl$) contribution cannot contain any interaction vertex but has to come from the free term in \refeq{shell-exp}. We find 
\ba
\left[\mathcal{S}^2_{\Psi}\right]_{\rm free} (\vk) &= a_0\fnl\left(-\frac{13}{21}\d(\vk)+\frac{43}{135}\left(\d^2\right)(\vk)-\frac{1699}{13230}\left(\G_{2}\right)(\vk)\right)\int_{\vp}\PshellPsi(p) +\text{h.d.}\,, \label{eq:Psi-displacement}
\\
\left[\mathcal{S}^2_{\Psi\d}\right]_{\rm free} (\vk) &= \left[\frac{13}{21}\delta(\vk)+\frac{478}{135} \delta^2(\vk)+\frac{79}{2205}\mathcal{G}_2(\vk)\right]\int_{\vp} \PshellPsi(p) +\text{h.d.}\,, \label{eq:Psid-displacement}
\\
\left[\mathcal{S}_{\psi\d^2}^2\right]_{\rm free}(\vk) &=a_0\fnl \Psi(\vk) \int_{\vp} \Pshell(p)+2a_0\fnl \d(\vk)\int_{\vp}\PshellPsi 
\\
&+\frac{68}{21}a_0\fnl\left[\Psi\d\right](\vk) \int_{\vp} \Pshell(p)+\frac{47}{21}a_0\fnl \d^2 (\vk) \int_{\vp} \PshellPsi(p)+\text{h.d.}\,,
\vs
\left[\mathcal{S}_{\rm \Psi\G_2}^2\right]_{\rm free}(\vk) &= a_0\fnl\left[-\frac{4}{3}\d(\vk) -\frac{31}{21} \d^2(\vk) -\frac{1}{21}\G_2(\vk)\right]\int_{\vp} \PshellPsi(p)+\text{h.d.}\,. \label{eq:PsiG2-displacement}
\ea
We notice they contribute both to Gaussian [via $P^\L_{\rm L,\Psi}$] and also non-Gaussian operators [via $ \PlinL(p)$].
We stress again that the operator $\Psi$ is calculated at the Lagrangian point $\Psi(\vq)=\Psi(\vx-\mathbf{s})$ and expanded up to third-order in displacements (total perturbative order 4; see \refapp{displacements}). In addition, we highlight that the first-order operator $\Psi$ contributes to other (leading-in-derivative) terms due to the Lagrangian displacements, differently than $\d$ that only contributes to higher-derivative terms due to mass and momentum conservation. We present the details of this calculation in \refapp{SOeval}. We emphasize that terms that break the Galilean invariance are cancelled, as required by the equivalence principle, and the coefficients of the operator $\d^{(n)}$ are the same at first ($n=1$) and second order ($n=2$) in perturbation theory. Specifically, the operators in \refeqs{Psi-displacement}{PsiG2-displacement} are calculated up to second order; in particular, $\d$ here corresponds to $\d^{(1)}+\d^{(2)}$.

After having calculated the shell contributions, we can derive the RGE for spin-0 non-Gaussianities and a generic scaling index $\Delta$ using \refeq{S2forallorder}, leading to
\ba
\frac{db_\d}{d\L} &=\frac{db_\d}{d\L}\Big{|}_{\rm free}-a_0\fnl\left[-\frac{13}{21}b_{\Psi}+\frac{13}{21}b_{\Psi\d}+2b_{\Psi\d^2}-\frac{4}{3}b_{\Psi\G_2}\right]\left(\frac{\Lambda}{k_{\rm NG}}\right)^\D\left(\frac{H_0}{\L}\right)^2\frac{3\;\Omega_m}{2\;T(\L)}\frac{d\sigma_\L^2}{d\L}\,, \label{eq:bd_spin0}
\\
\frac{db_{\d^2}}{d\L} &=\frac{db_{\d^2}}{d\L}\Big{|}_{\rm free}-a_0\fnl\left[\frac{43}{135}b_{\Psi}+\frac{478}{135}b_{\Psi\d}+\frac{47}{21}b_{\Psi\d^2}-\frac{31}{21}b_{\Psi\G_2} +b^{\{\d^2\}_{\rm NG}}_{n=4}\right]
\vs
&\hspace{7cm}\times \left(\frac{\Lambda}{k_{\rm NG}}\right)^\D\left(\frac{H_0}{\L}\right)^2\frac{3\;\Omega_m}{2\;T(\L)}\frac{d\sigma_\L^2}{d\L} \,, \label{eq:bd2_spin0}
\\
\frac{db_{\G_2}}{d\L} &=\frac{db_{\G_2}}{d\L}\Big{|}_{\rm free}-a_0\fnl\left[-\frac{1699}{13230}b_{\Psi}+\frac{79}{2205}b_{\Psi\d}-\frac{1}{21}b_{\Psi\G_2} +b^{\{\G_2\}_{\rm NG}}_{n=4}\right]
\vs
&\hspace{7cm}\times \left(\frac{\Lambda}{k_{\rm NG}}\right)^\D\left(\frac{H_0}{\L}\right)^2\frac{3\;\Omega_m}{2\;T(\L)}\frac{d\sigma_\L^2}{d\L}\,, \label{eq:bG2_spin0}
\ea
where $\frac{db}{d\L}\Big{|}_{\rm free}$ refer to the running of Gaussian coefficients without any interactions described in \cite{Rubira:2023vzw}. We also have included $b^{\{O\}_{\rm NG}}_{n=4}$, the contributions of fourth-order non-Gaussian operators to the $b_O$, which are beyond the scope of this work.\footnote{Notice that $\frac{db}{d\L}\Big{|}_{\rm free}$ includes $b^{\{O\}_{\rm G}}_{n=4}$, the contribution of fourth-order Gaussian operators to $O$. In general an operator of order $n$ contributes to all operators that are order $n-2$ \cite{Rubira:2024tea}.}
We find that the running of Gaussian coefficients due to non-Gaussian operators is doubly suppressed by $(H_0/\L)^2$ and $(\L/k_{\rm NG})^\Delta$. 

For the non-Gaussian operators, we find, including both the free and interaction terms,
\ba
\frac{d b_\Psi}{d \L} &=-a_0\fnl b_{\Psi\d^2}\frac{d\sigma^2_\L}{d\L}-4a_0\fnl b_{\d^2}\left(\frac{k_{\rm NG}}{\Lambda}\right)^\D\frac{d\sigma_\L^2}{d\L}\,, \label{eq:bpsi}
\\
\frac{d b_{\Psi\d}}{d \L} &= -a_0\fnl \left[\frac{68}{21}b_{\Psi\d^2} +b^{\{\Psi\d\}_{\rm NG}}_{n=4} \right] \frac{d\sigma^2_\L}{d\L}
\vs
&\quad -a_0\fnl\left[\frac{272}{21}b_{\d^2}+12b_{\d^3}-\frac{8}{3}b_{\G_2\d}+b^{\{\Psi\d\}_{\rm G}}_{n=4}\right]\left(\frac{k_{\rm NG}}{\Lambda}\right)^\D\frac{d\sigma_\L^2}{d\L}\,, \label{eq:bpsid}
\ea
where $b^{\{O\}_{\rm G}}_{n=4}$ is the contribution of fourth-order Gaussian operators to $b_O$.
We see that the running of non-Gaussian coefficients due to Gaussian operators is enhanced by $(k_{\rm NG}/\Lambda)^\Delta$. We summarize those findings in \reftab{cCoeff}, with the coefficients $s_{O'}^O$ of \refeq{S2forallorder}. 
We further discuss solutions for those equations in \refsec{results}.
\begin{table}[ht]
    %\begin{minipage}{.7\linewidth}
	%\caption{}
	\centering
    \scalebox{0.7}{
	{ \begin{tabular}{|c|c|c|c||c|c|c|c||c|c|c||}
			\hline
			%\multicolumn{1}{|c||}{\rotatebox[origin=c]{45}{$O$ $O'$}} & $\d$ & $\d^2$ & $\G_2$ & $\d^3$ & $\G_3$ & $\Gamma_3$ & $\d\G_2$  \\
			$s_{O'}^O$ & $\d^2$ & $\d^3$&$\d\G_2$ & $\Psi$&$\Psi\d$ & $\Psi\d^2$ & $\Psi\G_2$ & $\Tr\Psi\Pi^{[1]}$ & $\d\Tr\Psi\Pi^{[1]}$& $\Tr\Psi\Pi^{[2]}$\\
			\hline\hline
			$\d$ & $ 68/21$& 3& $-4/3$& $-13/21$ &$13/21$& $2$& $-4/3$& $34/21$& $1$& $34/21$\\
			\hline
			$\d^2$ & $ 8126/2205$& $68/7$&$- 376/105$&$43/135 $&$478/135$& $47/21$& $-31/21$& $124/315$& $178/105$& $14347/6027$ \\
			\hline
			$\G_2$ & $254/2205$&-&$ 116/105$&$-1699/13230$&$79/2205$&-& $-1/21$& $-661/4410$& $4/35$& $-241/735$ \\
			\hline \hline
            $\Psi$ & $4$& -& -&- &-& 1& -& -& -& -\\ \hline
            $\d\Psi$ & $272/21$& $12$& $-8/3$&- &-& $68/21$& -& -& -& -\\
            \hline \hline
            $\Tr\Psi\Pi^{[1]}$ & $64/105$& -& $16/15$& - &-& -& -& -& 8/105& 58/305\\
            \hline
	\end{tabular}}}
	%	\end{minipage}%
	\caption{The $s_{O'}^O$ coefficients from \refeq{S2forallorder}, that describe how the $O'$ operators (columns, up to third-order, including only terms that source leading-in-derivative operators) contribute to $O$ operators (rows, up to second order). This generalizes Table~1 of \cite{Rubira:2024tea} to include non-Gaussian contributions, for both spin-0 and spin-2 types.}
	\label{tab:cCoeff}
\end{table}

\paragraph{Spin-2 non-Gaussianity.}
We proceed to spin-2 non-Gaussianity using again the squeezed limit \refeq{squeezed_limit} of the $\Knl$ kernels and now projecting onto the Legendre polynomial of order two,
\begin{equation}
\label{eq:squeezed_limit2}
\Knl(\vp_{\rm shell},\vk)\xrightarrow{k\ll p_{\rm shell}} a_{2}\left(\frac{k}{p_{\rm shell}}\right)^{\Delta}  \mathcal{P}_2\left(\hat{\vk}\cdot\hat{\vp}_{\rm shell}\right) \,.
\end{equation}
The evaluation of the shell integrals for spin-2 field is included in more details in \refapp{spin2eval}, but we summarize them here

\ba
\left[ \mathcal{S}_{\d^2} \right]_{\rm int} (\vk) &=a_2\fnl \frac{64}{105}\Tr\left[\Psi\Pi^{[1]}\right](\vk)\int_{\vp}\left(\frac{k_{\rm NG}}{p}\right)^\D P_{\rm shell}(p)+\text{h.d.} \,,
\\
\left[ \mathcal{S}_{\d\G_2} \right]_{\rm int} (\vk) &=a_2\fnl \frac{16}{15}\Tr\left[\Psi\Pi^{[1]}\right](\vk)\int_{\vp}\left(\frac{k_{\rm NG}}{p}\right)^\D P_{\rm shell}(p)+\text{h.d.}\,,
\ea
for the contribution of the Gaussian terms via the new interaction and 
\ba
\left[\mathcal{S}^2_{\rm Tr[\Psi\Pi^{[1]}]}\right]_{\rm free} (\vk) &=a_2\fnl\left[\frac{34}{21}\d(\vk)+\frac{124}{315}\d^2(\vk)-\frac{661}{4410}\mathcal{G}_2(\vk)\right]\int_{\vp}\PshellPsi(p) +\text{h.d.}\,,
\\
\left[\mathcal{S}^2_{\rm Tr[\d\Psi\Pi^{[1]}]}\right]_{\rm free} (\vk) &=a_2\fnl\left[\d(\vk)+\frac{178}{105}\d^2(\vk)+\frac{4}{35}\mathcal{G}_2(\vk)\right]\int_{\vp}\PshellPsi(p)\,,
\vs
&\quad +a_2\fnl\frac{8}{105}\Tr\left[\Psi\Pi^{[1]}\right] (\vk)\int_{\vp}\Pshell(p) +\text{h.d.}
\\
\left[\mathcal{S}^2_{\rm Tr[\Psi\Pi^{[2]}]}\right]_{\rm free} (\vk) &=a_2\fnl\left[\frac{34}{21}\d(\vk)+\frac{14347}{6027}\d^2(\vk)-\frac{241}{735}\mathcal{G}_2(\vk)\right]\int_{\vp}\PshellPsi(p)
\vs
&\quad+a_2\fnl\frac{58}{105}\Tr\left[\Psi\Pi^{[1]}\right](\vk)\int_{\vp}\Pshell(p)+\text{h.d.}\,,
\ea
for the contribution of the non-Gaussian operators via the free term in \refeq{shell-exp}.
We again note the same kind of features pointed out for the spin-0 scenario, such as the need to include non-Gaussian operators in \refeq{Generalised-Bias-spin-two} to have a basis that is complete under renormalization. Moreover, differently than for the spin-0 case, in which we have a first-order non-Gaussian operator, we find that spin-2 non-Gaussianities start to contribute at second-order via $\Tr\left[\Psi\Pi^{[1]}\right]$.

Putting everything together, we find the RGE for the Gaussian operators to be
\ba
\frac{db_\d}{d\L} &=\frac{db_\d}{d\L}\Big{|}_{\rm free}  -a_2\fnl\left[\frac{34}{21}b_{\Psi\Pi^{[1]}}+b_{\d\Psi\Pi^{[1]}}+\frac{34}{21}b_{\Psi\Pi^{[2]}}\right]\vs
&\hspace{7cm}\times\left(\frac{\Lambda}{k_{\rm NG}}\right)^\D\left(\frac{H_0}{\L}\right)^2\frac{3\;\Omega_m}{2\;T(\L)}\frac{d\sigma_\L^2}{d\L} \,, \label{eq:bd_spin2}
\\
\frac{db_{\d^2}}{d\L} &=\frac{db_{\d^2}}{d\L}\Big{|}_{\rm free} -a_2\fnl\left[\frac{124}{315}b_{\Psi\Pi^{[1]}}+\frac{178}{105}b_{\d\Psi\Pi^{[1]}}+\frac{14347}{6027}b_{\Psi\Pi^{[2]}}+b^{\{\d^2\}_{\rm NG}}_{n=4}\right]
\vs
&\hspace{7cm}\times\left(\frac{\Lambda}{k_{\rm NG}}\right)^\D\left(\frac{H_0}{\L}\right)^2\frac{3\;\Omega_m}{2\;T(\L)}\frac{d\sigma_\L^2}{d\L} \,, \label{eq:bd2_spin2}
\\
\frac{db_{\G_2}}{d\L} &=\frac{db_{\G_2}}{d\L}\Big{|}_{\rm free} -a_2\fnl\left[-\frac{661}{4410}b_{\Psi\Pi^{[1]}}+\frac{4}{35}b_{\d\Psi\Pi^{[1]}}-\frac{241}{735}b_{\Psi\Pi^{[2]}}+b^{\{\G_2\}_{\rm NG}}_{n=4}\right]
\vs
&\hspace{7cm}\times\left(\frac{\Lambda}{k_{\rm NG}}\right)^\D\left(\frac{H_0}{\L}\right)^2\frac{3\;\Omega_m}{2\;T(\L)}\frac{d\sigma_\L^2}{d\L} \,, \label{eq:bG2_spin2} 
\ea
where again $\frac{db}{d\L}\Big{|}_{\rm free}$ represents the running of Gaussian coefficients due to the free theory described in \cite{Rubira:2023vzw} and for the non-Gaussian operator
\ba
\frac{db_{\Psi\Pi^{[1]}}}{d\L} &=-a_2\fnl\left[\frac{8}{105}b_{\d\Psi\Pi^{[1]}}+\frac{58}{305}b_{\Psi\Pi^{[2]}}+b^{\{\Psi\Pi^{[1]}\}_{\rm NG}}_{n=4}\right]\frac{d\sigma^2_\L}{d\L} 
\vs
&\quad-a_2\fnl\left[\frac{64}{105}b_{\d^2}+\frac{16}{15}b_{\d\G_2}+b^{\{\Psi\Pi^{[1]}\}_{\rm G}}_{n=4}\right]\left(\frac{k_{\rm NG}}{\Lambda}\right)^\D\frac{d\sigma_\L^2}{d\L} \,.
\ea
Those RGE are again summarized by \refeq{S2forallorder}, with the coefficients in \reftab{cCoeff}.  Results for those equations will be discussed in the following section.

%%%%%%%%%%%%%%%%%%%%%%%%%%%%%%%%%%%%%%%%%%%%%%%%%%%%%%%%%%%%%%%%%%%%%%%%%%%
%%%%%%%%%%%%%%%%%%%%%%%%%%%%%%%%%%%%%%%%%%%%%%%%%%%%%%%%%%%%%%%%%%%%%%%%%%%
\section{Results}
\label{sec:results}

After having derived the RGE for the spin-0 and spin-2 scenarios, we discuss the solutions of these equations in this section. 
We start with the evolution of the Gaussian bias parameters. We then focus on the evolution of the non-Gaussian parameters, splitting the discussion into local and non-local PNG.

\subsection{Gaussian coefficients} \label{sec:Gauss_sol}

The evolution of Gaussian coefficients (and the truncation of the RGE) was first developed for the free theory in \cite{Rubira:2023vzw}. Here we discuss how this is affected by the interaction term and the new non-Gaussian bias operators, focussing on spin-0 local PNG, since similar conclusions also hold for other PNG types. We can write \refeqs{bd_spin0}{bG2_spin0} as 
\ba
\frac{db_\d}{d\L} &= -\left[\frac{68}{21}b_{\d^2}(\L)+b_{n=3}^{\ast\{\d\}_{\rm G}}\right]\frac{d \s^2_\L}{d \L} \vs
&\hspace{2cm}-a_0\fnl\left[-\frac{13}{21}b_{\Psi}+\frac{13}{21}b_{\Psi\d}+b_{n=3}^{\ast\{\d\}_{\rm NG}}\right] \left(\frac{H_0}{\L}\right)^2\frac{3\;\Omega_m}{2\;T(\L)}\frac{d\sigma_\L^2}{d\L}\,; \label{eq:drun}
\\
\frac{db_{\d^2}}{d\L} &= -\left[\frac{8126}{2205}b_{\d^2}(\L) + b_{n=3+4}^{\ast\{\d^2\}_{\rm G}}  \right]\frac{d \s^2_\L}{d \L}
\vs
&\hspace{2cm}-a_0\fnl\left[\frac{43}{135}b_{\Psi}+\frac{478}{135}b_{\Psi\d}+b^{*\{\d^2\}_{\rm NG}}_{n=3+4}\right] \left(\frac{H_0}{\L}\right)^2\frac{3\;\Omega_m}{2\;T(\L)}\frac{d\sigma_\L^2}{d\L}\,; \label{eq:d2run}
\\
\frac{db_{\G_2}}{d\L} &= -\left[  \frac{254}{2205 }b_{\d^2}(\L) + b_{n=3+4}^{\ast\{\G_2\}_{\rm G}} \right]\frac{d \s^2_\L}{d \L} 
\vs
&\hspace{2cm}-a_0\fnl\left[-\frac{1699}{13230}b_{\Psi}+\frac{79}{2205}b_{\Psi\d}  +b^{*\{\G_2\}_{\rm NG}}_{n=3+4}\right] \left(\frac{H_0}{\L}\right)^2\frac{3\;\Omega_m}{2\;T(\L)}\frac{d\sigma_\L^2}{d\L} \,, \label{eq:Grun}
\ea
noticing again the different $\L$ dependence of non-Gaussian sources relative to the Gaussian sources discussed in \cite{Rubira:2023vzw}. We also use $b^\ast = b(\L_\ast)$ to denote the parameters evaluated at a fixed renormalization scale $\L_\ast$ with 
\ba
b_{n=3}^{\{\d\}_{\rm G}} = 3b_{\d^3}-\frac{4}{3}b_{\G_2\d}   \quad &\textrm{and} \quad 
b_{n=3}^{\{\d\}_{\rm NG}} = 2b_{\Psi\d^2}-\frac{4}{3}b_{\Psi\G_2} \,, \label{eq:hobd} 
\\
b_{n=3+4}^{\{\d^2\}_{\rm G}} = \frac{{68}}{7}b_{\d^3}  - \frac{376}{105}b_{\G_2\d} + b_{n=4}^{\{\d^2\}}    \quad &\textrm{and} \quad 
b_{n=3+4}^{\{\d^2\}_{\rm NG}} =  \frac{68}{21}b_{\Psi\d^2}-\frac{31}{21}b_{\Psi\G_2} +b^{\{\d^2\}_{\rm NG}}_{n=4}\,, 
\\
b_{n=3+4}^{\{\G_2\}_{\rm G}} = \frac{116}{105}b_{\G_2\d} + b_{n=4}^{\{\G_2\}}    \quad &\textrm{and} \quad 
b_{n=3+4}^{\{\G_2\}_{\rm NG}} = \frac{20}{21}b_{\Psi\G_2} +b^{\{\G_2\}_{\rm NG}}_{n=4}\,, \label{eq:hobG2}
\ea
with $b_{n=3}^{\{O\}}$ ($b_{n=4}^{\{O\}}$) representing contributions from third(fourth)-order operators to $O$, and $b_{n=3+4}^{\{O\}} = b_{n=3}^{\{O\}} + b_{n=4}^{\{O\}}$ from either Gaussian or non-Gaussian operators. We generically refer to {\it all} those higher-order bias parameters as $b_{\rm H.O.}$ when discussing solutions to the ODE. As pointed out by \cite{Rubira:2023vzw}, and verified below, these higher-operator terms constitute an important source for the running of lower-order operators, but one can safely neglect their running in the RGE.

\begin{figure}[t]
    \centering
    \includegraphics[width=\linewidth]{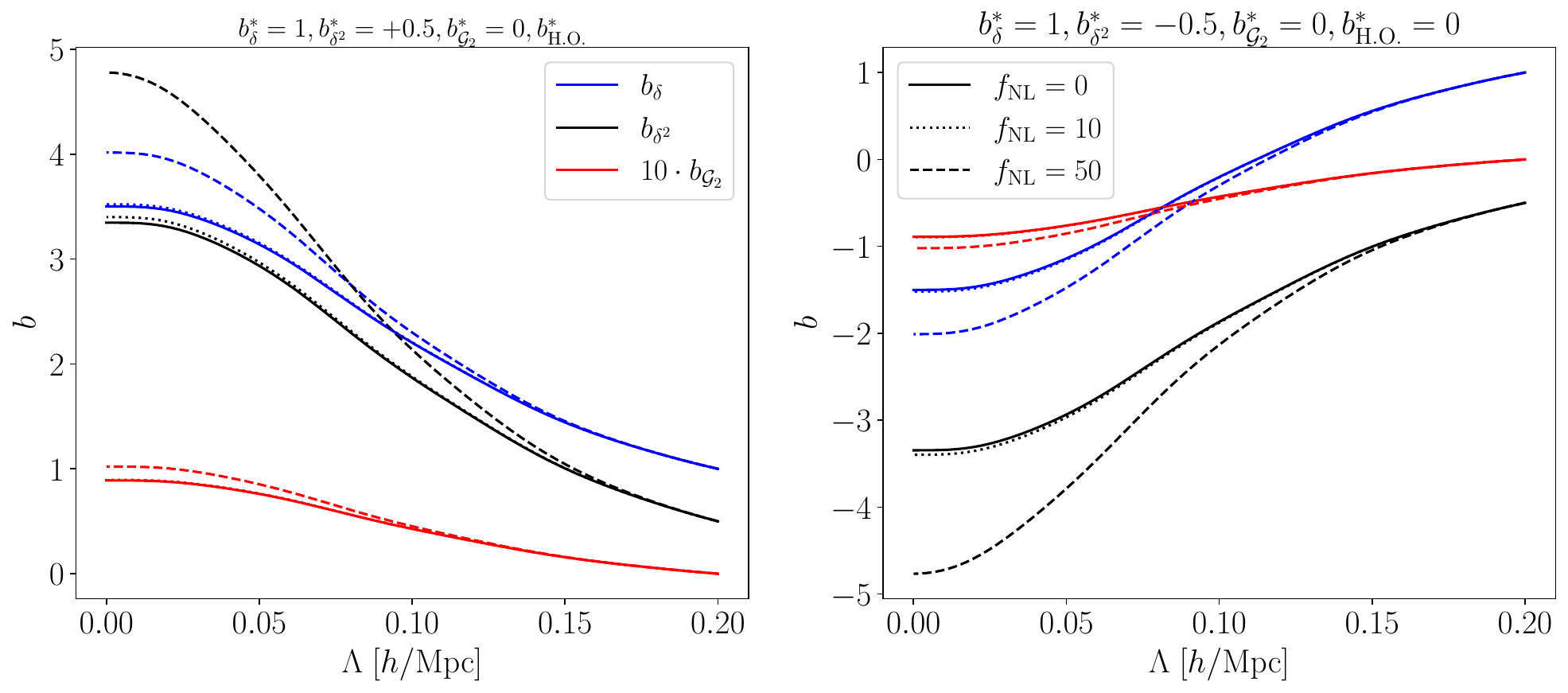}
    \caption{We show the running of the Gaussian bias parameters $b_\d$ and $b_{\d^2}$ and $b_{\G_2}$ for spin-0 and local ($\Delta = 0$) PNG. We present results for $\fnl=10$ (dotted) and $\fnl=50$ (dashed) and for two different initial conditions in the left and right panels. Furthermore we assume all higher-order bias terms to be zero ($b^*_{\rm H.O.} = 0$).}
    \label{fig:bias-Gaussian}
\end{figure}
We display in \reffig{bias-Gaussian} the evolution of the Gaussian bias operators for two different initial conditions (left and right panels) and for different values of $\fnl$ for spin-0 local PNG ($\Delta = 0$, discussed in more detail in the next section). We neglect higher-order operators fixing $b^*_{\rm H.O.} = 0$.
Notice the nonlinear dependence on $\fnl$: while the influence of $\fnl=10$ on the running of the Gaussian bias parameters is modest, it becomes significant for $\fnl = 50$. This is due to a back-reaction effect: the Gaussian parameters sensitively affect the non-Gaussian bias coefficients and those, in turn, contribute to the running of the Gaussian coefficients. The overall effect of interactions therefore involves higher powers of $\fnl$. In other words, despite the fact that the RGE are linear in $\fnl$ and in the biases, their solutions are not.\footnote{A linear analysis based on the RHS of \refeqs{bd_spin0}{bG2_spin0} and \refeqs{bd_spin2}{bG2_spin2} would lead to an estimate for the running of the Gaussian parameters from the renormalization scale $\L_\ast$ to another scale $\L_{\rm min}$ due to non-Gaussian parameters  to be (setting $a_L = 1$) 
\ba 
\Delta b_{\rm G}\Big{|}_{\rm lin} \propto & \,\fnl\int_{\L_{\rm min}}^{\Lambda_\ast} \frac{\L^2\,d\L}{2\pi^2} \left[ \left(\frac{H_0}{\L}\right)^2\frac{3\;\Omega_m}{2\;T(\L)}\frac{d\sigma_\L^2}{d\L} \right] 
%\vs
%& \hspace{-1cm} 
= \fnl \frac{3}{2}\Omega_m H_0^2 \int_{\L_{\rm min}}^{\Lambda_\ast} \frac{\L^2\,d\L}{2\pi^2}  \frac{\Plin(\L)}{\L^2} \frac{1}{T(\L)} 
\simeq 10^{-5}\, \fnl \,.
\label{eq:Hubblesupp} 
\ea
In the last equality we assumed $\L_\ast = 0.25 \,h/$Mpc and we integrate from $\L_{\rm min} = 0$. This $10^{-5}$  suppression can be misleading since it is only valid {\it at linear-order in} $\fnl$ and neglects the coupled ODE system.}
Therefore, when considering $\fnl \gtrsim 10$, one has to take into account the contribution of PNG to the running of Gaussian coefficients and solve the joint set of RGE as we do throughout this section (unless explicitly stated otherwise). 

Furthermore, another point evident from \reffig{bias-Gaussian} is that the scales in which the PNG effect for the running is relevant are different than the scales where the Gaussian running is important. While the latter are determined by the scale at which $\Pshell$ varies, the dynamical range of PNGs is proportional to $P^{\Psi}_{\rm shell}$ [see \refeq{Ppsishell}]. For the case of $\D = 0$ shown in \reffig{bias-Gaussian}, we find PNG to be relevant on larger scales.

%%%%%%%%%%%%%%%%%%%%%%%%%%%%%%%%
\subsection{Local (spin-0) PNG} \label{sec:localPNGolution}

\begin{figure}[t]
    \centering
    \includegraphics[width=0.9\linewidth]{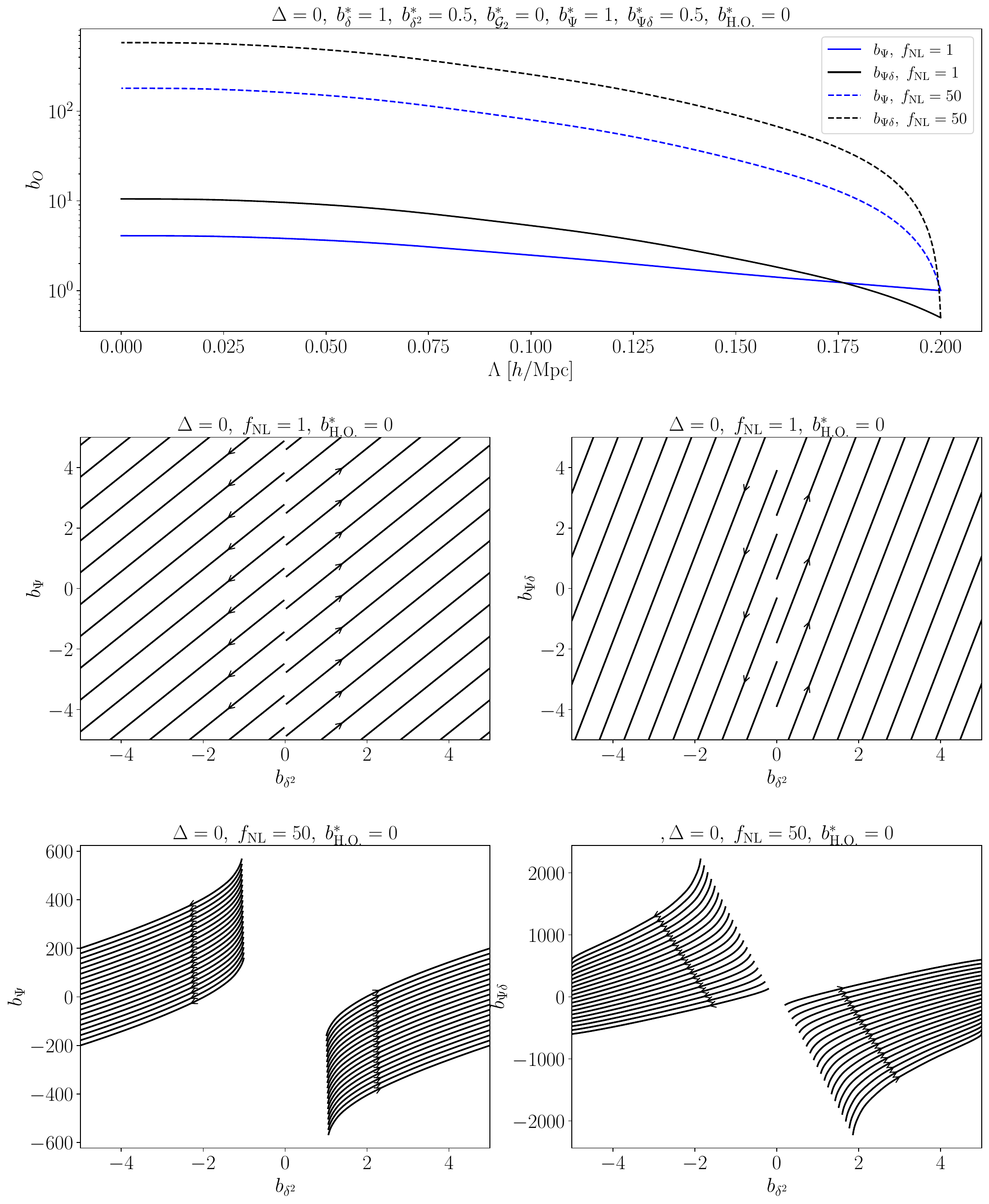}
    \caption{\emph{Top row:} the running of the bias parameters $b_\Psi$ (blue) and $b_{\Psi\d}$ (black) for local-type spin-0 PNG for different values of $\fnl^{\rm loc}$ (solid and dashed) and assuming all higher-order bias terms $b^*_{\rm H.O.} = 0$ at $\Lambda_\ast$. \emph{Middle and bottom rows:} the RG flow of $b_\Psi$ and $b_{\Psi\d}$ as a function of $b_{\d^2}$ for $\fnl = 1$ and $\fnl = 50$ respectively. Each line displays different initial conditions (with all other bias parameters fixed to zero at $\L \to 0$). The arrows point towards $\L \to 0$.
    }
    \label{fig:bias-local}
\end{figure}
We now turn to the running of the non-Gaussian coefficients. For local ($\D=0$) spin-0 non-Gaussianity, \refeqs{bpsi}{bpsid} reduce to
\ba
\frac{d b_\Psi}{d \L} &=-a_0\fnl b^{*\{\Psi\}_{\rm NG}}_{n=3}\frac{d\sigma^2_\L}{d\L}-4a_0\fnl b_{\d^2}\frac{d\sigma^2_\L}{d\L} \label{eq:dbPsi_local}\,,
\\
\frac{d b_{\Psi\d}}{d \L}&=-a_0\fnl\left[\frac{272}{21}b_{\d^2}+b^{*\{\Psi\d\}_{\rm G}}_{n=3+4}+b^{*\{\Psi\d\}_{\rm NG}}_{n=3+4}\right]\frac{d\sigma^2_\L}{d\L} \label{eq:dbPsid_local} \,,
\ea
in which we fixed the higher-order operators at $\L_\ast$ with
\ba
b^{*\{\Psi\}_{\rm NG}}_{n=3} &= b_{\Psi\d^2}^\ast\,,
\vs 
b^{*\{\Psi\d\}_{\rm G}}_{n=3+4} &= 12b_{\d^3}^\ast-\frac{8}{3}b_{\G_2\d}^\ast+b^{*\{\Psi\d\}_{\rm G}}_{n=4}\,,
\vs
b^{*\{\Psi\d\}_{\rm NG}}_{n=3+4} &= \frac{68}{21}b_{\Psi\d^2}^* +b^{*\{\Psi\d\}_{\rm NG}}_{n=4}\,.
\ea
We discuss this approximation later in this section.\footnote{We highlight that these equations can be solved analytically after neglecting the running of Gaussian coefficients due to non-Gaussian terms. In that case, we can change variables in \refeqs{dbPsi_local}{dbPsid_local} writing the equations in terms of the variance $\sigma^2(\L)$ 
\ba
\frac{d b_\Psi}{d \sigma^2} &=-a_0\fnl b^{*\{\Psi\}_{\rm NG}}_{n=3}-4a_0\fnl b_{\d^2} \,, \label{eq:bpsi_fixed}
\\
\frac{d b_{\Psi\d}}{d \sigma^2}&=-a_0\fnl\left[\frac{272}{21}b_{\d^2}+b^{*\{\Psi\d\}_{\rm G}}_{n=3+4}+b^{*\{\Psi\d\}_{\rm NG}}_{n=3+4}\right] \,, \label{eq:bpsid_fixed}
\ea
which can be directly integrated to find a solution similar to those provided in Sec.~2.4 of \cite{Rubira:2023vzw}. However, as mentioned in \refsec{Gauss_sol}, the coupling between the Gaussian and non-Gaussian coefficients is relevant and \refeqs{drun}{Grun} do not admit the same variable transformation. Therefore, we numerically solve the entire coupled ODE system.}

\begin{figure}[t]
    \centering
    \includegraphics[width=\linewidth]{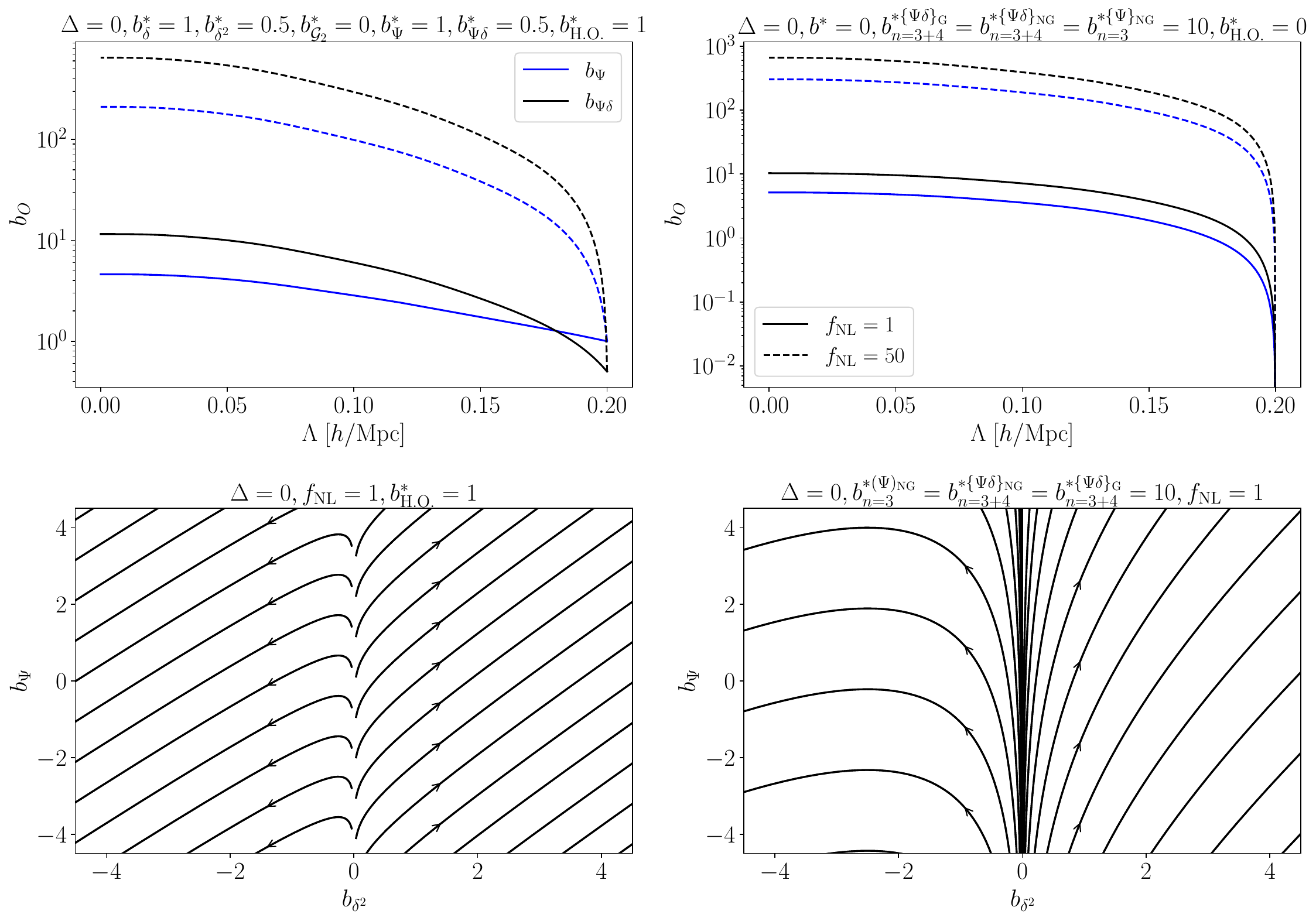}
    \caption{Similar to \reffig{bias-local}, but with non-zero higher-order bias operators. On the left, we fix all higher-order bias operators $b^*_{\rm H.O.} = 1$. On the right we set {\it all} initial conditions $b^*=0$ and all higher order bias parameters to zero, except  $b^{*\{\Psi\}_{\rm NG}}_{n=3} = b^{*\{\Psi\d\}_{\rm G}}_{n=3+4} = b^{*\{\Psi\d\}_{\rm NG}}_{n=3+4} = 10$. We notice in the right panels that even when setting the initial values of $b_\Psi, b_{\Psi\d}$ to zero, a nontrivial running is induced by non-zero higher order terms.}
    \label{fig:bias-local-static}
\end{figure}
We display in the top panels of \reffig{bias-local} some examples for the evolution of the local spin-0 bias parameters with $\fnl = 1$ (solid) and $\fnl = 50$ (dashed), which are still in agreement with structure formation probes.
We fixed the initial conditions at $\L_\ast$ and neglect all higher-order operators $b_{\rm H.O.} = 0$ ($b^{*\{\Psi\}_{\rm NG}}_{n=3} = b^{*\{\Psi\d\}_{\rm G}}_{n=3+4} = b^{*\{\Psi\d\}_{\rm NG}}_{n=3+4} = 0$). Again, we fix $a_0 = 1$.
The middle (bottom) panels show the RG flow of the those bias parameters as a function of $b_{\d^2}$ for $\fnl = 1$ (50). Lower values of $\fnl$ lead to a rather simple RG structure, similarly to \cite{Rubira:2023vzw} while higher values of $\fnl$ (bottom panels) lead to a more complex structure in the RG flow due to the back-reaction of the Gaussian-non-Gaussian coupling for the parameters running discussed in \refsec{Gauss_sol}.

We next turn to the impact of higher-order operators, whose coefficients we have set to zero so far.
We consider in the panels of \reffig{bias-local-static} scenarios with non-zero higher-order biases. In the left, we take {\it all} higher-order bias parameters $b^*_{\rm H.O.} = 1$ at $\L_\ast \simeq 0.2 h/$Mpc.    
We see that the RG flow of the non-Gaussian operators present a more complex structure (bottom panels) compared to the bottom panels of \reffig{bias-local}.
The right panels consider the scenario in which all first- and second-order bias parameteres are zero and only $b^{*\{\Psi\d\}_{\rm G}}_{n=3+4}+b^{*\{\Psi\d\}_{\rm NG}}_{n=3+4},\;b^{*\{\Psi\}_{\rm NG}}_{n=3}$ is non-zero in \refeq{bpsid_fixed}. We find in that case that despite all parameters starting at zero, the higher-order operators source $\Psi$ and $\Psi\d$  via \refeq{bpsi_fixed} and \refeq{bpsid_fixed}. We find then substantial differences in the bias evolution when compared to the scenario in which higher-order bias parameters are zero (\reffig{bias-local}).

\begin{figure}[t]
    \centering
    \includegraphics[width=\linewidth]{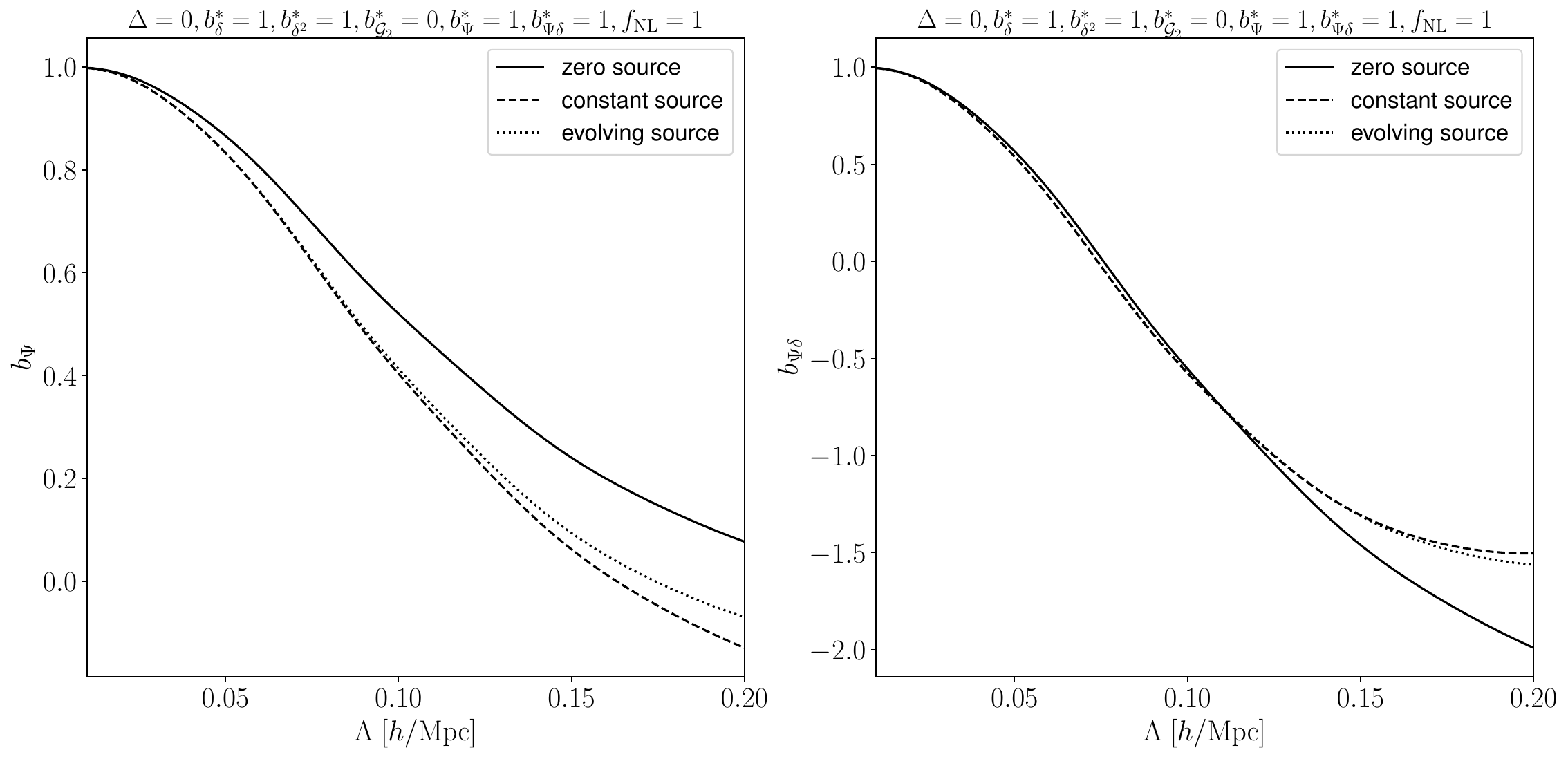}
    \caption{Running of the non-Gaussian bias parameters $b_\Psi$ (left panel) and $b_{\Psi\d}$ (right panel) neglecting the higher-order bias parameters (solid), assuming them as constants (dashed) or considering the exponential ansatz \refeq{bansatz} (dotted). We set the initial conditions of the RGE at $\L_\ast =10^{-5} \,h/$Mpc. We neglected here the non-Gaussian counterpart  related to the running of the Gaussian parameters, since for $f_{\rm NL}=1$ it is suppressed.}
    \label{fig:bias-local-ansatz}
\end{figure}

Finally, we discuss the truncation of the RGE hierarchy by neglecting the \emph{running} of higher-order bias terms.
We display in \reffig{bias-local-ansatz} the solutions for local PNG bias operators when neglecting higher-order operators (solid) and when taking them as constant-in-$\L$ factors (dashed), now setting the initial conditions at  $\L_\ast =10^{-5}\,h/$Mpc. We notice a substantial difference when considering those terms, especially for $b_\Psi$. In fact, this is in agreement with the results of \cite{Rubira:2023vzw} and there is no reason to expect that those contributions are suppressed.
Similar to Eq.~(2.40) of \cite{Rubira:2023vzw}, we also include as dotted lines in \reffig{bias-local-ansatz} the results for considering a heuristic ansatz
\bea \label{eq:bansatz}
 b_{n=3+4}^{\{O\}}(\s^2) = b^{*\{O\}}_{n=3+4}\, e^{-s^{\{O\}}(\s^2 - \s^2_\ast)} \,, 
\eea
for the evolution of higher-order bias parameters, with $s^{\{O_G\Psi\}}$ the equivalent of the shell prefactors for higher order non-Gaussian bias parameters and $b^{*\{O\}}_{n=3+4}$ the higher order bias. They are fixed at $s^{\{O_G\Psi\}}=b^{*\{O\}}_{n=3+4}=1$. We can see that both for $b_{\Psi}$ and $b_{\Psi\d}$ the difference of the dynamical ansatz \refeq{bansatz} for a constant higher-order bias operator is minimal, especially on large scales.  In summary, we find similar results to those reported in \cite{Rubira:2023vzw}, in which third and fourth-order operators are found to be relevant for the evolution of first and second-order operators, but it is sufficient (on large scales) to approximate those terms as $\L$-independent constants. Thus, this offers a self-consistent way to truncate the bias hierarchy within the RG flow. 

%%%%%%%%%%%%%%%%%%%%%%%%%%%%%%%%
\subsection{Non-local (spin-0 and spin-2) PNG}

We now focus on non-local ($\Delta \neq 0$) PNG both for spin-0 and spin-2 fields.
For spin-0, we have for the non-Gaussian operators
\ba
\frac{d b_\Psi}{d \L} &=-a_0\fnl b^{*}_{\Psi\d^2}\frac{d\sigma^2_\L}{d\L}-4a_0\fnl b_{\d^2}\left(\frac{k_{\rm NG}}{\Lambda}\right)^\Delta \frac{d\sigma_\L^2}{d\L}\,,
\\
\frac{d b_{\Psi\d}}{d \L} &=-a_0\fnl\left[\frac{272}{21}b_{\d^2}+b^{*\{\Psi\d\}_{\rm G}}_{n=3+4}\right]\left(\frac{k_{\rm NG}}{\Lambda}\right)^\Delta \frac{d\sigma_\L^2}{d\L} -a_0\fnl b^{*\{\Psi\d\}_{\rm NG}}_{n=3}\frac{d\sigma^2_\L}{d\L}\,,
\ea
and for the spin-2 case, the running of $\Tr\left[\Psi\Pi^{[1]}\right]$ is given by
\ba
\frac{d b_{\Psi\Pi^{[1]}}}{d \L}=-a_2\fnl b^{*\{\Psi\Pi^{[1]}\}_{\rm NG}}_{n=4}\frac{d\sigma^2_\L}{d\L}-a_2\fnl\left[\frac{64}{105}b_{\d^2}+ b^{*\{\Psi\Pi^{[1]}\}_{\rm G}}_{n=3+4}\right]\left(\frac{k_{\rm NG}}{\Lambda}\right)^\Delta\frac{d\sigma_\L^2}{d\L} \,.
\ea
Note that now $\left(k_{\rm NG}/\L\right)^\Delta$ enhances the running of non-Gaussian coefficients due to Gaussian contributions for $\Delta >0 $. 
The term proportional solely to $d\sigma^2_\L/d\L$, arising due to the free contribution in the action is therefore suppressed compared to the enhanced term.
Therefore, the running of non-Gaussian $\Delta>0$ PNG is mainly driven by Gaussian terms due to the interaction term in the action.
Furthermore, note the extra dependence with $\L$ compared to the local type described in the former section.

\begin{figure}[t]
    \centering
    \includegraphics[width=\linewidth]{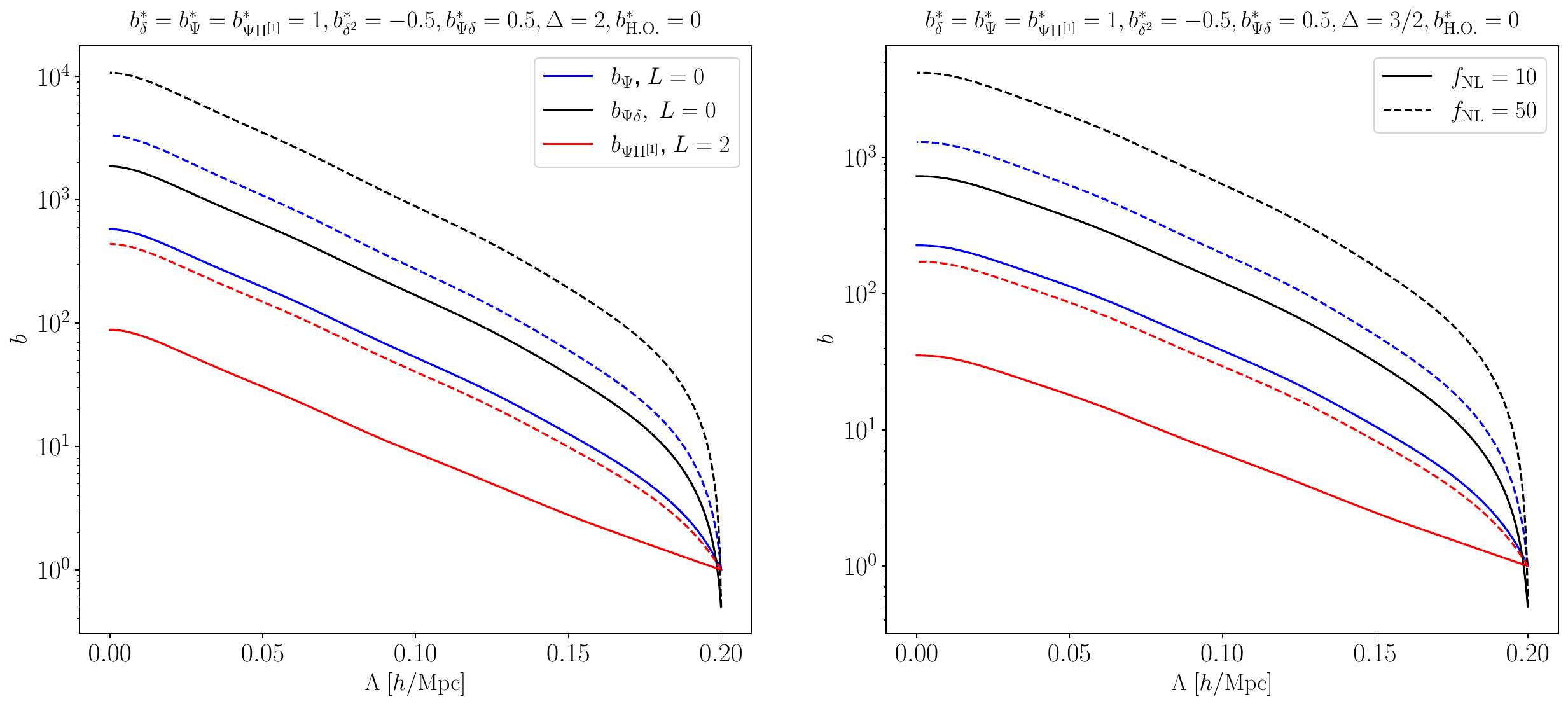}
    \caption{\emph{Left:} running of all the non-Gaussian parameters for $\D=2$ and $\fnl=10 \,(50)$ in solid (dashed) lines for spin $L=0$ and $L=2$ operators. \emph{Right:} same, but for $\D=3/2$. We present both spins together for convenience, but the respective ODE are solved separately since they do not mix.}
    \label{fig:bias-allspins}
\end{figure}
\begin{figure}[t]
    \centering
    \includegraphics[width=\linewidth]{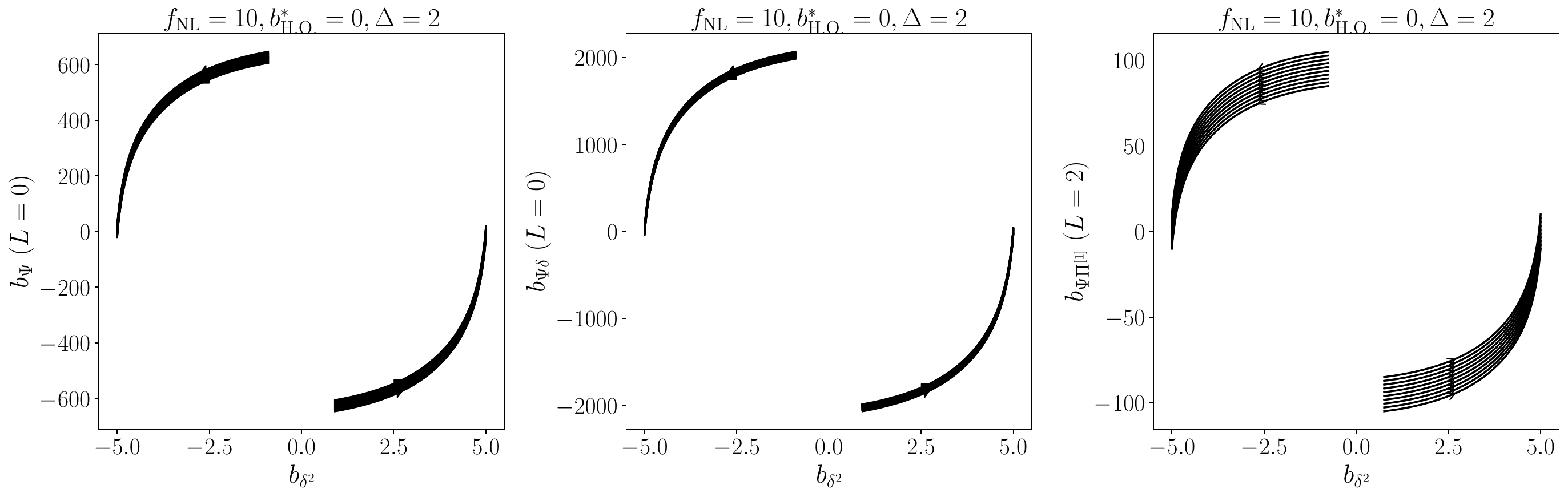}\\
    \includegraphics[width=\linewidth]{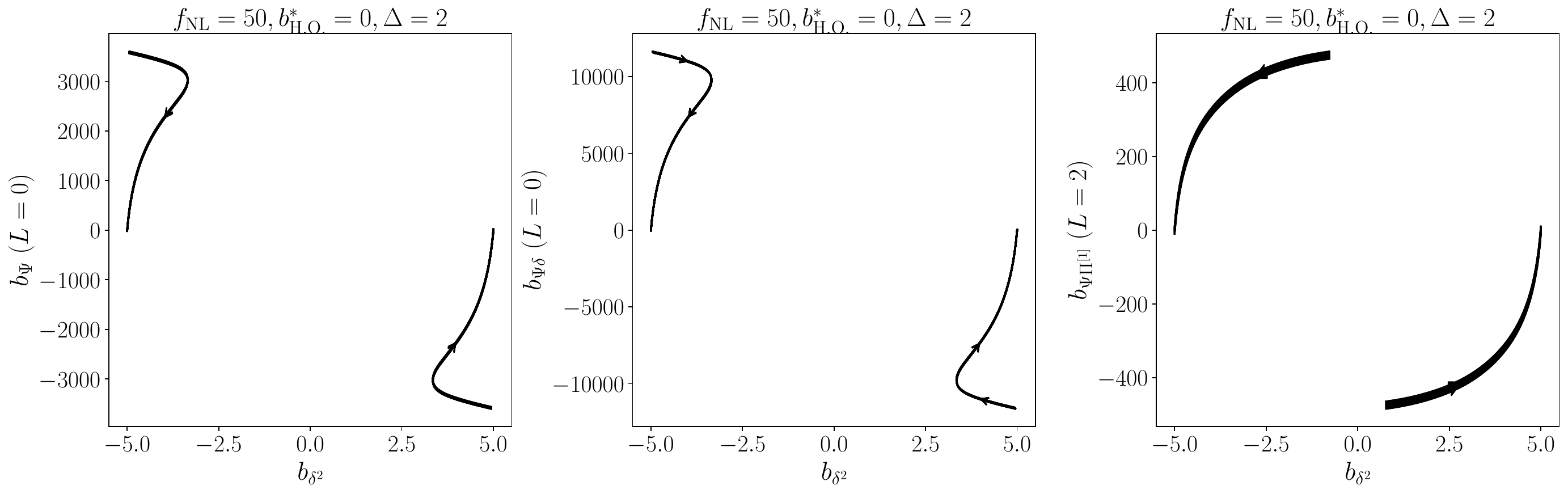}
    \includegraphics[width=\linewidth]{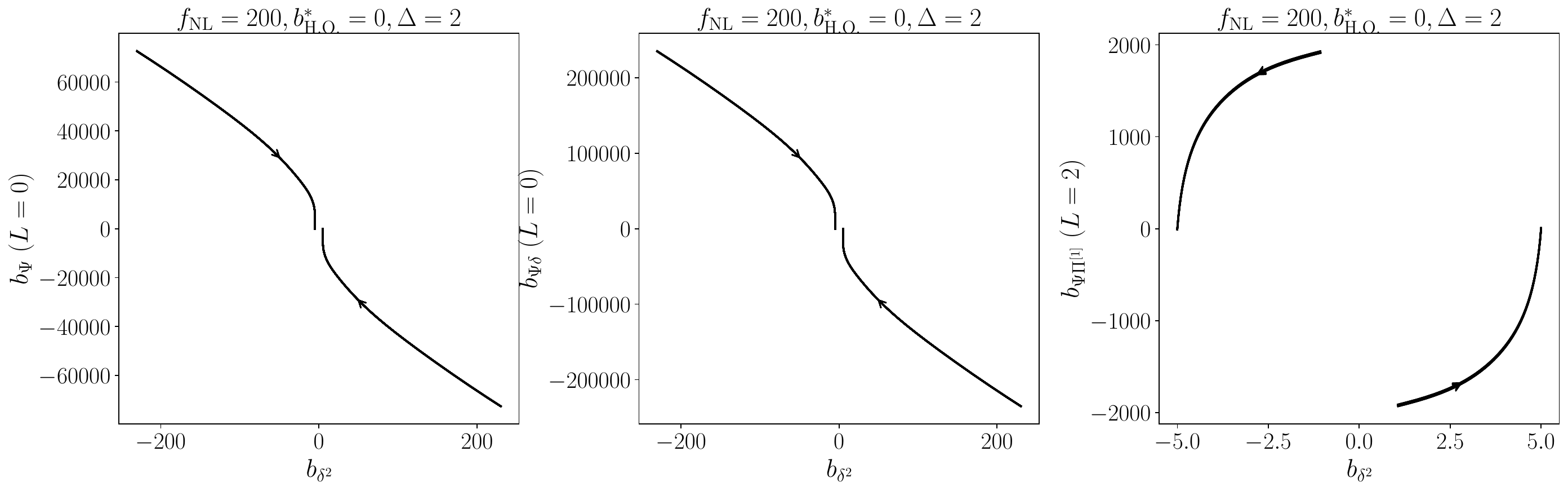}
    \caption{Running of non-Gaussian parameters for $\D=2$ and $f_{\rm NL}=10$ (top), $50$ (middle)  and 200 (bottom). Note the change in the regime as $\fnl$ increases: for low $\fnl$ the $b_{\d^2}$ term on the RHS of \refeq{d2run} drives the evolution of $b_{\d^2}$, but as $\fnl$ increases, $b_{\Psi}$ and $b_{\Psi \d}$ become relevant. Each line displays different initial conditions with all other bias parameters fixed to zero at $\L \to 0$, the direction indicated by the arrows. }
    \label{fig:RG-allspins-D2}
\end{figure}

We display the numerical solution for the $\Delta = 2$, $L=0,2$ cases in the left panel of \reffig{bias-allspins}, with the evolution of $b_\Psi$ (first-order spin-0), $b_{\Psi \d}$ (second-order spin-0) and $b_{\Psi \d}$ (second-order spin-2) bias parameters as a function of $\L$. 
We fix $\fnl = 10$ (solid) and $\fnl = 50$ (dashed), values that are still allowed by CMB \cite{planckcollaboration2019planck} and LSS data. 
Note that the running of the $\D=2$ bias is enhanced due to the extra $\left(k_{\rm NG}/\L\right)^\Delta$ factor, as compared to the local scenario described in \refsec{localPNGolution} and \reffig{bias-local}. 
The RG flow of those parameters as a function of $b_{\d^2}$ are displayed in \reffig{RG-allspins-D2} for $\fnl = 10,\,50$ and $200$ for some representative initial conditions. We find that increasing $\fnl$ leads to a more rich RG flow structure, since the coupling with Gaussian parameters start to be significant: we change from a weakly coupled scenario (in which we could in principle solve separately for Gaussian and non-Gaussian terms) to a strongly coupled scenario in which both ODE systems have to be solved together. For this last case, we caution that the $\fnl^2$ contributions neglected in this work could start to play a significant role. Notice that most of the trajectories in \reffig{RG-allspins-D2}  converge toward an attractor trajectory as we evolve towards the IR ($\L \to 0$).

\begin{figure}[t]
    \centering
    \includegraphics[width=\linewidth]{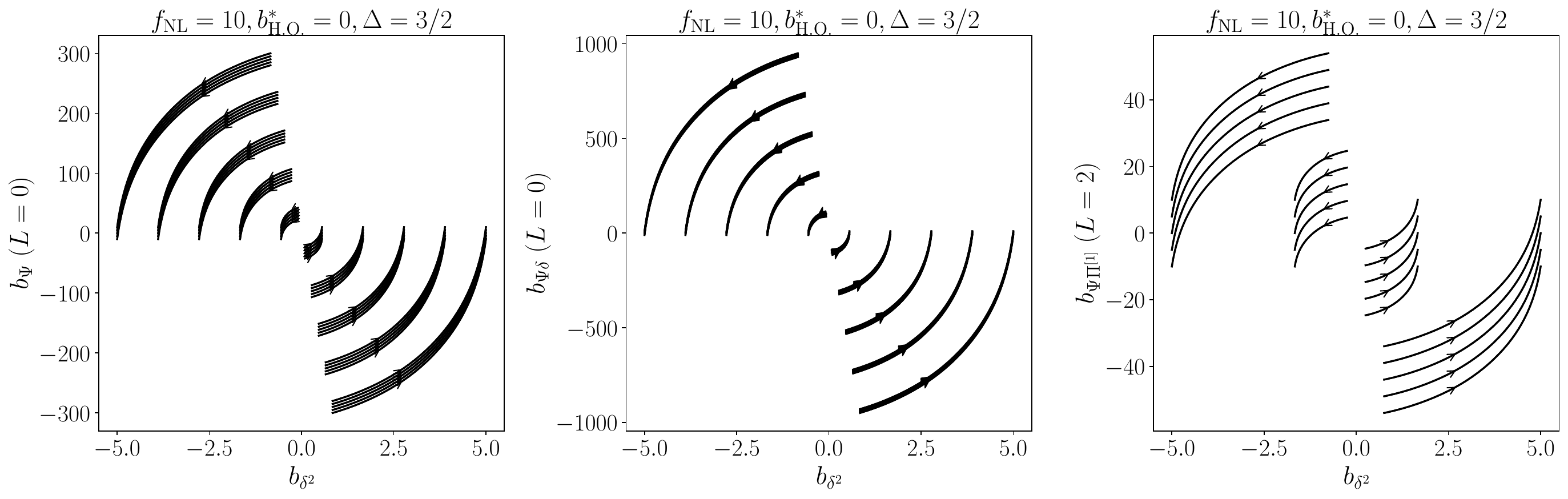}
    \includegraphics[width=\linewidth]{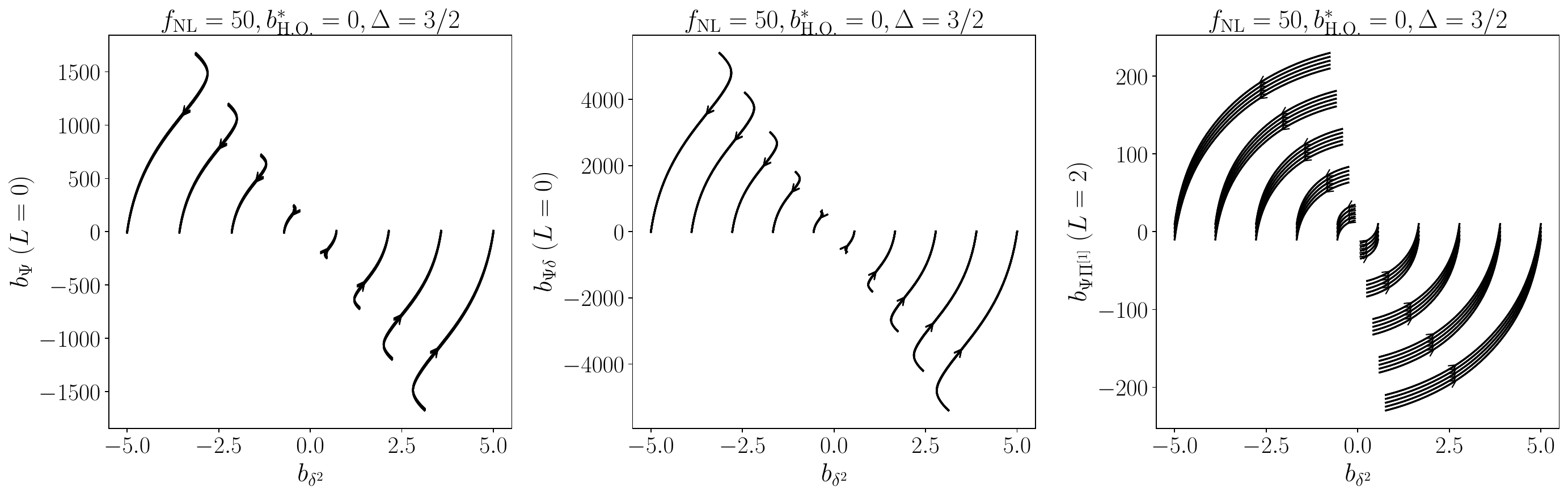}
    \includegraphics[width=\linewidth]{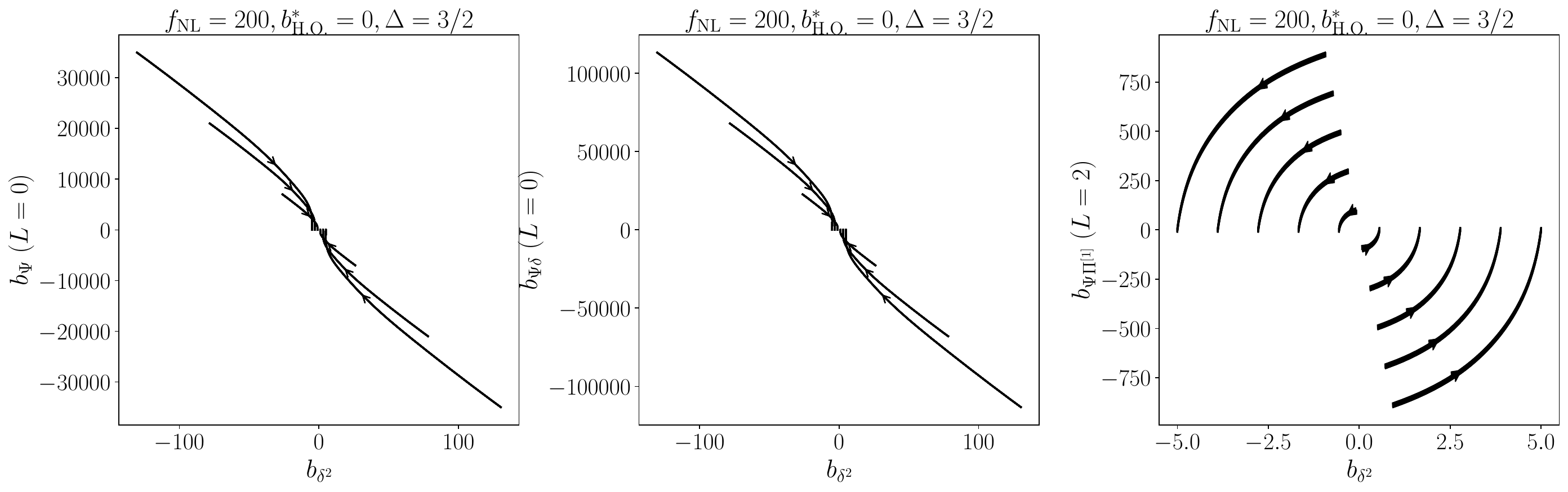}
    \caption{Same as \reffig{RG-allspins-D2} but for $\D = 3/2$. Here we have added more trajectories starting at different values of $b_{\d^2}$ of the RG flow to better illustrate the flow in the RG phase space.}
    \label{fig:RG-allspins-D3half}
\end{figure}

 Moreover, to exemplify other PNG cases, we also display in the right panel of \reffig{bias-allspins} the evolution of the bias parameters as a function of $\L$ for the $\Delta = 3/2$ case for different $\fnl$, as we did for $\D=2$. The RG-flow for $\Delta = 3/2$ is displayed in  \reffig{RG-allspins-D3half}. In the $\D=3/2$ case, we included an enlarged set of initial conditions, leading to trajectories through different regions of the phase space. We see a very similar behaviour to the $\D = 2$ scenario.
 For both $\D = 2$ and $\D = 3/2$ we can see a change in the main force driving the RG flow as we increase $\fnl$: for low $\fnl$ the $b_{\d^2}$ term on the RHS of \refeq{d2run} seems to drive the evolution of $b_{\d^2}$, but as $\fnl$ increases the $b_{\Psi}$ and $b_{\Psi \d}$ contributions become relevant.
 Moreover, we notice that the running of the spin-2 $b_{\Psi\Pi^{[1]}}$ parameter is substantially slower compared to the other spin-0 parameters, since $s_{\Psi\Pi^{[1]}, \d^2} = 64/105$ is smaller compared to the other coefficients in \reftab{cCoeff}. In summary, we find a richer structure for the RG evolution when compared to the solely Gaussian case of \cite{Rubira:2023vzw}. This could lead to relevant changes in the fixed point structure of the system, a subject that we plan to address in a future project.

\begin{figure}[t]
    \centering
    \includegraphics[width=0.6\linewidth]{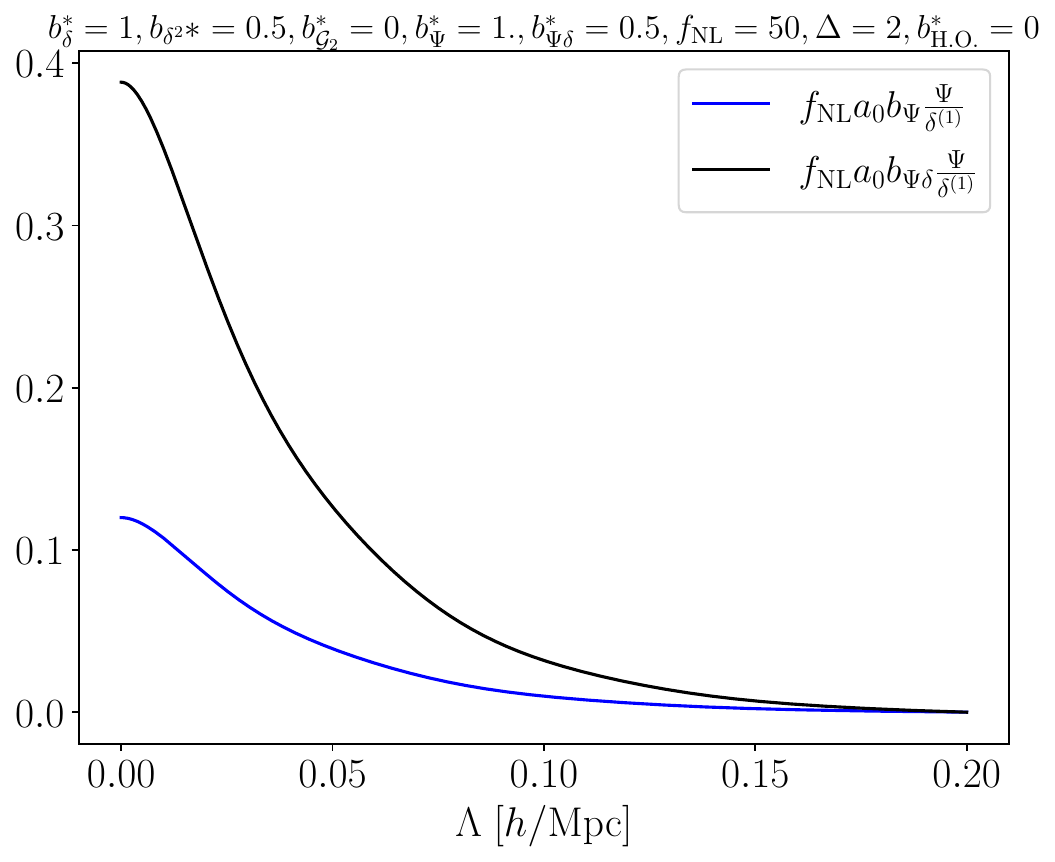}
    \caption{The running of the non-Gaussian contributions to the $n$-point galaxy statistics $f_{\rm NL}a_0b_{\Psi}\frac{\Psi_\L}{\dlin_\L}$ and $f_{\rm NL}a_0b_{\Psi\delta}\frac{\Psi_\L}{\dlin_\L}$ evaluated at different $\L$, such that we can compare the contribution of the RG running for those terms relative to the linear power spectra. We assume all quantities are calculated at a wavelength $k\lesssim k_{\rm eq}$ such that $T(k)= 1$. We observe a moderate running in contrast to the ``bare'' bias operators $b_{\Psi}$ and $b_{\Psi\d}$. Notice that this running is to be absorbed by other parameters, such that the $n$-point function becomes $\L$-independent.}
    \label{fig:deltab}
\end{figure}

It is evident that for both $\Delta=2$ and $\Delta=3/2$ cases the bias parameters grow a few orders of magnitude due to the enhancement mentioned above. One may then wonder about the observational consequences of such high values for the bias parameters. Notice that the ultimate observable in the case of the $2$-point function at tree-level is (see \refapp{pkbk} for a complete derivation)
\ba
    \left<\d_g(\vk)\d_g(\vk')\right>_{\rm tree }'&=(b_\d^\L)^2 \left<\dlin_\L(\vk)\dlin_\L(\vk')\right>'
    + 2 a_0\fnl b_\d^\L b_\Psi^\L \left<\dlin_\L(\vk)\Psi_\L(\vk')\right>'
    \vs
    &\quad  + \left(a_0\fnl\right)^2(b_\Psi^\L)^2 \left<\Psi_\L(\vk)\Psi_\L(\vk')\right>' \,.
    \ea
    For $\Delta \gtrsim 1$, and as long as $\fnl$ is not extremely large, the last term is numerically suppressed. The main PNG contribution is therefore
\be \label{eq:b_contribution}
a_0\fnl b_\d^\L b_\psi^\L \left<\dlin_\L(\vk)\Psi_\L(\vk')\right>' = a_0\fnl b_\d^\L b _\Psi^\L \left(\frac{k}{k_{\rm NG}}\right)^\Delta \frac{1}{M(k)}\PlinL(k)\,.
\ee
We aim to compare the term to the linear Gaussian term $(b_\d^\L)^2 \left<\dlin(\vk)\dlin(\vk')\right>'$ as a function of $\L$. We focus on $\D = 2$, for which the only $k$-dependent part is the transfer function $T(k)$ inside $M(k)$. In that case, if we focus on wavenumbers smaller than the equality scale, $k<k_{\rm eq}$, we have $T(k)\sim 1$ and \refeq{b_contribution} becomes $k$ independent.
We display in \reffig{deltab} the ratios\footnote{Notice that in that case the ratio of operators is
meant to display the same $(k/k_{\rm NG})^\D\;M^{-1}(k)$ dependence as the one in \refeq{b_contribution} }
\be b_{\Psi}^\L \frac{\Psi_\L}{\dlin_\L} \,,
\quad \textrm{and} \quad b_{\Psi\d}^\L \frac{\Psi_\L}{\dlin_\L}\,,
\ee
to estimate the relative size of this contribution. We find that, despite the running in the $b^\L_O$ varying by a few orders of magnitude, the relative contribution of this PNG $\L$-running term only changes by a small factor relative to the linear bias parameter. Ultimately, the $n$-point functions are independent of $\L$ by construction, and the $\L$-dependence part of $b_\Psi^\L$ and $b_{\Psi \d}^\L$ described in \reffig{deltab} is absorbed by the running of other parameters, including $b_\d^\L$.
Thus, \reffig{deltab} shows that the apparently large running of $b_\Psi^\L$, for example, can in fact be absorbed by a modest running of $b_\d^\L$.

\begin{figure}[t]
    \centering
    \includegraphics[width=\linewidth]{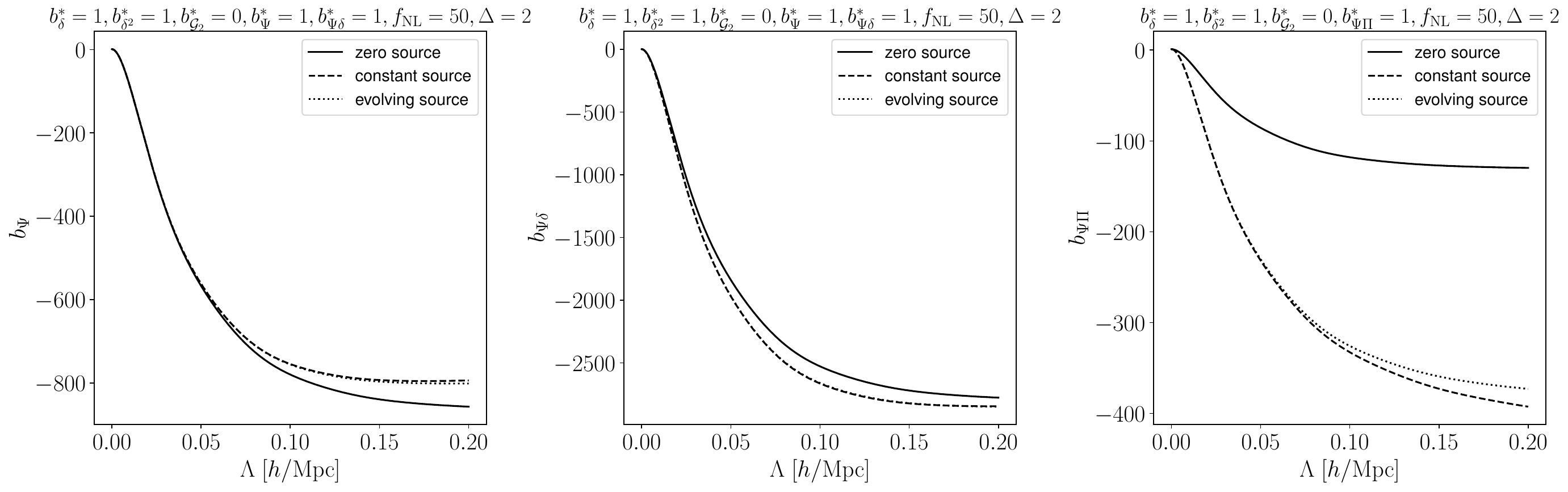}
    \caption{Similar to \reffig{bias-local-ansatz} but for the $\Delta=2$ scenario with $\fnl=50$. In order to solve the equations with the analytical ansatz \refeq{bansatz}, we approximate the running of non-Gaussian bias as dominated by the part that involves the enhancing factor $\propto \left(k_{\rm NG}/\L\right)^\D$ and we neglect the contribution to the evolution controlled by just $d\sigma^2_\L/d\L$.}
    \label{fig:bias-D=2-ansatz}
\end{figure}

Finally, we discuss the truncation of the RG equations for non-local PNG, similarly to \refsec{localPNGolution}. We display in \reffig{bias-D=2-ansatz} the same scenarios considered in \reffig{bias-local-ansatz}, but now for $\Delta=2$ $L=0,2$ cases. We find a similar conclusion: the inclusion of higher-order terms substantially changes the running of the lower-order parameters. However, the ansatz \refeq{bansatz} for the running of those higher-order parameters does not significantly contribute to lower-order operators. Taking those contributions as constant is therefore sufficient and the truncation in terms of the order of the coefficients still works.

%%%%%%%%%%%%%%%%%%%%%%%%%%%%%%%%%%%%%%%%%%%%%%%%%%%%%%%%%%%%%%%%%%%%%%%%%%%
%%%%%%%%%%%%%%%%%%%%%%%%%%%%%%%%%%%%%%%%%%%%%%%%%%%%%%%%%%%%%%%%%%%%%%%%%%%

%%%%%%%%%%%%%%%%%%%%%%%%%%%%%%%%%%%%%%%%%%%%%%%%%%%%%%%%%%%%%%%%%%%%%%%%%%%
%%%%%%%%%%%%%%%%%%%%%%%%%%%%%%%%%%%%%%%%%%%%%%%%%%%%%%%%%%%%%%%%%%%%%%%%%%%
\section{Discussion and conclusions}
\label{sec:disc}

In this work, we have extended the RG-LSS framework of \cite{Rubira:2023vzw,Rubira:2024tea} to include interaction terms from primordial non-Gaussianities. We have considered a cubic vertex \refeq{nonGoverdensity}, and work to linear order in $\fnl$ throughout. We have found interaction terms \refeq{Sintsplit} that can be added directly to the Lagrangian. The new interaction can be easily incorporated using the Wilson-Polchinski framework to understand the running of bias parameters when including PNG, providing a master formula \refeq{S2forallorder} valid at linear order in $\fnl$. We identify the main diagrams [see \refeq{shell-exp} and \refapp{cubic}] from the interaction term that contribute to the running of the bias.
In addition to the well-known squeezed limit of the interaction kernel ({\it scattering} contribution), which has been the basis for the literature on the effect of PNG on LSS so far, we identify a new {\it annihilation} contribution which is in a different kinematic regime than the squeezed limit. We argue that it will be absorbed by a new stochastic contribution to the partition function, but defer a detailed investigation of this new contribution to future work.

Using the squeezed limit of the interaction kernel, parametrized as $\propto \fnl (k_\ell/k_s)^\Delta$, we derive the RGE describing the running of the bias parameters for operators induced by spin-0 and 2 interactions during inflation,
and show that non-Gaussian bias terms are in fact generated by the RG flow. 
The RGE for the bias evolution are still linear, as in the case for Gaussian initial conditions. 
We provide solutions for the bias running for some specific scenarios (spin-0 local and equilateral and spin-2 equilateral). These solutions in fact scale non-linearly with the interaction strength $\fnl$. This also applies to the Gaussian bias coefficients, i.e. those present in the bias expansion without PNG.
  Moreover, the running of non-Gaussian coefficients also depends on the Gaussian terms and is enhanced for larger values of $\Delta$. 

We find that the solutions for the evolution of the coefficients present a rich structure (see e.g. \reffig{RG-allspins-D2}), much richer than the case for bias evolution with Gaussian initial conditions, which could substantially change the discussion around fixed points.

The RG solutions predict the value of the PNG bias coefficients, once fixed at a given scale, at any other scale. This evolution can be compared with measurements of the PNG bias coefficients as a function of scale (e.g. in simulations at the field level \cite{2021JCAP...08..029B}), which provides an independent validation of the EFT description of galaxy statistics. Should PNG be detected in actual data, this validation can also be applied there. Moreover, the evolution of the non-Gaussian bias coefficients with scale is important to put priors on these coefficients in field-level inference analyses of PNG \cite{Andrews:2022nvv}, which necessarily work at a finite scale \cite{Schmidt:2020viy}. Whether the RGE is able to ressum part of diagrams improving the theoretical prediction is something we leave for a future project.

In the future it would also be interesting to generalize the interaction terms to include higher-order vertices, such as $g_{\rm NL}$ and higher-in-$\fnl$ contributions. Going to higher-order in PNG including new types of vertices could eventually induce ($\L$-dependent) corrections to $\fnl$ itself, similarly to what happens in the context of QFT. Also, the structure of stochasticity when including non-Gaussianities changes, as we point out in \refapp{cubic}.
A related issue is the ambiguity of the kernel $K_{\rm NL}$ beyond the squeezed limit when given only the bispectrum as constraint \cite{Schmidt:2010gw,Scoccimarro_2012}.
We leave a more in-depth study of these issues for a future work.

%%%%%%%%%%%%%%%%%%%%%%%%%%%%%%%%%%%%%%%%%%%%%%%%%%%%%%%%%%%%%%%%%%%%%%%%%%%
%%%%%%%%%%%%%%%%%%%%%%%%%%%%%%%%%%%%%%%%%%%%%%%%%%%%%%%%%%%%%%%%%%%%%%%%%%%
\acknowledgments
HR is supported by the Deutsche Forschungsgemeinschaft under Germany's Excellence Strategy EXC 2094 `ORIGINS'. (No.\,390783311). CN acknowledges support from DAAD through the scholarship programme ``Master studies for all academic disciplines.'' We thank Mathias Garny, Julia Stadler, Beatriz Tucci and Leander Thiele for discussions. 
\appendix

%%%%%%%%%%%%%%%%%%%%%%%%%%%%%%%%%%%%%%%%%%%%%%%%%%%%%%%%%%%%%%%%%%%%%%%%%%%
%%%%%%%%%%%%%%%%%%%%%%%%%%%%%%%%%%%%%%%%%%%%%%%%%%%%%%%%%%%%%%%%%%%%%%%%%%%
\section{Non-Gaussianities via direct coupling with the current} \label{app:couplingtoJ}

In this section we will show that at \textit{linear} order in $\fnl$ our description including an interaction term in the partition function is equivalent to using non-Gaussian statistics in the matter density directly. From the partition function \refeq{non-GaussZ}, and writing the Gaussian measure explicitly, we have
\ba
&\Z[J_{\rm m,\L},J_{\rm g,\L}]=\int\Del\dlin_\L\exp\left\{-\frac{1}{2}\int_{\vk}^\L \frac{|\dlin_\L|^2}{\PlinL(k)}+S_{\rm int}\right\}
\vs
&\hspace{3cm}\times \exp\left\{\int_{\vk}J_{m,\L}(\vk)\delta[\dlin_\L](-\vk)+\int_{\vk}J_{g,\L}(\vk)\delta_{g}[\dlin_\L](-\vk)\right\}\,.
\ea
Note that the $J$-independent term can be written as
\ba
&-\frac{1}{2}\int_{\vk}^\L \frac{|\dlin_\L|^2}{\PlinL(k)}+S_{\rm int}=-\frac{1}{2}\int_{\vk}^\L \frac{\dlin_\L(\vk)}{\PlinL(k)} \left(\dlin_\L(-\vk) \right.
\\
&\hspace{3cm}\left.-2\fnl\int_{\vp_1,\vp_2}\diracpi(\vk-\vp_{12})\Knl(\vp_1,\vp_2)\frac{M(k)}{M(p_1)M(p_1)}\dlin_\L(\vp_1)\dlin_\L(\vp_2)\right)\,. \nonumber
\ea
After applying the transformation
\be
\Tilde{\d}_{\L}^{(1)}(\vk)=\dlin_\L(\vk)-\fnl\int_{\vp_1,\vp_2}\diracpi(\vk-\vp_{12})\Knl(\vp_1,\vp_2)\frac{M(k)}{M(p_1)M(p_1)}\dlin_\L(\vp_1)\dlin_\L(\vp_2) \,,
\ee
the inverse transformation can be written as
\be
\dlin_\L(\vk)=\Tilde{\d}_{\L}^{(1)}(\vk)+\fnl\int_{\vp_1,\vp_2}\diracpi(\vk-\vp_{12})\Knl(\vp_1,\vp_2)\frac{M(k)}{M(p_1)M(p_1)}\Tilde{\d}_{\L}^{(1)}(\vp_1)\Tilde{\d}_{\L}^{(1)}(\vp_2)+\O(\fnl^2) \,,
\ee
and we get
\be
-\frac{1}{2}\int_{\vk}^\L \frac{|\dlin_\L|^2}{\PlinL(k)}+S_{\rm int}=-\frac{1}{2}\int_{\vk}^\L \frac{|\Tilde{\d}_{\L}^{(1)}|^2}{\PlinL(k)}.
\ee
Furthermore, at linear order in $\fnl$ we see that\footnote{The Jacobian of this transformation does not contribute at $\O(\fnl)$.}
\be \label{eq:Dtransform}
\int \Del\Tilde{\d}_{\L}^{(1)}\Big|\frac{\Del\dlin_\L}{\Del\Tilde{\d}_{\L}^{(1)}}\Big|\P[{\Tilde{\d}_{\L}^{(1)}}]=\int \Del\Tilde{\d}_{\L}^{(1)}\P[{\Tilde{\d}_{\L}^{(1)}}]+\O(\fnl^2) \,.
\ee

In conclusion, the partition function for the field $\Tilde{\d}_{\L}^{(1)}$ will have an interaction-less $J$-dependent form
\be
\Z[J_{\rm m,\L},J_{\rm g,\L}]=\int\Del\Tilde{\d}_{\L}^{(1)}\P[\Tilde{\d}_{\L}^{(1)}]\exp\left\{\int_{\vk}J_{\rm m,\L}(\vk)\delta[\Tilde{\d}_{\L}^{(1)}](-\vk)+\int_{\vk}J_{\rm g,\L}(\vk)\delta_{\rm g}[\Tilde{\d}_{\L}^{(1)}](-\vk)\right\}\,,
\ee
and the non-Gaussianities are coupled to the current directly.
An example of how an operator constructed from $\Tilde{\d}_{\L}^{(1)}$ will look like is
\ba
&\d^2[\Tilde{\d}_{\L}^{(1)}](\vk)=(\Tilde{\d}^2)(\vk)+2\fnl\int_{\vp_1,\vp_2,\vp_3}\diracpi(\vk-\vp_{123})\left[\Knl(\vp_2,\vp_3)\frac{M(p_1)}{M(p_2)Mp_ 3)} \right.
\\
&\hspace{7cm}\times\left.\Tilde{\d}_{\L}^{(1)}(\vp_1)\Tilde{\d}_{\L}^{(1)}(\vp_2)\Tilde{\d}_{\L}(\vp_3)\right]+\O(\fnl^2)\,. \nonumber
\ea

%%%%%%%%%%%%%%%%%%%%%%%%%%%%%%%%%%%%%%%%%%%%%%%%%%%%%%%%%%%%%%%%%%%%%%%%%%%
%%%%%%%%%%%%%%%%%%%%%%%%%%%%%%%%%%%%%%%%%%%%%%%%%%%%%%%%%%%%%%%%%%%%%%%%%%%
\section{The $n$-point statistics from the non-Gaussian partition function} \label{app:pkbk}
%%%%%%%%%%%%%%%%%%%%%%%%%%%%%%%%%%%%%%%%%%%%%%%%%%%%%%%%%%%%%%%%%%%%%%%%%%%
%%%%%%%%%%%%%%%%%%%%%%%%%%%%%%%%%%%%%%%%%%%%%%%%%%%%%%%%%%%%%%%%%%%%%%%%%%%

In this section we show  power spectra and bispectra generated from the non-Gaussian partition function \refeq{non-GaussZ}.
We start evaluating the power spectrum taking derivatives of the partition function w.r.t the currents (either $J_{ m, \L}$ or $J_{g, \L}$). For the galaxy power spectrum
\ba
 \langle \delta_{g}(\vk)\delta_{g}(\vk')\rangle &= \frac{1}{\Z_\L[0]}\frac{\d^2\Z}{\d J_{g, \L}\d J_{g, \L}}=\int \Del\dlin_{\L} \P[\dlin_\L ]\; \delta_{g}(\vk)\delta_{g}(\vk') e^{S_{\rm int}} \,. 
 \ea
We see then that by using the bias expansion \refeq{dgPNG} we get
\ba
    \left<\d_g(\vk)\d_g(\vk')\right>_{\rm tree }'&=(b_\d^\L)^2 \left<\dlin(\vk)\dlin(\vk')\right>'
    + 2a_0\fnl b_\d^\L b_\Psi^\L \left<\dlin(\vk)\Psi_\L(\vk')\right>'
    \vs
    &\quad  + \left(a_0\fnl\right)^2(b_\Psi^\L)^2 \left<\Psi_\L(\vk)\Psi_\L(\vk')\right>' \,,
\ea
where we notice that non-Gaussianities appear already at tree-level power spectrum for matter tracers.
We can similarly calculate the (tree-level) matter-galaxy cross power spectrum
\ba
    & \left<\d_m(\vk)\d_g(\vk')\right>_{\rm tree}' = \int \Del\dlin_{\L}\P[\dlin_\L]\,\, \d_m(\vk) \d_g(\vk')\left[1+S_{\rm int}+\O(\fnl^2)\right]
    \vs
    &\quad = \int \Del\dlin_{\L}\P[\dlin_\L]\dlin_\L(\vk)\left(b_\d^\L\dlin_\L(\vk)+\fnl b_\Psi^\L \Psi[\dlin_\L](\vk)\right)\vs&=\quad b_\d \left<\dlin(\vk)\dlin(\vk')\right>' +\fnl b_{\Psi}\left<\dlin(\vk)\Psi_\L(\vk')\right>'\,.
\ea

We now move to the bispectra analysis, defined via three derivatives of the partition function. For the matter bispectrum we have
\ba
\langle \delta(\vk_1)\delta(\vk_2)\delta(\vk_3)\rangle &=\frac{1}{\Z_\L[0]}\frac{\d^3\Z}{\d J_{ m, \L}\d J_{ m, \L}\d J_{ m,\L}}
\vs
&=\int \Del \dlin \P[\dlin_\L ]\; \delta(\vk_1)\delta(\vk_2)\delta(\vk_3) e^{S_{\rm int}} \,,
\ea
and by expanding the exponential we get
\ba
&\langle \delta(\vk_1)\delta(\vk_2)\delta(\vk_3)\rangle =\int \Del \dlin_\L \P[\dlin_\L ]\; \dlin_\L(\vk_1)\dlin_\L(\vk_2)\dlin_\L(\vk_3)\left(1+S_{\rm int}+\dots\right)
\vs
&= \fnl\int_{\vp_1,\vp_2, \vp_3}\diracpi(\vp_{123}) \PlinL^{-1}\left(\vp_1\right)\Knl\left(\vp_3,\vp_2\right)\frac{M(\vp_1)}{M\left(\vp_3\right)M(\vp_2)}
\vs
&\times \Big{\langle}\dlin_\L(\vk_1)\dlin_\L(\vk_2)\dlin_\L(\vk_3)\dlin_\L\left(\vp_1\right)\dlin_\L\left(\vp_2\right)\dlin_\L\left(\vp_3\right)\Big{\rangle}\dots
\vs
&=0+2\fnl\frac{M(k_1)}{M(k_2)M(k_3)}\PlinL(k_2)\PlinL(k_3)\diracpi(\vk_{123})+\text{perms}+\O\left[\fnl^2,\left(\dlin_\L\right)^ 4\right]\,,
\ea
where we imply two cyclic permutations on all 3-momenta. We again see that for the Gaussian case ($\fnl=0$), the bispectrum vanishes at linear order of all fields.

The galaxy bispectrum is defined as
\ba
\langle \delta_{g}(\vk_1)\delta_{g}(\vk_2)\delta_{g}(\vk_3)\rangle &=\frac{1}{\Z_\L[0]}\frac{\d^3\Z}{\d J_{g, \L}\d J_{g, \L}\d J_{g,\L}}
\vs
&=\int \Del \dlin \P[\dlin_\L ]\; \delta_{g}(\vk_1)\delta_{g}(\vk_2)\delta_{g}(\vk_3) e^{S_{\rm int}} \,,
\ea
and following a similar procedure, we find at tree-level
\ba
&\left<\d_g(\vk_1)\d_g(\vk_2)\d_g(\vk_3)\right>_{\rm tree} = \int \Del\dlin_{\L}\P[\dlin_\L]\d_g(\vk_1)\d_g(\vk_2)\d_g(\vk_3)(1+S_{\rm int}) \nonumber\\
&=b_\d^3 B_{111}(\vk_1,\vk_2,
\vk_3)\vs
& +\left(b_\d+a_0\fnl b_\Psi^\L \left(\frac{k_1}{k_{\rm NG}}\right)^\Delta \frac{1}{M(k_1)}\right)\left[b_\d+a_0\fnl b_\Psi^\L \left(\frac{k_2}{k_{\rm NG}}\right)^\Delta \frac{1}{M(k_2)}\right]\vs&
\left[\frac{34}{21}b_\d+2b_{\d^2}+a_0\fnl b_{\Psi\d}\left(\left(\frac{k_1}{k_{\rm NG}}\right)^\Delta \frac{1}{M(k_1)}+ \left(\frac{k_2}{k_{\rm NG}}\right)^\Delta \frac{1}{M(k_2)} \right)\right]\PlinL(k_1)\PlinL(k_2)\,.
\ea
One can similar calculate cross terms for matter and galaxy. 

%%%%%%%%%%%%%%%%%%%%%%%%%%%%%%%%%%%%%%%%%%%%%%%%%%%%%%%%%%%%%%%%%%%%%%%%%%%
\section{The cubic vertex} \label{app:cubic}
%%%%%%%%%%%%%%%%%%%%%%%%%%%%%%%%%%%%%%%%%%%%%%%%%%%%%%%%%%%%%%%%%%%%%%%%%%%

In this appendix, we study the structure of the cubic interaction defined in \refeq{Sintdef}, which we repeat here for convenience
\ba
  S_{\rm int}[\dlin_\L] &=  \fnl\int_{\vp_1,\vp_2, \vp_3}\diracpi(\vp_{123}) \Knl\left(\vp_2,\vp_3\right)\frac{M(p_1)}{M(p_2)M(p_3)}\frac{\dlin_\L\left(\vp_1\right)}{\PlinL\left(p_1\right)}\dlin_\L\left(\vp_2\right)\dlin_\L\left(\vp_3\right) \vs
  &= \raisebox{-0.0cm}{\includegraphicsbox[scale=.7]{figs/diags/diag_vertex0.pdf}} \,,\label{eq:intdef2}
\ea
in which the doubled lines represent the $\dlin_\L\left(\vp_1\right)/\PlinL\left(\vp_1\right)$ leg.
When evaluating the interaction at $S_{\rm int}[\dlin_\L+\dlinshell]$, we can use \refeq{Sintsplit}
\bea 
&&S_{\rm int}[\dlin_\L+\dlinshell]=S_{\rm int}[\dlin_\L]+S_{\rm int}^{(1)_{\rm shell}}[\dlin_\L,\dlinshell] \label{eq:Sintdef2}
\\
&& \hspace{4cm}+S_{\rm int}^{(2)_{\rm shell}}[\dlin_\L,\dlinshell]+S_{\rm int}^{(3)_{\rm shell}}[\dlinshell]\,, \nonumber
\eea
in which $S_{\rm int}^{(\ell)_{\rm shell}}$ refer to the number of external shell legs $\ell$ in the interaction. We now describe each one of those terms in more details.

\subsection{The \texorpdfstring{$S_{\rm int}^{(1)_{\rm shell}}$}{S31shell} term}  
We start from the contribution with one single leg in the shell 
\ba
S_{\rm int}^{(1)_{\rm shell}}[\dlin_\L,\dlinshell] &=  \fnl\int_{\vp_1,\vp_2, \vp_3}\diracpi(\vp_{123}) \Knl\left(\vp_2,\vp_3\right)\frac{M(p_1)}{M(p_2)M(p_3)} \label{eq:ints1}
\\
&\hspace{1cm}\times \left[ \frac{\dlinshell(\vp_1)}{P_{\rm shell}(p_1)}\dlin_\L\left(\vp_2\right)\dlin_\L\left(\vp_3\right) + 2\,\frac{\dlin_\L\left(\vp_1\right)}{\PlinL\left(p_1\right)}\dlinshell\left(\vp_2\right)\dlin_\L\left(\vp_3\right) \right]  \,, \nonumber
\vs
  &= \raisebox{-0.0cm}{\includegraphicsbox[scale=.7]{figs/diags/diag_vertex1.pdf}} + \raisebox{-0.0cm}{\includegraphicsbox[scale=.7]{figs/diags/diag_vertex1cut.pdf}} \,. \nonumber
\ea
These diagrams vanish by the momentum constrain $\diracpi(\vp_{123})$ when taking two of the legs with momentum much smaller than the third one that is on the shell.

\subsection{The \texorpdfstring{$S_{\rm int}^{(3)_{\rm shell}}$}{S33shell} term}  
We skip for now the $S_{\rm int}^{(2)_{\rm shell}}$ term, focusing on the contribution with three external legs in the shell
\ba
S_{\rm int}^{(3)_{\rm shell}}[\dlinshell] &=  \fnl\int_{\vp_1,\vp_2, \vp_3}\diracpi(\vp_{123}) \Knl\left(\vp_2,\vp_3\right)\frac{M(p_1)}{M(p_2)M(p_3)} \label{eq:ints3}
\\
&\hspace{4cm} \times  \frac{\dlinshell(\vp_1)}{P_{\rm shell}(p_1)}\dlinshell\left(\vp_2\right)\dlinshell\left(\vp_3\right)    \nonumber
\vs
  &= \raisebox{-0.0cm}{\includegraphicsbox[scale=.7]{figs/diags/diag_vertex3.pdf}}  \,. \nonumber
\ea
For this diagram contractions, we have two options.
First, if it is contracted with another operator $O^{(n),(3)_{\rm shell}}$, leading to higher-order contributions of the form $\left[\int_{\vp}P_{\rm shell}(p)\right]^2$, that is a two-loop term and is not relevant to the order of expansion we are looking at in this work. 
The second option is to contracted with $O^{(n),(1)_{\rm shell}}$, which forces the other two remaining shell lines to contract among themselves internally in the vertex. 
This contribution contracted with the shell from another operator gives a tadpole that evaluates to
\ba
\langle \dlinshell(\vk) \times S_{\rm int}^{(3)_{\rm shell}}  \rangle_{\rm shell}&= 
\vs
& \hspace{-2cm} \diracpi(\vk) \fnl \left[M(k)\int_{\vp}\Knl(\vp,-\vp)\frac{P_{\rm shell}(p)}{M^2(p)}+ 2\frac{P_{\rm shell}(k)}{M(k)}\int_{\vp}\Knl(-\vk,\vp)\right] \label{eq:tadpoleSint3}
\\
&= \raisebox{-0.0cm}{\includegraphicsbox[scale=1.0]{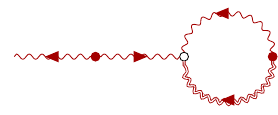}} + \raisebox{-0.0cm}{\includegraphicsbox[scale=1.0]{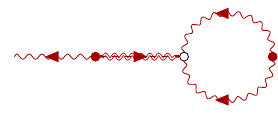}}  \,, \nonumber
\ea
which is cancelled by noticing that the expected value of the single linear density field is no longer zero
\ba
\label{eq:1ptfunction}
\langle \dlin (\vk) \rangle &=\frac{1}{\Z[0]}\frac{\d \Z}{\d J_{\rm m}(\vk)}\Big|_{J_{\rm m}=0}=\int \Del \dlin_\L \P[\dlin_\L ]\;\dlin (\vk)e^{S_{\rm int}}
\vs
&=\diracpi(\vk) \fnl \left[ M(k)\int_{\vp}\frac{\Knl(\vp,-\vp)}{M^2(p)}\Plin(p)+2\frac{\Plin(k)}{M(k)}\int_{\vp}\Knl(-\vk,\vp) \right]\;,
\ea
where we neglected terms $\O(\fnl^2)$. We can then change the definition of the linear density to subtract this part as
\be
 \dlin (\vk) \to  \dlin (\vk) - \langle \dlin (\vk) \rangle \,. \label{eq:linredef}
\ee
After this subtraction, $S_{\rm int}^{(3)_{\rm shell}}$ only provides higher-order in $\lambda$ contributions.

\subsection{The \texorpdfstring{$S_{\rm int}^{(2)_{\rm shell}}$}{S32shell} term}  \label{app:Sint2shell}

Therefore, we are left with the $S_{\rm int}^{(2)_{\rm shell}}$ diagrams, which as we will see leads to the single relevant contribution to the RG flow.
There are of two types of $S_{\rm int}^{(2)_{\rm shell}}$ diagrams, either with a cut in a shell propagator or with a cut in a linear external leg:
\ba
S_{\rm int}^{(2)_{\rm shell}}[\dlin_\L,\dlinshell] &=  \fnl\int_{\vp_1,\vp_2, \vp_3}\diracpi(\vp_{123}) \Knl\left(\vp_2,\vp_3\right)\frac{M(p_1)}{M(p_2)M(p_3)} 
\vs
&\hspace{1cm}\times \left[2\, \frac{\dlinshell(\vp_1)}{P_{\rm shell}(p_1)}\dlinshell\left(\vp_2\right)\dlin_\L\left(\vp_3\right) + \frac{\dlin_\L\left(\vp_1\right)}{\PlinL\left(p_1\right)}\dlinshell\left(\vp_2\right)\dlinshell\left(\vp_3\right) \right]   
\vs
&\equiv  \left[S_{\rm int}^{(2)_{\rm shell}}\right]_{\rm scatter} + \left[S_{\rm int}^{(2)_{\rm shell}}\right]_{\rm annihilation} \label{eq:ints2}
\\
  &= \raisebox{-0.0cm}{\includegraphicsbox[scale=.7]{figs/diags/diag_vertex2.pdf}} + \raisebox{-0.0cm}{\includegraphicsbox[scale=.7]{figs/diags/diag_vertex2cut.pdf}} \,. \nonumber
\ea
The $\left[S_{\rm int}^{(2)_{\rm shell}}\right]_{\rm scatter}$ diagram with a cut in the shell propagator has the arguments of the $\Knl$ kernel in the squeezed limit. It therefore represents the {\it scattering} of a UV mode by a IR mode. This term will be the main contribution to the running of non-Gaussian terms.

The $\left[S_{\rm int}^{(2)_{\rm shell}}\right]_{\rm annihilation}$ diagram, with a double line in the external propagator, was already mentioned in \cite{Assassi:2015EFT} where it was shown that it couples to the stochastic part of the energy momentum tensor. When dealing with the bias, the literature has so far focused on the squeezed limit of the interaction kernel (e.g., \cite{Assassi:2015fma}) as the relevant limit for the galaxy statistics and neglected this term. 
We formally show here (up to our knowledge) for the first time that this term leads to a stochastic contribution to galaxy clustering. 
This is to some extent expected since this contribution comes from the {\it annihilation} of two shell (small scale) modes into a large-scale mode.\footnote{For the $\left[S_{\rm int}^{(2)_{\rm shell}}\right]_{\rm scatter}$ diagram, only one of the shell (UV) modes couple through the kernel.}
Here the arguments of $\Knl$ have the same orders of magnitude, which does not allow us to use the squeezed limit expansion of this kernel. 

Given that it is sufficient to consider the $S_{\rm int}^{(2)_{\rm shell}}$ terms, we then start from the partition function
\ba
\Z[J_{\L}] &=\int \Del\dlin_\L \P[\dlin_\L]\int\Del\dlinshell \P[\dlinshell]\exp \left(\left[S_{\rm int}^{(2)_{\rm shell}}\right]_{\rm scatter} + \left[S_{\rm int}^{(2)_{\rm shell}}\right]_{\rm annihilation}\right)
\vs
&\hspace{2.7cm} \times \exp\left( \int_{\vk}J_\L(\vk)\sum_Ob_O^{\L'}O[\dlin_\L+\dlinshell](-\vk)    +\O\left[J^2\right] \right)
\vs
&=\int \Del\dlin_\L \int\Del\dlinshell  
\vs
& \quad \exp \left(-\frac{1}{2}\int_{\vk}\frac{\left(\dlin_\L(\vk)\right)^2}{\PlinL(k)}-\frac{1}{2}\int_{\vk}\frac{\left(\dlinshell(\vk)\right)^2}{P_{\rm shell}(k)} + \left[S_{\rm int}^{(2)_{\rm shell}}\right]_{\rm scatter} + \left[S_{\rm int}^{(2)_{\rm shell}}\right]_{\rm annihilation}\right) 
\vs
& \quad \times \exp\left( \int_{\vk}J_\L(\vk)\sum_Ob_O^{\L'}O[\dlin_\L+\dlinshell](-\vk)    +\O\left[J^2\right] \right)\,,
\ea
where we wrote the kinetic terms explicitly.
The $\left[S_{\rm int}^{(2)_{\rm shell}}\right]_{\rm scatter}$ integral, as said before, contributes to the squeezed limit corrections and we calculate them in \refapp{SOeval}. Our goal is to understand the effect of the $\left[S_{\rm int}^{(2)_{\rm shell}}\right]_{\rm annihilation}$ in the RG-flow. In order to gain more physical insights about this term we will apply a similar transformation as the one described in \refapp{couplingtoJ}:
\ba
\label{eq:non-G-trafo}
\Tilde{\d}_{\L}^{(1)}(\vk) &=\dlin_\L(\vk)-\fnl M(k)\int_{\vp_2,\vp_3}\diracpi(\vk-\vp_{23})\frac{\Knl(\vp_2,\vp_3)}{M(p_2)M(p_3)}\dlinshell(\vp_2)\dlinshell(\vp_3) \,,
\vs
\Tilde{\d}_{\rm shell}^{(1)}(\vk) &=\dlin_{\rm shell}(\vk)-\fnl M(k)\int_{\vp_2,\vp_3}\diracpi(\vk-\vp_{23})\frac{\Knl(\vp_2,\vp_3)}{M(p_2)M(p_3)}\dlinshell(\vp_2)\dlinshell(\vp_3) \,,
\ea
with the integration measure of the path-integral [see \refeq{Dtransform}]
\ba
\Del\dlin_\L &=\Del\Tilde{\d}^{(1)}_{\L}+\O(\fnl^2)\,,
\vs
\Del\dlin_{\rm shell} &=\Del\Tilde{\d}_{\rm shell}^{(1)}+\O(\fnl^2)\,.
\ea
We notice that with this transformation the annihilation part is absorbed in the field redefinition
\ba
-\frac{1}{2}\int_{\vk}\frac{\left(\dlin_\L(\vk)\right)^2}{\PlinL(k)}+\left[S_{\rm int}^{(2)_{\rm shell}}\right]_{\rm annihilation}=-\frac{1}{2}\int_{\vk}\frac{\left(\Tilde{\d}^{(1)}_{\L}\right)^2(\vk)}{\PlinL(k)} \,,
\ea
while the scattering term will remain unchanged at linear order in $\fnl$
\be
\left[S_{\rm int}^{(2)_{\rm shell}}\right]_{\rm scatter}\left(\dlin_\L,\dlinshell\right)=\left[S_{\rm int}^{(2)_{\rm shell}}\right]_{\rm scatter}\left(\Tilde{\d}_{\L}^{(1)},\Tilde{\d}_{\rm shell}^{(1)}\right)\,.
\ee
The kinetic term with the shell transforms as
\ba
-\frac{1}{2}\int_{\vk}\frac{\left(\dlinshell(\vk)\right)^2}{P_{\rm shell}(k)}=-\frac{1}{2}\int_{\vk}\frac{\left(\Tilde{\d}_{\rm shell}^{(1)}\right)^2(\vk)}{P_{\rm shell}(k)}-S_{\rm int}^{(3)_{\rm shell}}(\Tilde{\d}_{\rm shell}^{(1)})\,,
\ea
and the remaining three-shell interaction will exactly cancel the interaction vertex \refeq{ints3}.
After the field redefinition and grouping the terms together we have
\ba
\Z[J_\L] &=\int \Del\Tilde{\d}^{(1)}_{\L}\P[\Tilde{\d}^{(1)}_{\L}]\Del\Tilde{\d}^{(1)}_{\rm shell}\P[\Tilde{\d}^{(1)}_{\rm shell}]\exp\left(\left[S_{\rm int}^{(2)_{\rm shell}}\right]_{\rm scatter}\right)
\vs
&\hspace{1cm}\times \exp\left(\int_{\vk}J_\L(\vk)\sum_Ob_O^{\L'}O[\Tilde{\d}^{(1)}_{\L}+\Tilde{\d}^{(1)}_{\rm shell}](-\vk) +\O(J_\L^2)\right)\,,
\ea
in which we observe that the annihilation interaction is absorbed by the field transformation leaving only the scattering vertex. Furthermore, the terms coupled to the current now contain a non-Gaussian contribution: we have to solve the inverse transformation of \refeq{non-G-trafo} to get $\dlin$ in terms of $\Tilde{\d}^{(1)}$ at linear order in $\fnl$ and the resulting operators coupled to the current will then be of the form $O[\dlin]=O[\dlin[\Tilde{\d}^{(1)}]]\equiv O[\Tilde{\d}^{(1)}]$. Since we work only at linear order in $\fnl$ the effect of the scattering vertex is the same as before the transformation because we only consider coupling with the Gaussian counterpart of the $\Tilde{\d}_{\L}^{(1)}$ [couplings with the non-Gaussian will give terms of $\O(\fnl^2)$].

We can immediately see that the field redefinition \refeq{non-G-trafo} will not change the shell corrections $[\Shell_O^2]_{\rm free}$ defined in \refeq{shell-exp} since
\bea
[\mathcal{S}_{\d^2}^2]_{\rm free}\supset M(0) \int_{\vp}\frac{\Knl(\vp,-\vp)}{M^2(p)}P_{\rm shell}(p)=0 \,,
\eea
and therefore the annihilation term does not source any $J^1$ corrections to the bias parameters.
Despite not changing the result for the $J^1$ terms calculated in this work, the annihilation interactions induce the running of stochastic coefficients (see \cite{Rubira:2024tea} for a description of those terms for the Gaussian case).
As seen in \cite{Rubira:2024tea}, the main source of stochasticity (or $J^m$ terms with $m>1$) come from shell correlators such as 
\ba
\Shell_{O_1, O_2}^{22} [\dlin_\L] (\vk_1, \vk_2) = 
 \sum_{n_1\geq 2, n_2 \geq 2}  \left\< O_1^{(n_1),(2)_{\rm shell}} [\dlin_\L,\dlinshell](\vk_1)  O_2^{(n_2),(2)_{\rm shell}} [\dlin_\L,\dlinshell](\vk_2) \right\>_{\rm shell} \,. 
 \ea
We have then, for instance, a new non-Gaussian contribution
\bea
\mathcal{S}_{\d\d^2}^{22}(\vk, \vk')=M(k)\diracpi(\vk+\vk')\int_{\vp}\frac{\Knl(\vp,-\vp)}{M^2(p)} \left[P_{\rm shell}(p)\right]^2 + \O[\dlin_\L] \,,
\eea
while for the Gaussian case this term is zero \cite{Rubira:2024tea}.
For local non-Gaussianity, with $\Knl =1$, this becomes
\ba
\mathcal{S}_{\d\d^2}^{22}(\vk, \vk')=2M(k)\diracpi(\vk+\vk')P_\vphi(\L)\frac{d\sigma^2_\L}{d\L}\lambda \,,
\ea
leading to a non-zero contribution to the running of stochastic terms.
We leave a more detailed analysis of stochasticity in the presence of non-Gaussianities for a future work.

%%%%%%%%%%%%%%%%%%%%%%%%%%%%%%%%%%%%%%%%%%%%%%%%%%%%%%%%%%%%%%%%%%%%%%%
\section{Evaluation of \texorpdfstring{$\Shell_{O}$}{SO}} \label{app:SOeval}
%%%%%%%%%%%%%%%%%%%%%%%%%%%%%%%%%%%%%%%%%%%%%%%%%%%%%%%%%%%%%%%%%%%%%%%

In this appendix we evaluate the $\Shell_{O}$ in the presence of PNG interactions. For a complete calculation without interactions, see \cite{Rubira:2023vzw}. We expand the shell integrals 
\ba
\Shell_{O} [\dlin_\L] (\vk) &=  \sum_{\ell\geq 0} \left(\Shell_{O} [\dlin_\L] (\vk)\right)_{(\ell)_{\rm legs}}\,,
\label{eq:nlegs}
\ea
in term of the number of external legs $\ell$ (and therefore contributions to operators starting from $\ell$-th order) with 
\ba
    \left(\Shell_{O} [\dlin_\L] (\vk)\right)_{(\ell)_{\rm legs}} = \raisebox{-0.0cm}{\includegraphicsbox[scale=.9]{figs/diags/diag_S2_ell_legs.pdf}} + \raisebox{-0.0cm}{\includegraphicsbox[scale=.9]{figs/diags/NG_inter.pdf}} \,. \label{eq:shelldiaglegs}
    \ea
Note that the first type of diagrams correspond to the usual corrections from Gaussian initial conditions considered in \cite{Rubira:2023vzw}. The second type of diagrams arise due to the interacting non-Gaussian vertex presented in this work. In \refapp{spin0eval} and \refapp{spin2eval} we evaluate these respectively for spin-0 and spin-2 non-Gaussianities. 

\subsection{Displacements of \texorpdfstring{$\Psi$}{psi}} \label{app:displacements}

Before evaluating the shell integrals, we discuss one important point when evaluating expectation values including $\Psi$. 
It is important to evaluate the operator $\Psi$ at the (Lagrangian) position $\vq$ in the initial conditions that corresponds to a given final-time (Eulerian) position $\vx$. Physically, this is due to the fact that the PNG interactions are imprinted at initial time. Formally, it is required to cancel the non-Galilean-invariant displacement terms involving the gradient of the potential $\partial_i\Phi_g$ \cite{Assassi:2015EFT,Assassi:2015fma}.
So at the Lagrangian position $\vq=\mathbf{x}-\mathbf{s}$, with $\mathbf{s}$ the displacements, we can write
\be
\Psi(\vq)=\Psi(\vx-\mathbf{s}(\vq))\,.
\ee
Note that the displacement field admits a perturbative expansion
\be
\vdisp(\vq)=\vdisp_{(1)}(\vq)+\vdisp_{(2)}(\vq)+\vdisp_{(3)}(\vq)+\dots\,,
\ee
and is also evaluated at the Lagrangian position $\vq$, meaning that it should also be expanded by
\be
\vdisp(\vq)=\vdisp(\vx-\vdisp)=\vdisp(\vx)-\vdisp(\vq)\nabla\vdisp(\vx)+\dots\,.
\ee
With the above, we have at second-order in perturbation theory \cite{MSZ}
\bea
\label{eq:disp2}
\Psi^{[2]}(\vq)=\Psi(\vx)-s_{(1)}^i\partial_i\Psi(\vx)\;,
\eea
with
\be
\mathbf{s}_{(1)}=-\frac{\nabla}{\nabla^2}\dlin\;.
\ee
At third order, we have
\ba
\label{eq:dips3}
\Psi^{[3]}(\mathbf{\vq})&=\Psi(\mathbf{x}-\mathbf{s})
\vs
&=\Psi(\mathbf{x})-s_{(1)}^i\partial_i\Psi(\mathbf{x})-s_{(2)}^i\partial_i\Psi(\mathbf{x})+s^i_{(1)}\partial_i(s^j_{(1)})\partial_j\Psi(\mathbf{x})+\frac{1}{2}s_{(1)}^is_{(1)}^j\partial_i\partial_j\Psi(\mathbf{x})\;,
\ea
with
\be
\mathbf{s}_{(2)}=\frac{3}{14}\frac{\nabla}{\nabla^2}\G_2\;.
\ee
For the corrections of $\Psi$ we have to consider its expansion at fourth order, including terms up to third order in the displacements 
\ba
&\Psi^{[4]}(\vq)=\Psi^{[3]}(\vx)- s_{(3)}^i\partial_i\Psi+s_{(1)}^j\partial_js_{(2)}^i\partial_i\Psi+s_{(2)}^j\partial_js_{(1)}^i\partial_i\Psi-\frac{1}{2}s^j_{(1)}s^k_{(1)}\left(\partial_j\partial_k s^i_{(1)}\right)\partial_i\Psi
\vs
&+s^i_{(2)}s^j_{(1)}\partial_i\partial_j\Psi-s^k_{(1)}\partial_ks^j_{(1)}\partial_js^i_{(1)}\partial_i\Psi-s^k_{(1)}\partial_ks^i_{(1)}s^j_{(1)}\partial_i\partial_j\Psi-\frac{1}{6}s^i_{(1)}s^j_{(1)}s^k_{(1)}\partial_i\partial_j\partial_k\Psi \,, \label{eq:dips4}
\ea
with \cite{Baldauf:2020bsd}
\ba
&\vdisp_{(3)}=\frac{i}{6}\frac{\vk}{k^2}\int_{\vp_1,\vp,\vp_3}\diracpi(\vk-\vp_{123})\dlin(\vp_1)\dlin(\vp_2)\dlin(\vp_3)
\\
&\hspace{3cm}\times\left[ \frac{5}{7}\sigma^2_{\vp_1,\vp_2}\sigma^2_{\vp_1+\vp_2,\vp_3}-\frac{1}{3}\left(-2-3\sigma^2_{\vp_1,\vp_2}+2\frac{(\vp_1\cdot\vp_2)\;(\vp_2\cdot\vp_3)\;(\vp_1\cdot\vp_3)}{p_1^2p_2^2p_3^2}\right)\right]\,. \nonumber
\ea
Note that corresponding displacement contributions also exist for the Gaussian operators such as $\d$, $\G_2$, ..., but these are included in the perturbation theory kernels.

%%%%%%%%%%%%%%%%%%%%%%%%%%%%%%%%%%%%
\subsection{Spin-0 non-Gaussianity} \label{app:spin0eval}

\paragraph{Calculation of $\left(\mathcal{S}_O\right)_{(0)_{\rm legs}}$.}

For the Gaussian operators, it is impossible to source a zero-leg leading-order contribution by the structure of the vertex \refapp{cubic}, since those would scale as $\lambda^2$.

For $\Psi$, despite this operator being linear, it admits an expansion with respect to the displacements as in \refeqs{disp2}{dips4}. Then we see that at second order in $\Psi$
\ba
\left(\mathcal{S}_{\Psi}\right)_{(0)_{\rm legs}}(\vk)=-\diracpi(\vk)a_0\fnl\int_{\vp}\PshellPsi(p)\,,
\ea
which will vanish after subtracting the tadpole contribution  $\Psi \to \Psi + \diracpi(\vk)\int_{\vp}P_{\rm 1\Psi}(p)$ via \refeq{Onorm}. 

Non-Gaussian contributions involving $O_G\Psi$ also vanish after using \refeq{Onorm}, i.e. after subtracting the tadpole contribution
\ba
\left(\mathcal{S}_{\d\Psi}\right)_{(0)_{\rm legs}}(\vk)=\diracpi(\vk)a_0\fnl\int_{\vp}\PshellPsi(p)\,,
\ea
from $\d\Psi$ via $\d\Psi \to \d\Psi - \diracpi(\vk)\int_{\vp}P_{\rm 1\Psi}(p)$.

\paragraph{Calculation of $\left(\mathcal{S}_O \right)_{(1)_{\rm legs}}$.}
For operators that do not include $\Psi$, we have 
\ba
\left(\mathcal{S}_\d \right)_{(1)_{\rm legs}}(\vk)&=4\fnl\vphi_\L(\vk)\int_{\vp}F_2(\vk+\vp,-\vp)\Knl(\vk,\vp)\frac{M(|\vp+\vk|)}{M(p)}P_{\rm shell}(p)
\vs
&=4\fnl a_0\Psi_\L(\vk)\int_{\vp}F_2(\vk+\vp,-\vp)\left(\frac{k_{\rm NG}}{p}\right)^{\D}P_{\rm shell}(p) \,,
\\
\label{eq:non-G-delta2}
\left(\mathcal{S}_{\d^2} \right)_{(1)_{\rm legs}}(\vk)&=4\fnl a_0\Psi_\L(\vk)\int_{\vp}\left(\frac{k_{\rm NG}}{p}\right)^\D P_{\rm shell}(p) \,, 
\\
\left(\mathcal{S}_{\G_2} \right)_{(1)_{\rm legs}}(\vk)&=4\fnl a_0\Psi_\L(\vk)\int_{\vp}\sigma^2_{\vp+\vk,-\vp}\left(\frac{k_{\rm NG}}{p}\right)^\Delta P_{\rm shell}(p)+\text{h.d.} \,.
\ea
\refeq{non-G-delta2} is the leading source for the non-Gaussian bias coefficient $b_\Psi$, and coincides with the counter-term derived in \cite{Desjacques:2016bnm} and \cite{Assassi:2015fma}, with a factor 4 difference with respect to the latter, presumably originating from the possible permutations of the kernel and the number of equivalent contractions.
The terms with $F_2(\vk+\vp,-\vp)\propto \left(k/p\right)^2$ and $\sigma_{\vk+\vp,-\vp}^2\propto \left(k/p\right)^2$ both source the higher-derivative operators $\nabla^2\Psi_\L$. 

For operators including $\Psi$ we have to go up to third order in $\Psi$ including displacements \refeq{disp2}. 

Starting from the contribution of $\Psi$, we find
\ba
\left(\mathcal{S}_{\Psi}\right)_{(1)_{\rm legs}}(\vk)=-\frac{13}{21}a_0\fnl\dlin_\L(\vk)\int_{\vp}\PshellPsi(p) \,,
\ea
noticing that it contributes to $\d$. This is different than what is observed for $\d$, which only sources higher-derivative contributions as a consequence of mass and momenta conservation via the equations of motion. Therefore, for PNGs, we find also first-order operators acting as a source.
For the calculation of $\Psi\d$, we start with the undisplaced

field $\Psi$ in \refeq{disp2}, for which
\ba
\left(\mathcal{S}_{\d\Psi} \right)_{(1)_{\rm legs}}(\vk)&\supset\frac{34}{21}\fnl a_0\dlin_\L(\vk)\int_{\vp}\PshellPsi(p)\,.
\ea
The second term in \refeq{disp2} gives
\ba
\label{eq:FirstDisp}
s_{(1)}^i\left[\partial_i\Psi(\vx)\right]\delta(\vx)=\int_{\vp_{1},\vp_{2},\vp_{3}}\diracpi(\vk-\vp_{123})\frac{\vp_1\cdot\vp_2}{p_1^2}\dlin_\L(\vp_1)\Psi_\L(\vp_2)\delta[\dlin_\L](\vp_3) \,,
\ea
leading to [at linear order in $\delta[\dlin_\L](\vp_3)=\dlin_\L(\vp_3)$]
\ba
\left(\mathcal{S}_{\d\Psi}^2\right)_{(1)_{\rm legs}}(\vk)&\supset-\fnl a_0\dlin_\L(\vk)\int_{\vp}\PshellPsi(p)\,.
\ea
Therefore, we find that the displaced field in that case changes the overall prefactor of the $\left(\mathcal{S}_{\d\Psi}^2\right)_{(1)_{\rm legs}}$ calculation. After taking the displacements of $\Psi$ in account we find
\ba
\left(\mathcal{S}_{\d\Psi}\right)_{(1)_{\rm legs}}(\vk)&=\frac{13}{21}\fnl a_0\dlin_\L(\vk)\int_{\vp}\PshellPsi(p)\,.\label{eq:dpsioneleg}
\ea
For the other terms involving $\Psi$
\ba
\left(\mathcal{S}_{\d^2\Psi}\right)_{(1)_{\rm legs}}(\vk)&=a_0\fnl\Psi_\L(\vk)\int_{\vp}P_{\rm shell}(p)+2a_0\fnl\dlin_\L(\vk)\int_{\vp}\PshellPsi(p)\,,\label{eq:d2psioneleg}\\
\left(\mathcal{S}_{\G_2\Psi}\right)_{(1)_{\rm legs}}(\vk)&=- a_0\fnl\frac{4}{3}\dlin_\L(\vk)\int_{\vp}\PshellPsi(p)\,.\label{eq:G2psioneleg}
\ea

\paragraph{Calculation of $\left(\mathcal{S}_O\right)_{(2)_{\rm legs}}$.}
For terms that do not involve $\Psi$ and ignoring the terms sourcing higher derivative operators we get
\ba
\label{eq:non-G-delta22}
\left(\mathcal{S}_{\d^2} \right)_{(2)_{\rm legs}} (\vk)&=8\fnl\int_{\vp_1,\vp_2,\vp}\diracpi(\vk-\vp_{12})\dlin_\L(\vp_1)\dlin_\L(\vp_2)
\vs
&\quad \times\Knl(\vp_2,\vp)\frac{M(|\vp_2+\vp|)}{M(p_2)M(p)}\left[ F_2(\vp+\vp_2,\vp_1)+F(-\vp,\vp_1)\right]P_{\rm shell}(p)
\vs
&=\fnl a_0\left(\frac{272}{21}[\Psi\d](\vk)-4\left[s^i\partial_i\Psi\right](\vk)\right)\int_{\vp}\left(\frac{k_{\rm NG}}{p}\right)^\D P_{\rm shell}(p)+\text{h.d.}\;,
\\
\left(\mathcal{S}_{\d^3} \right)_{(2)_{\rm legs}}(\vk) &=12\fnl a_0[\Psi\d](\vk)\int_{\vp}\left(\frac{k_{\rm NG}}{p}\right)^\Delta P_{\rm shell}(p)\;,
\\
\left(\mathcal{S}_{\G_2\d}\right)_{(2)_{\rm legs}}(\vk) &=-\frac{8}{3}\fnl a_0[\Psi\d](\vk)\int_{\vp}\left(\frac{k_{\rm NG}}{p}\right)^\Delta P_{\rm shell}(p)+\text{h.d.}\;,
\ea
where we see that the prefactors of $\Psi$ in \refeq{non-G-delta2} and $s_{(1)}^i\partial_i\Psi$ in \refeq{non-G-delta22} match as expected. Adding them together, we get
\ba
4\left[\Psi(\vk)-\left[s^i\partial_i\Psi\right](\vk)\right]\int_{\vp}\left(\frac{k_{\rm NG}}{p}\right)^\D P_{\rm shell}(p)  \,,
\ea
corresponding then to $\Psi(\vq)$ and demonstrates again the importance of calculating $\Psi$ at the Lagrangian point $\vq$ in order to absorb all terms appearing from the dynamics.
Once more, \refeq{non-G-delta22} agrees (up to a factor 4) with the second-order counterterm for $\d^2$ calculated in \cite{Assassi:2015fma}.

For operators that involve $\Psi$ it is again crucial to calculate the displaced operator. If  we calculate the operator $\Psi$ or $\Psi\d$ at the point $\vx$ this will lead to terms violating the Galilean invariance.
For the operator $\Psi$, we have to consider the fourth-order expansion \refeq{dips4} to find
\ba
\left(\mathcal{S}_{\Psi}\right)_{(2)_{\rm legs}}(\vk)&=a_0\fnl\int_{\vp_1,\vp_2,\vp}\diracpi(\vk-\vp_{12})\dlin_\L(\vp_1)\dlin_\L(\vp_2)
\vs
&\hspace{1.5cm}\times\left(\frac{1}{210}-\frac{13}{42}\frac{\vp_1
\cdot\vp_2}{p_1^2}-\frac{13}{42}\frac{\vp_1
\cdot\vp_2}{p_2^2}-\frac{577}{1890}\frac{\left(\vp_1\cdot\vp_2\right)^2}{p_1^2p_2^2}\right)\PshellPsi(p)
\vs
&=a_0\fnl\left(-\frac{13}{21}\d^{(2)}(\vk)+\frac{43}{135}\left(\d^2\right)^{(2)}(\vk)-\frac{1699}{13230}\left(\G_{2}\right)^{(2)}(\vk)\right)\int_{\vp}\PshellPsi(p)+\text{h.d.}\,.
\ea
We calculate the corrections from $\Psi\d$ more carefully to show the cancellation of Galilean breaking terms. Evaluating it in the Eulerian position, we find:
\ba
\label{eq:non-disp}
&\left(\mathcal{S}_{\d\Psi|_x}\right)_{(2)\rm {legs}}=\fnl\int_{\vp_1,\vp_2}\diracpi(\vk-\vp_{12})\dlin_\L(\vp_1)\dlin_\L(\vp_2)\int \frac{p^2dp}{2\pi^2}\PshellPsi(p) 
\vs
&\hspace{2cm} \times \left[\frac{172}{105}+\frac{41}{42}\frac{\vp_1\cdot\vp_2}{p_1^2}+\frac{41}{42}\frac{\vp_1\cdot\vp_2}{p_2^2}+\frac{p^2}{6}\frac{\vp_1\cdot\vp_2}{p_1^2 p_2^2}+\frac{157}{315}\frac{\left(\vp_1\cdot\vp_2\right)^2}{p_1^2 p_2^2}\right]\,,
\ea
where we see terms such as $\frac{\vp_1\cdot\vp_2}{p_2^2}$ and $\frac{p^2}{6}\frac{\vp_1\cdot\vp_2}{p_1^2 p_2^2}$ that break Galilean symmetry, corresponding to operators such as $(\partial_i\Phi_g)^2$ and $\partial_i\Phi_g\partial^i\d$. Notice that after including displacements via the second term in \refeq{disp2}, we first have to expand \refeq{FirstDisp} at third order, and this term leads to
\ba
\label{eq:s1term}
&\left(\mathcal{S}_{\d\Psi}\right)_{2\rm {legs}}(\vk)\supset-\fnl\int_{\vp_{1},\vp_{2}}\diracpi(\vk-\vp_{12})\dlin_\L(\vp_1)\dlin_\L(\vp_2)
\vs
&\hspace{5cm}\times\int \frac{p^2 dp}{2\pi^2}\PshellPsi(p)\left[\frac{\vp_1\cdot\vp_2}{6p_1^2}+\frac{\vp_1\cdot\vp_2}{6p_2^2}+\frac{p^2}{3}\frac{\vp_1\cdot\vp_2}{p_1^2 p_2^2}\right] \,.
\ea
Now we also have to consider the third-order displacement expansion \refeq{dips3} that has terms such as 
\ba
\frac{1}{2}s_{(1)}^is_{(1)}^j\left[\partial_i\partial_j\Psi(\mathbf{x})\right]\delta(\vx)&=\frac{1}{2}\int_{\vp_1,\vp_2,\vp_3,\vp_4}\diracpi(\vk-\vp_{1234})\frac{\vp_1\cdot\vp_3}{p_1^2}\frac{\vp_2\cdot\vp_3}{p_2^2}
\vs
&\hspace{3cm}\times \dlin_\L(\vp_1)\dlin_\L(\vp_2)\Psi_\L(\vp_3)\dlin_\L(\vp_4) \,,
\ea
that leads to 
\ba
&\left(\mathcal{S}_{\d\Psi}\right)_{2\rm {legs}}(\vk)\supset
\fnl\int_{\vp_{1},\vp_{2}}\diracpi(\vk-\vp_{12})\dlin_\L(\vp_1)\dlin_\L(\vp_2)
\int \frac{p^2 dp}{2\pi^2}\PshellPsi(p)\left[\frac{p^2}{6}\frac{\vp_1\cdot\vp_2}{p_1^2 p_2^2}\right] \,.
\ea
We see that by displacing the field $\Psi$ we exactly cancel the non-Galilean term $\frac{p^2}{6}\frac{\vp_1\cdot\vp_2}{p_1^2 p_2^2}$ that would violate the Galilean-invariance. By consistently carrying out the rest of the calculations one can show that the other violating terms will be generated with the correct pre-factor in order to be absorbed in the already displaced operators $\d,\;\d^2,\;\G_2$.
In conclusion, after including the expansion of $\Psi$ up to third order in displacements \refeq{dips3} we have
\ba
\label{eq:displaced}
\left(\mathcal{S}_{\d\Psi}\right)_{(2)_{\rm legs}}(\vk) &=\fnl a_0\left[\frac{13}{21}\delta^{(2)}(\vk)+\frac{478}{135}\left(\delta^2\right)^{(2)}(\vk)+\frac{79}{2205}\mathcal{G}_2^{(2)}(\vk)\right]\int_{\vp}\PshellPsi(p)+\text{h.d.} \,,
\\
\left(\mathcal{S}_{\d^2\Psi}\right)_{(2)_{\rm legs}} (\vk)&=\fnl a_0\left(2\d^{(2)}(\vk)+\frac{47}{21}\left(\d^2\right)^{(2)}(\vk)\right)\int_{\vp}\PshellPsi(p) 
\vs
&\quad +a_0\fnl\left(\frac{68}{21}[\Psi\d](\vk)-s^i\partial_i\Psi(\vk)\right)\int_{\vp}P_{\rm shell}(p)+ \text{h.d.}\,,
\\
\left(\mathcal{S}_{\G_2\Psi}\right)_{(2)_{\rm legs}}(\vk) &=a_0\fnl\left[-\frac{4}{3}\d^{(2)}(\vk)-\frac{31}{21}\left[\d^2\right]^{(2)}(\vk)+\frac{-1}{21}\G_2^{(2)}(\vk)\right]\int_{\vp}\PshellPsi(p)+\text{h.d.}\,.
\ea
Notice here that the prefactors of $\d^{(2)}$ exactly match with those of $\d^{(1)}$ calculated in \refeqs{dpsioneleg}{G2psioneleg}, as a consequence of the Galilean invariance we mentioned before. Moreover, no $\Psi$ term is sourced with two legs, as expected since this is a linear operator.

%%%%%%%%%%%%%%%%%%%%%%%%%%%%%%%%
\subsection{Spin-2 non-Gaussianity} \label{app:spin2eval}

\paragraph{Calculation of $\left(\mathcal{S}_{O}\right)_{(1)_{\rm legs}}$.}
For these kind of terms, we again have to consider $\Psi_{ij}(\vq)$ at second order in displacements. Collecting the contributions, we get
\ba
\left(\mathcal{S}_{\rm Tr[\Psi\Pi^{(1)}]}\right)_{(1)_{\rm legs}}(\vk) &=a_2\fnl\frac{34}{21}\d^{(1)}(\vk)\int_{\vp}\PshellPsi(p) \,, \label{eq:psipioneleg}
\\
\left(\mathcal{S}_{\rm \d Tr[\Psi\Pi^{(1)}]}\right)_{(1)_{\rm legs}}(\vk) &=a_2\fnl\d^{(1)}(\vk)\int_{\vp} \PshellPsi(p) \,, \label{eq:dpsipioneleg}
\\
\left(\mathcal{S}_{\rm Tr[\Psi\Pi^{(2)}]}\right)_{(1)_{\rm legs}}(\vk) &=a_2\fnl\frac{34}{21}\d^{(1)}(\vk)\int_{\vp} \PshellPsi(p) \,. \label{eq:psipi2oneleg}
\ea

\paragraph{Calculation of $\left(\mathcal{S}_{O}\right)_{(2)_{\rm legs}}$.} 
We first consider operators that do not include $\rm Tr[\Psi\Pi^{(i)}]$. The terms that source leading-in-derivatives operators are
\ba
\left(\mathcal{S}_{\d^2}\right)_{(2)_{\rm legs}}(\vk)&=16a_2\fnl\int_{\vp_1,\vp_3}\diracpi(\vk-\vp_1-\vp_3)\dlin_\L(\vp_1)\Psi_\L(\vp_3)\left(\frac{-2}{105}+\frac{2}{35}\frac{\left(\vp_3\cdot\vp_1\right)^2}{p_3^2 p_1^2}\right)
\vs
&\hspace{5cm}\times\int_{\vp}\left(\frac{k_{\rm NG}}{p}\right)^{\D'} P_{\rm shell}(p)+\text{h.d.}
\vs
&=a_2\fnl\frac{64}{105}\Tr\left[\Psi\Pi^{[1]}\right](\vk)\times\int_{\vp}\left(\frac{k_{\rm NG}}{p}\right)^{\D'} P_{\rm shell}(p)+\text{h.d.}\;,
\\
\left(\mathcal{S}_{\d\G_2}\right)_{(2)_{\rm legs}}(\vk)&=a_2\fnl\frac{16}{15}\Tr\left[\Psi\Pi^{[1]}\right](\vk)\times\int_{\vp}\left(\frac{k_{\rm NG}}{p}\right)^{\D'} P_{\rm shell}(p)+\text{h.d.}\,.
\ea

Calculating now for the terms that include $\rm Tr[\Psi\Pi^{(i)}]$, we find again that we have to consider $\Psi_{ij}(\vq)$ at third order in displacements (similarly to $\Psi\d$ in the spin-0 case). Collecting the contributions, we get
\ba
\left(\mathcal{S}_{\rm Tr[\Psi\Pi^{(1)}]}\right)_{(2)_{\rm legs}}(\vk) &=a_2\fnl\left[\frac{34}{21}\d^{(2)}(\vk)+\frac{124}{315}\left[\d^2\right]^{(2)}(\vk)-\frac{661}{4410}\G_2^{(2)}(\vk)\right]\int_{\vp}\PshellPsi(p)\vs 
& \hspace{9cm}+\text{h.d.} \,,
\\
\left(\mathcal{S}_{\rm Tr[\d\Psi\Pi^{(1)}]}\right)_{(2)_{\rm legs}}(\vk)&=a_2\fnl\left[\d^{(2)}(\vk)+\frac{178}{105}\left(\d^2\right)^{(2)}(\vk)+\frac{4}{35}\G_2^{(2)}(\vk)\right]\int_{\vp}\PshellPsi(p)
\vs
&\quad +a_2\fnl\frac{8}{105}\Tr\left[\Psi\Pi^{[1]}\right](\vk)\int_{\vp}P_{\rm shell}(p)+ \text{h.d.} \,,
\\
\left(\mathcal{S}_{\rm Tr[\Psi\Pi^{(2)}]}\right)_{(2)_{\rm legs}}(\vk)&=a_2\fnl\left[\frac{58}{105}\Tr\left[\Psi\Pi^{[1]}\right](\vk)\right]\int_{\vp}P_{\rm shell}(p)
\vs
&\hspace{-3cm} + a_2\fnl\left[\frac{34}{21}\d^{(2)}(\vk)+\frac{14347}{6027}\left(\d^2\right)^{(2)}(\vk)-\frac{241}{735}\G_2^{(2)}(\vk)\right]\int_p \frac{p^2dp}{2\pi}\PshellPsi(p) +\text{h.d.} \,.
\ea
Notice again that the prefactors exactly match with those calculated in \refeqs{psipioneleg}{psipi2oneleg}.

%%%%%%%%%%%%%%%%%%%%%%%%%%%%%%%%%%%%%%%%%%%%%%%%%%%%%%%%%%%%%%%%%%%%%%%%%%%
%%%%%%%%%%%%%%%%%%%%%%%%%%%%%%%%%%%%%%%%%%%%%%%%%%%%%%%%%%%%%%%%%%%%%%%%%%%

%%%%%%%%%%%%%%%%%%%%%%%%%%%%%%%%%%%%%%%%%%%%%%%%%%%%%%%%%%%%%%%%%%

%%%%%%%%%%%%%%%%%%%%%%%%%%%%%%%%%%%%%%%%%%%%%%%%%%%%%%%%%%%%%%%%%%%%%%%%%%%

%%%%%%%%%%%%%%%%

%%%%%%%%%%%%%%%%%%%%%%%%%%%%%%%%%%%%%%%%%%%%%%%%%%%%%%%%%%%%%%%%%%%%%%%%%%%
%%%%%%%%%%%%%%%%%%%%%%%%%%%%%%%%%%%%%%%%%%%%%%%%%%%%%%%%%%%%%%%%%%%%%%%%%%%

\bibliographystyle{JHEP}
\bibliography{main}

\end{document}